\documentclass[12pt,a4paper]{article}
\usepackage[top=1.25in, bottom=1.25in, left=1.05in, right=1.05in]{geometry}
\usepackage[UKenglish]{babel}
\usepackage[T1]{fontenc}

\usepackage{natbib}
\usepackage{enumitem}
\usepackage{setspace}
\usepackage[section]{placeins}		% floats not to next section

\usepackage{algorithm}
\usepackage{algpseudocode}

\usepackage{amsmath,amssymb,amsfonts,amsthm,bbm,mathtools}
\usepackage[subnum]{cases}	% multi-case equations with separate number for each
\allowdisplaybreaks			% breaks "align" in multiple pages
\makeatletter
\newcommand*{\distas}[1]{\mathbin{\overset{#1}{\kern\z@\sim}}}	
\makeatother
\newcommand*\abs[1]{\left|#1\right|}		% absolute value
\theoremstyle{definition}
\theoremstyle{remark}
\newtheorem*{remark}{Remark}

\usepackage{hyperref}
\hypersetup{unicode=true, colorlinks=true, urlcolor=blue, linkcolor=blue, filecolor=blue, citecolor=blue}
\usepackage{booktabs}

\usepackage{xcolor,epsfig,graphicx} 
\graphicspath{{./figures/}}

\usepackage{longtable,float,array,multirow,multicol,rotating}
\usepackage{threeparttable}

\usepackage[width=0.92\linewidth, font={footnotesize}]{caption}
\usepackage{subcaption}

\definecolor{fireenginered}{rgb}{0.81, 0.09, 0.13}
\definecolor{navyblue}{rgb}{0.0, 0.0, 0.5}
\definecolor{bostonred}{rgb}{0.8, 0.0, 0.0}

\usepackage{dsfont}
\def\I {\mathbb{I}}
\def\Id {\mathbf{I}}
\def\bO {\mathbf{0}}
\def\R {\mathds{R}}

\def\P {\mathbb{P}}

\def\bX {\mathbf{X}}

\def\br {\mathbf{r}}

\def\by {\mathbf{y}}
\def\bx {\mathbf{x}}
\def\bw {\mathbf{w}}

\def\bv {\mathbf{v}}
\def\bz {\mathbf{z}}
\def\balpha   {\boldsymbol{\alpha}}
\def\bbeta    {\boldsymbol{\beta}}
\def\bepsilon {\boldsymbol{\epsilon}}
\def\bzeta    {\boldsymbol{\zeta}}
\def\boldeta  {\boldsymbol{\eta}}

\def\biota    {\boldsymbol{\iota}}
\def\bmu      {\boldsymbol{\mu}}

\def\bvarpi   {\boldsymbol{\varpi}}

\def\btau     {\boldsymbol{\tau}}
\def\bupsilon {\boldsymbol{\upsilon}}
\def\bphi     {\boldsymbol{\phi}}
\def\bpsi     {\boldsymbol{\psi}}
\def\bomega   {\boldsymbol{\omega}}

\def\Sp {\mathds{S}_{++}}
\newtheorem{proposition}{Proposition}
\newtheorem{lemma}{Lemma}

\title{\vspace{-60pt} \textbf{A Bayesian Dynamic Latent Space Model for Weighted Networks}
}

\author{
Roberto Casarin\thanks{Ca' Foscari University of Venice, Italy.
\color{blue}\texttt{r.casarin@unive.it}}
\and
Matteo Iacopini\thanks{Luiss University, Italy.
\color{blue}\texttt{miacopini@luiss.it}}
\and
Antonio Peruzzi\thanks{Ca' Foscari University of Venice, Italy.
\color{blue}\texttt{antonio.peruzzi@unive.it}}
}

\date{\today}

\begin{document}

\maketitle

\onehalfspacing

%%%%%%%%%%%%%%%%%%%%%%%%%%%%%%%%%%%%%%%%%%%%%%%%%%%%%%%%
\begin{abstract}
\begin{singlespace}
A new dynamic latent space eigenmodel (LSM) is proposed for weighted temporal networks. The model accommodates integer-valued weights, excess of zeros, time-varying node positions (features), and time-varying network sparsity. The latent positions evolve according to a vector autoregressive process that accounts for lagged and contemporaneous dependence across nodes and features, a characteristic neglected in the LSM literature. A Bayesian approach is used to address two of the primary sources of inference intractability in dynamic LSMs: latent feature estimation and the choice of latent space dimension. We employ an efficient auxiliary-mixture sampler that performs data augmentation and supports conditionally conjugate prior distributions. A point-process representation of the network weights and the finite-dimensional distribution of the latent processes are used to derive a multi-move sampler in which each feature trajectory is drawn in a single block, without recursions. This sampling strategy is new to the network literature and can significantly reduce computational time while improving chain mixing. To avoid trans-dimensional samplers, a Laplace approximation of the partial marginal likelihood is used to design a partially collapsed Gibbs sampler. Overall, our procedure is general, as it can be easily adapted to static and dynamic settings, as well as to other discrete or continuous weight distributions.

\vspace*{10pt}
\end{singlespace}
\end{abstract}
%%%%%%%%%%%%%%%%%%%%%%%%%%%%%%%%%%%%%%%%%%%%%%%%%%%%%%%%%%%

\clearpage
%%%%%%%%%%%%%%%%%%%%%%%%%%%%%%%%%%%%%%%%%%%%%%%%%%%%%%%%%%%

\section{Introduction}
\label{sec:introduction}

\subsection{Scope and Contribution}

The last two decades have seen an increasing interest in the statistical modelling of network data, leading to the development of several classes of network models \citep{sosa2021review}. Each of these classes captures different aspects of networks, such as hidden node clustering or the factors driving connectivity patterns.
An important class consists of latent space models \citep[LSM,][]{hoff2002latent, hoff2005bilinear, hoff2008modeling}, which have become a widely adopted framework in a broad range of disciplines, including media studies \citep{casarin2025media}, neuroscience \citep{durante2017nonparametric, wang2025establishing}, political science \citep{ParkBA1147, yu2021spatial}, and social science \citep{wang2023joint}.

In an LSM, each node $i \in\{1,\dots,N\}$ is represented by a vector of latent features, $\bx_i \in\R^d$, $d\ll N$, in a low-dimensional Euclidean space. The probability or the strength of a connection between nodes $i$ and $j$ depends on a function $f(\bx_i,\bx_j)$ of the nodes' latent features.
\citet{hoff2002latent} proposed distance-based models, where the probability of a connection is inversely related to the distance between nodes' latent features, and inner-product models, where the connection probability is linked to the angle of the nodes' latent features. Later, \citet{hoff2008modeling} considered eigenmodels\footnote{Eigenmodels can accommodate both positive and negative relationships between entities, unlike distance models. Moreover, although latent distance models may appear simple to understand, the term $f(\bx_i,\bx_j) = \bx_{i}' \bx_{j}$ in eigenmodels is conceptually appealing being interpretable as a random effect.} as a generalization of the latter with interaction probabilities driven by $f(\bx_i,\bx_j) =\bx_i'\Xi\bx_j$, where $\Xi$ is a diagonal matrix encoding the level of \textit{homophily} along each latent dimension (similarity in one latent feature rather than another may imply a higher interaction probability).

Originally, LSMs were designed to model static binary networks, although more general types of network data may be available. Dynamic weighted networks provide a refined picture of the time-varying strength of the relationships among agents. To study this class of data, we propose a novel dynamic latent space eigenmodel for time series of count-valued networks, in which each edge encodes the number of interactions between any two nodes, and we provide a novel and efficient MCMC algorithm for Bayesian inference.

Our proposal is motivated by three relevant issues. First, most existing approaches analyse binary data \citep[e.g., see][]{hoff2005bilinear} whereas weighted networks have received little attention, and efficient estimation schemes are lacking. Furthermore, count networks frequently display an excess of zeros.
Second, single-move Metropolis-Hastings algorithms (MH) for posterior inference on the latent variables are usually implemented for distance-based dynamic LSMs \citep[e.g., see][]{sewell2015latent} which may be highly inefficient.
Third, the complexity and performance of an LSM are driven by the dimension of the latent space, $d$, which is typically fixed or chosen using an information criterion without any uncertainty quantification \citep{loyal2023eigenmodel}.

The main contributions of this article can be summarised as follows. We propose a new dynamic zero-inflated latent space eigenmodel for count networks, a class that has been less explored due to the significant computational challenges posed by the lack of conjugacy in Poisson regression models. We address the computational issues of distance-based dynamic models by adopting a more general eigenmodel specification. Then, we leverage the auxiliary mixture sampler of \citep{fruhwirth2006auxiliary,fruhwirth2009improved} to obtain a conditionally linear Gaussian likelihood. To the best of our knowledge, this is the first time such a scheme has been used in LSM and, more generally, in multivariate count time-series modelling.

Traditional LSMs impose independence between nodes and features (called \textit{node-wise} parametrisation), thereby ignoring important characteristics of real-world data, in which actors may share similar features for geographical, cultural, or other reasons. To account for them, we propose a \textit{feature-wise} parametrisation that allows for contemporaneous dependence among nodes in the latent space.

From a computational perspective, we improve the speed and mixing by sampling directly from the smoothed distribution of node-specific latent features, without loops \citep{mccausland2011simulation}. We leverage the marginal data augmentation of \citet{zens2024ultimate} to enhance the mixing of the probability of structural zeros. A partially collapsed Gibbs sampler \citep{van2008partially} is designed to make inference on the latent space dimension $d$ without requiring any trans-dimensional move. The efficiency of the proposed methodology is assessed via an extensive simulation study.  Three real-life applications to United Nations voting, international trade, and brain connectivity illustrate the potential of our modelling approach. It is worth emphasising that the proposed methodologies and MCMC algorithms can be easily simplified to the static case, which represents an additional contribution of this article.

\subsection{Related literature}

The majority of LSMs in the statistics and machine learning literature are static and analyze a single, binary network.
Over the last decade, several methods have been proposed to investigate temporal networks.
\citet{sarkar2005dynamic} were among the first to design a dynamic LSM for binary networks. Later, \citet{durante2014nonparametric} proposed a Bayesian nonparametric approach to infer latent coordinates evolving in continuous time via Gaussian processes. In a distance-based model, \citet{sewell2015latent,sewell2016latent} used a single-move MH step to infer the path of the time-varying features, whereas \citet{turnbull2023sequential} designed a sequential Monte Carlo procedure. Recently, \citet{pavone2025phylogenetic} developed a phylogenetic latent space model to characterize the generative process of the nodes' feature vectors.

Models for count networks are more scant in the literature. In a static context, \citet{lu2025zeroPois} proposed two zero-inflated Poisson latent space specifications. The former provides a latent-space representation of the matrix of structural zeros, while the latter enables node clustering and offers a strategy for imputing non-observed interactions.
In the context of temporal networks, \citet{artico2023dynamic} introduced a dynamic distance-based LSM for count networks in both continuous- and discrete-time settings. Their approach relies on a likelihood approximation to implement an unscented Kalman filter for the latent features, and on an EM algorithm to estimate static parameters. More recently, \citet{kaur2024latent} use on a Poisson log-linear autoregressive model augmented by an LSM, whereas \citet{casarin2025media} proposed a dynamic Poisson latent distance model featuring Markov-switching dynamics.

Our contribution to this literature is a novel dynamic zero-inflated latent space eigenmodel for time series of count-valued networks (a static model is obtained as a special case). Different from existing approaches, our proposal models the link probability as a function of similarity, which provides a more general framework than a distance-based approach \citep[see][for a proof]{hoff2008modeling}.

One of the main challenges in Bayesian estimation of Poisson regressions is the lack of (conditionally) conjugate prior distributions, which requires tailored MH or Hamiltonian Monte Carlo algorithms. This challenge is exacerbated when a Markovian stochastic process is assumed to drive the Poisson intensity, leading to a nonlinear, non-Gaussian state-space model.
Existing methods based on single-move MH steps to sample the latent intensity path are both computationally intensive and potentially slow-mixing.

Recently, \citet{king2025warped} proposed a semiparametric approach to model multivariate time series of counts, by discretising a nonlinear transform of a latent Gaussian dynamic linear model.
One of the main drawbacks of this approach is a computational bottleneck arising from sampling latent variables from a multivariate Gaussian distribution truncated to an $N$-dimensional hypercube, which becomes increasingly severe as the probability mass in the region diminishes.
\citet{d2023efficient} proposed a P{\'o}lya-Gamma data augmentation scheme for Poisson regression, leveraging the limiting connection between the Negative Binomial and the Poisson distribution. The resulting approximation is simple, but its computational cost can be very high. A further approximation step is introduced; as highlighted by the authors, the accuracy and computational speed nevertheless depend heavily on tuning parameters that are difficult to set.
An alternative approach has been developed by \citet{fruhwirth2006auxiliary} and \citet{fruhwirth2009improved}, who introduced the improved auxiliary mixture sampler (IAMS) as a two-step data augmentation strategy for Poisson regression models. Theoretical properties of the sampler and possible improvement have been discussed in \citet{gardini2026note}.
The advantage of this approach is that the final augmented model has a linear-Gaussian (conditional) likelihood, allowing conjugate Gaussian prior distributions to be defined over the model's parameters, which is particularly appealing for high-dimensional multiple-equation models.

The choice of the latent dimension $d$ has been a crucial issue in the literature \citep[e.g., see][]{handcock2007model, lu2025zeroPois}. Some works assume a fixed small $d$ (e.g., $d = 2$ or $d = 3$) to enable a visual representation of the latent features \citep{gollini2016joint, sewell2016latent}, whereas others use an information criterion to select the optimal value \citep[e.g., see][]{handcock2007model}.
However, methods to properly estimate $d$ have been considered only recently.
\citet{loyal2025spike} proposed to infer the latent space dimension by specifying a static latent space eigenmodel. Then, they specified an upper bound on the dimension of the latent space, $\underline{d}$, and assumed a spike-and-slab prior distribution for each diagonal element of $\Xi$, with a spike component corresponding to a Dirac mass at zero and a shrinking sequence of slab probabilities to induce a stochastically decreasing ordering of the prior probability of $d$. However, a closed-form expression for the prior distribution of $d$ is not easily obtained.
A point estimate of the latent space dimension is indirectly obtained as the mode of the posterior distribution of the allocation variable introduced to resolve the spike-and-slab mixture.
Instead, \citet{gwee2025latent} proposed a way to shrink the latent dimensions of a static distance-based LSM by exploiting a multiplicative gamma process prior for the precision of the latent features. Despite its effectiveness, this methodology is highly sensitive to prior hyperparameters and computationally expensive for large networks. Furthermore, this strategy requires thresholding the variance of the latent coordinates to obtain an estimate of $d$.

\subsection{Structure and notation}

The rest of the article is organised as follows. The new model specifications are introduced and described in Section~\ref{sec:model}.
Then, Section~\ref{sec:inference} presents Bayesian inference and posterior approximation and briefly describe the performance of the proposed methods through simulated experiments which hare extensively covered in Section C of the Supplement. Real-world dataset applications are in Section~\ref{sec:application}. Section~\ref{sec:conclusion} concludes and discusses possible extensions. The proofs of the results are reported in Section A of the Supplement.

In the following, lowercase letters denote scalars, boldface lowercase letters denote (column) vectors, and uppercase letters denote matrices.
For a column vector $\bx \in \R^m$, let $\bx'$ be its transpose and $\|\bx\| = \|\bx\|_2 = (\sum_{i=1}^m x_i^2)^{1/2}$ the $L_2$-norm.
Moreover, $X = \operatorname{diag}(\bx) \in\R^{m\times m}$ denotes a diagonal matrix having the elements of the vector $\bx$ on the main diagonal.
Let $\Sp^k$ be the set of all $k \times k$ positive definite matrices and let $\I(x \in S)$ be the indicator function taking value $1$ if $x\in S$ and $0$ otherwise. Let $\Id_k$ denote the $k$-dimensional identity matrix and $\mathbf{O}_{n \times m}$ the $n \times m$ null matrix, $\biota_k$ and $\mathbf{0}_k$ the $k$-dimensional unit and null vectors, respectively. Finally, we denote with $\delta_{c}(x)$ a Dirac mass at $c$.
We denote with $\mathcal{P}oi(\cdot\mid \lambda)$ the Poisson distribution with intensity $\lambda \in\R_+$, with $\mathcal{N}(\cdot\mid\mu, \sigma^2)$ the Gaussian distribution with mean $\mu\in\R$ and variance $\sigma^2 \in\R_+$, with $\mathcal{N}_k(\cdot\mid \boldsymbol{\mu}, \Sigma)$ the multivariate Gaussian distribution with mean vector $\boldsymbol{\mu} \in\R^k$ and covariance matrix $\Sigma \in \Sp^k$, with $\mathcal{MN}_{k,p}(\cdot \mid M, \Sigma, \Upsilon)$ the matrix-variate Gaussian distribution with mean matrix $M \in\R^{k \times p}$, row covariance matrix $\Sigma \in \Sp^k$ and column covariance matrix $\Upsilon \in \Sp^p$.

%%%%%%%%%%%%%%%%%%%%%%%%%%%%%%%%%%%%%%%%%%%%%%%%%%%%%%%%%%%
\section{A Dynamic Zero-Inflated Counts Eigenmodel}
\label{sec:model}

\subsection{Model}

Consider an observed time series of counts, $\{ Y_t \}_{t\geq 0}$, with $Y_t \in \mathbb{N}^{N\times N}$ being a count-valued matrix.
We propose a dynamic latent space model for $Y_t$, which is based on the following specification for each $(i,j)$th entry of $Y_t$:
\begin{subequations}
\begin{align}
    \label{eq:model_yt}
    y_{ij,t} \mid \lambda_{ij,t}, z_{ij,t} & \overset{ind}{\sim} p_{ij,t}(z_{ij,t}) \mathcal{P}oi\big( y_{ij,t} \mid \lambda_{ij,t} \big) + (1-p_{ij,t}(z_{ij,t})) \delta_{\{0\}}(y_{ij,t}) \\
    \label{eq:model_lambdat}
    \log(\lambda_{ij,t}) & = \alpha_i + \alpha_j + \bx_{i:,t}' \Xi \bx_{j:,t} \\
    \label{eq:model_xt}
    X_{t} & = \widetilde\Phi X_{t-1}\Phi' + H_{t} \qquad H_t \overset{iid}{\sim} \mathcal{MN}_{N, d}(H_{t}\mid\mathbf{O}_{N, d}, \widetilde\Upsilon, \Upsilon) \\
    \label{eq:model_zt}
    z_{ij,t} & = \bbeta_i'\bv_{i:,t} + \bbeta_j'\bv_{j:,t} + \upsilon_{ij,t}, \qquad 
    \upsilon_{ij,t} \overset{iid}{\sim} \mathcal{N}(\upsilon_{ij,t}\mid0,1),
\end{align}
\end{subequations}
where $p_{ij,t}(z_{ij,t}) = \P(z_{ij,t} > 0)$, and $z_{ij,t}$ corresponds to the latent utility variable in the utility function formulation of the probit \citep{albert1993bayesian}, $\bv_{i:,t} \in\R^L$ is a vector of node-specific observed covariates, $\bx_{i:,t} \in \R^d$ is a vector of node-specific unobserved dynamic features, $X_t = (\bx_{1:,t}, \dots, \bx_{N:,t})' \in \R^{N \times d}$  is a matrix of latent variables and $\Xi = \operatorname{diag}(\xi_1,\dots,\xi_d) \in\R^{d\times d}$ is a diagonal matrix of parameters. The inner product in eq.~\eqref{eq:model_lambdat} captures the cosine similarity and postulates that an edge between two nodes has a higher (count) weight, the more similar their features are.
We assume $\Xi = \Id_d$ and discuss possible extensions to a real-valued diagonal matrix in Section \ref{sec:conclusion}.
Eq.~\eqref{eq:model_xt} is a matrix autoregressive process \citep{chen2021autoregressive}, where $\widetilde\Phi\in\R^{N\times N}$ and $\widetilde\Upsilon\in\Sp^N$ are respectively the row autoregressive and row covariance matrices, while $\Phi\in\R^{d\times d}$ and $\Upsilon\in\Sp^d$ are respectively the column autoregressive and column covariance matrices. A matrix representation of eqs. \eqref{eq:model_yt}-\eqref{eq:model_zt} is available in Section A of the Supplement.

The latent space model in equations \eqref{eq:model_yt} to \eqref{eq:model_xt} represents a nonlinear and non-Gaussian state space model.
It belongs to the class of latent eigenmodels, which includes latent class and latent distance models as special cases \citep{hoff2008modeling}. As a result, it can represent both stochastic equivalence (i.e., nodes can be divided into groups such that members of the same group have similar patterns of relationships) and homophily (i.e., relationships between nodes with similar characteristics are stronger), owing to the fact that it provides a low-rank approximation to the sociomatrix, and is therefore able to represent a wide array of patterns in the data. This specification is more tractable and computationally more efficient than a Gaussian Process or other non-parametric specifications.

Finally, eq.~\eqref{eq:model_zt} is the random utility representation of a probit model for the (prior) probability of sampling from a random Poisson term \citep[e.g., see][]{zens2024ultimate}.

\subsection{Data Augmentation}
We resolve the mixture of the zero-inflated Poisson in eq.~\eqref{eq:model_yt} by introducing the auxiliary allocation variable $w_{ij,t} = \I(z_{ij,t} > 0)$, such that $p(y_{ij,t}, w_{ij,t} \mid \lambda_{ij,t}, z_{ij,t}) = p(y_{ij,t} \mid w_{ij,t} ,\lambda_{ij,t}) \times p(w_{ij,t} \mid z_{ij,t})$ is equal to
\begin{align}
    &p(y_{ij,t} \mid w_{ij,t}, \lambda_{ij,t}) = \left( \mathcal{P}oi\big( y_{ij,t} \mid \lambda_{ij,t} \big)\right)^{w_{ij,t}}\left(  \delta_{\{0\}}(y_{ij,t})\right)^{1-w_{ij,t}} \label{eq:prior_wt1},\\  &p(w_{ij,t} \mid z_{ij,t}) = p_{ij,t}(z_{ij,t})^{w_{ij,t}}(1-p_{ij,t}(z_{ij,t}))^{1-w_{ij,t}}.
    \label{eq:prior_wt2}
\end{align}
For $w_{ij,t} = 1$, one obtains the Poisson latent eigenmodel. One of the main issues in Poisson regression models is the lack of conjugate priors for the latent intensity parameters, $\lambda_{ij,t}$, which induces a computational bottleneck. This issue is further exacerbated in dynamic settings, where the entire path $\{ \lambda_{ij,t} \}_{t=1}^T$ must be estimated.

To address this issue, we exploit the improved auxiliary mixture sampler (IAMS) introduced by \citet{fruhwirth2009improved} and \citet{fruhwirth2006auxiliary} to transform our nonlinear non-Gaussian observation equation into a (conditionally) linear Gaussian one.
This approach is a double data augmentation scheme that exploits the property of the Poisson distribution, with intensity $\lambda_t$, which is the distribution of the total number of jumps of a Poisson process with the same intensity over the unit interval $[0,1]$.

\begin{lemma}
\label{lemma:auxiliary_mixture}
Define $\btau_{ij,t} = (\tau_{ij, 1t}, \tau_{ij, 2t})$ and $\br_{ij,t} = (r_{ij, 1t},r_{ij, 2t})$ and assume $w_{ij,t}=1$, then the Poisson distribution in eq.~\eqref{eq:model_yt} is the marginal distribution of $p(y_{ij,t}, \btau_{ij,t}, \br_{ij,t} \mid w_{ij,t}=1,  \bx_{i:,t},\bx_{j:,t}, \alpha_i,\alpha_j)$, which can be approximated  by $q(y_{ij,t}, \btau_{ij,t}, \br_{ij,t} \mid d, w_{ij,t}=1, \bx_{i:,t},\bx_{j:,t},\alpha_i,\alpha_j)$, defined as: % for the complete data likelihood of the term $y_{ij,t}$:
\begin{align}
    % \notag
	\label{eq:complete_likelihood_yijt}
    & g_1(\tau_{ij,1t})\prod_{k=1}^{R(y_{ij,t})} c_k^{\I(r_{ij,1t}=k)} 
    \Big( g_2(\tau_{ij,2t})\prod_{k=1}^{R(y_{ij,t})} c_k^{\I(r_{ij,2t}=k)} \Big)^{\I(y_{ijt}>0)},
\end{align}
where $g_1(\tau_{ij,1t})$ and $g_2(\tau_{ij,2t})$ are the densities of $\mathcal{N}( \! -\log(\tau_{ij, 1t}) \mid \log(\lambda_{ij,t}) + \mu_{r_{ij,1t}}(1), \, \sigma^2_{r_{ij,1t}}(1))$ and $\mathcal{N}( \! -\log(\tau_{ij, 2t}) \mid \log(\lambda_{ij,t}) + \mu_{r_{ij,2t}}(y_{ij,t}), \, \sigma^2_{r_{ij,2t}}(y_{ij,t}))$, respectively, with $\mu_{r_{ij,st}}(\cdot)$, $\sigma^2_{r_{ij,st}}(\cdot)$, $s=1,2$, and $c_k(\cdot)$ the auxiliary mixture locations, scales, and weights.
\end{lemma}

The optimal values of the location, scale, and weights of the auxiliary mixture are tabulated in \citet{fruhwirth2006auxiliary} and \citet{fruhwirth2009improved}. Let us define the collections of auxiliary variables $\btau = \{ \btau_{ij,t} \}_{ijt}$ and $\br = \{ \br_{ij,t} \}_{ijt}$, the collections of parameters and latent variables $\balpha = (\alpha_1,\dots,\alpha_N)'$ and $\bx = (\bx_{::,1}',\dots,\bx_{::,T}')'$ respectivelly with $\bx_{::,t} = \operatorname{vec}({X_t})$. Let
$\mathcal{Q}_t = \{(i,j): i\in\{1,\dots,N\}, \: j\in\{i+1,\dots,N\}, w_{ij,t} = 1\}$ be the set of edges allocated to the Poisson component. From eq.~\eqref{eq:prior_wt1}, \eqref{eq:prior_wt2}, \eqref{eq:complete_likelihood_yijt} and eq.~\eqref{eq:model_yt}, we obtain the complete data likelihood
\begin{equation}
\begin{aligned}
    p(\by, \btau, \br, \bw \mid \bx,\balpha)  = & \prod_{t=1}^{T}\prod_{(i,j)\in\mathcal{Q}_t}  p(y_{ij,t}, \btau_{ij,t}, \br_{ij,t} \mid w_{ij,t}, \bx_{i:,t},\bx_{j:,t}, \alpha_i,\alpha_j) \\
    & \quad \times \prod_{(i,j)\notin\mathcal{Q}_t} \delta_{\{0\}}(y_{ij,t}) \: \times \prod_{i<j}  p(w_{ij,t}).
	\label{eq:complete_likelihood}
\end{aligned}
\end{equation}

\subsection{Parametrisation}

The $N \times d$-dimensional Gaussian matrix autoregressive process in eq.~\eqref{eq:model_xt} allows for contemporaneous dependence between nodes and features, but is overparametrised. To address the dimensionality issue while maintaining meaningful interpretation of the model's results, we propose two parsimonious parameterisations.

A specification frequently adopted in the literature assumes independence across nodes and  postulates that the node-specific latent features, $\bx_{i:,t} = (x_{i,1,t},\dots,x_{i,d,t})' \in\R^d$ in the $i$th row of $X_t$, follow a vector autoregressive process,
\begin{align}
    \bx_{i:,t} \mid \bx_{i:,t-1}, \Phi,\Upsilon & \overset{ind}{\sim} \mathcal{N}_d(\bx_{i:,t} \mid \Phi \bx_{i:,t-1}, \Upsilon).
    \label{eq:model_xt_node}
\end{align}
This construction follows from the restriction $\widetilde\Phi = \Id_N$ and $\widetilde\Upsilon = \Id_N$ in eq.~\eqref{eq:model_xt} and we label it as \textit{node-wise} specification \citep[see also][among the others]{sewell2015latent,sarkar2005dynamic}. Correlation among the latent features via $\Upsilon$ is undesirable, as we would expect the latent features to be as orthogonal as possible to maximise the information content of the latent space and avoid redundancy. For this reason, the LSM literature usually assumes $\Upsilon = \sigma^2 \Id_d$. 

We argue that this specification is inconsistent with the properties of real-world networks. Nodes may exhibit similar features (and thus be more likely to connect) for intrinsic reasons not captured by an LSM model with independent factors. For instance, people in the same household are more likely to share similar (latent) features due to biological and cultural heritage \citep[see][for a similar argument]{pavone2025phylogenetic}. Thus, we propose a \textit{feature-wise} specification that allows for arbitrary contemporaneous dependence across nodes. It assumes a feature-specific vector autoregressive process $\bx_{:\ell,t} = (x_{1,\ell,t},\dots,x_{N,\ell,t})' \in\R^N$ for the $\ell$th column of $X_t$, and independence across features, that is
\begin{align}
   \bx_{:\ell,t} \mid \bx_{:\ell,t-1}, \widetilde\Phi,\widetilde\Upsilon & \overset{ind}{\sim} \mathcal{N}_N(\bx_{:\ell,t} \mid \widetilde\Phi \bx_{:\ell,t-1}, \widetilde\Upsilon).
\label{eq:model_xt_feature}
\end{align}
This parametrisation is obtained from eq.~\eqref{eq:model_xt} by assuming $\Phi = \Id_d$ and $\Upsilon = \Id_d$. By assuming dependence, the effects of the omitted common factors can be captured by the covariance matrix $\widetilde\Upsilon$.

We propose using the \textit{feature-wise} specification and, in the following section, provide a numerical comparison of the two methods.
It is worth emphasising that the difference between the specifications in eq.~\eqref{eq:model_xt_feature} and eq.~\eqref{eq:model_xt_node} is made only at the level of the state transition distribution. In either case, posterior inference is performed by sampling the entire trajectory over time of each node-specific vector, $\bx_{i:,:} = (\bx_{i:,1}',\dots,\bx_{i:,T}')' \in\R^{dT}$, as described in the next section.
This result is obtained by first assuming that the vector stacking all the nodes and features at each time, $\bx_{::,t}$, has a joint Gaussian distribution (conditionally on its past $\bx_{::,t-1}$), then deriving the joint prior for the entire trajectory, $\bx$, and finally obtaining the joint prior of $\bx_i$ conditionally on the remaining $\bx_{-i:,:}$.
An advantage of this sampling strategy, coupled with a joint sampling of the entire temporal path, is the design of a sampler that samples $N$ times from a $dT$-dimensional distribution, where $d \ll N$, making it the most efficient choice compared to alternative samplers for the latent vectors.

%%%%%%%%%%%%%%%%%%%%%%%%%%%%%%%%%%%%%%%%%%%%%%%%%%%%%%%%%%%
\section{Bayesian Inference}
\label{sec:inference}

\subsection{Parameter prior specification}

We assume a multivariate Gaussian prior distribution for the intercepts $\balpha \sim \mathcal{N}_N( \balpha \mid \mathbf{0}_N, \underline{\sigma}^2_\alpha \Id_N)$, and $\bbeta_i \overset{iid}{\sim} \mathcal{N}_L( \bbeta_i \mid  \mathbf{0}_L, \underline{\sigma}^2_\beta \Id_L)$, $i=1,\dots,N$, and a multivariate Gaussian prior for the initial latent vector $\bx_{::,0} \sim \mathcal{N}_{dN}(\bx_{::,0} \mid \underline{\bx}_{::,0}, \underline{\Omega}_0)$.
Concerning the parameters driving the dynamics of the latent features, we assume multivariate Gaussian priors for the non-zero elements of $\widetilde\Phi$ and $\Phi$, that is
\begin{equation}
   \widetilde\bphi = \operatorname{vec}(\widetilde\Phi) \sim \mathcal{N}_{N^2}(\widetilde\bphi \mid \underline{\bmu}_{\widetilde\bphi}, \underline{\Sigma}_{\widetilde\bphi}), \qquad
   \bphi = \operatorname{vec}(\Phi) \sim \mathcal{N}_{d^2}(\bphi \mid \underline{\bmu}_{\bphi}, \underline{\Sigma}_{\bphi}),
\end{equation}
in the \textit{feature-wise} and \textit{node-wise} specifications, respectively. Finally, for the covariance matrix, we assume a different prior distribution in the two cases. Specifically, an inverse Wishart prior is assumed for the low-dimensional $\Upsilon$ in the \textit{node-wise} case, $\Upsilon \sim \mathcal{IW}_d( \Upsilon \mid \underline{\nu}, \underline{\Omega})$.
Instead, for the high-dimensional $\widetilde\Upsilon$ in the \textit{feature-wise} case, we define the auxiliary variables $\rho_{\Omega}$ and $\boldsymbol{\varpi} = \{\varpi_{ij}: 1 \leq i \leq j \leq N\}$ and assume a graphical horseshoe prior \citep{li2019graphical} for $\widetilde\Omega = \widetilde\Upsilon^{-1}$, that is
\begin{align}
    p(\widetilde\Omega, \rho_{\widetilde\Omega}, \boldsymbol{\varpi}) & \propto \mathcal{C}^+(\rho_{\widetilde\Omega} \mid 0,1) \prod_{i<j} \mathcal{N}(\omega_{ij} \mid 0, \varpi_{ij}^2 \rho_{\widetilde\Omega}^2) \mathcal{C}^+(\varpi_{ij} \mid 0,1) \times \I(\widetilde\Omega \in\Sp^N),
\end{align}
where $\mathcal{C}^+$ denotes the half Cauchy distribution. We then exploit the mixture representation of the latter to perform data augmentation via the introduction of the inverse gamma distributed auxiliary variables $\eta_{ij}^\varpi$ and $\eta^\rho$
{\footnotesize
\begin{align*}
    \varpi_{ij}^2 \mid \eta_{ij}^\varpi & \overset{ind}{\sim} \mathcal{IG}\Big(\varpi_{ij}^2 \mid  \frac{1}{2}, \frac{1}{\eta_{ij}^\varpi} \Big),\,\,
    \eta_{ij}^\varpi  \overset{iid}{\sim} \mathcal{IG}\Big( \eta_{ij}^\varpi \mid \frac{1}{2}, 1 \Big), \,\,
    \rho^2_{\widetilde\Omega} \mid \eta^\rho  \sim \mathcal{IG}\Big(\rho^2_{\widetilde\Omega} \mid \frac{1}{2}, \frac{1}{\eta^\rho} \Big), \,\,
    \eta^\rho  \sim \mathcal{IG}\Big(  \eta^\rho \mid \frac{1}{2}, 1 \Big).
\end{align*}}%
The advantage of the graphical horseshoe prior over the traditional inverse Wishart is that the former allows shrinking irrelevant precision elements to zero, thereby enabling valid inference in high-dimensional settings, such as for $\widetilde\Upsilon$. Conversely, since $d \ll N$, this approach is not necessary for $\Upsilon$.

Regarding the latent space dimension $d$, we choose a prior distribution and derive a posterior distribution that fully quantifies uncertainty. This is in stark contrast to the current literature, which relies on information criteria to select $d$ post-estimation.
As a prior, we assume a discrete distribution supported on $\mathcal{D} = \{ 1,\dots,\underline{d} \}$, for a pre-specified upper bound on the latent space dimension, $\underline{d}$. Besides our default choice, the discrete uniform prior $\mathcal{U}(d \mid \mathcal{D})$, one may consider other distributions, such as a truncated geometric distribution, $\mathcal{TG}eo(d\mid \theta, \mathcal{D})$.

\subsection{Posterior sampling}
The posterior distribution is approximated by Monte Carlo methods. In this section, we propose a partially collapsed Gibbs sampler \citep[PCG,][]{van2008partially} that iterates over the following steps: A) sample the latent space dimension and features \textit{jointly} from $p(d,\bx,\bx_{::,0} \mid \by,\btau,\br,\bw,\balpha,\Phi,\Upsilon)$; B) sample the latent variables and parameters given the space dimension and features from $p(\btau,\br,\bw,\balpha,\Phi,\Upsilon \mid \by,\bx,\bx_{::,0}, d)$.

Standard approaches to parameter tuning, such as tuning the dimension of the parameter space $d$ via the Gibbs sampler, require designing trans-dimensional moves that enable the chain to transition between parameter spaces of different sizes.
However, constructing such moves is very difficult in complex models such as ours, leading to computationally expensive samplers with poor mixing. We circumvent these issues by first deriving a factorization of the full-conditional joint posterior of $d$, $\mathbf{x}$, and $\bx_{::,0}$  in A), and then by applying Lemma \ref{lemma:auxiliary_mixture} and Laplace approximation.

\begin{proposition}
\label{proposition:joint_posterior_dx}
The joint posterior distribution of the latent space dimension and latent features factorises as $p(d,\bx, \bx_{::,0}  \mid \by,\btau,\br,\bw,\balpha,\Phi,\Upsilon) = p(d \mid \by,\btau,\br,\bw,\balpha,\Phi,\Upsilon) p(\bx, \bx_0 \mid d,\btau,\br,\bw,\balpha,\Phi,\Upsilon)$ and the first term approximates as
\begin{equation}
\begin{aligned}
    p(d & \mid \by,\btau,\br,\bw,\balpha,\Phi,\Upsilon) \approx \frac{q(\by, \btau,\br \mid d, \bw, \balpha,\Phi,\Upsilon)p(d)}{\sum_{\ell=1}^{\underline{d}} q(\by, \btau,\br \mid \ell, \bw, \balpha,\Phi,\Upsilon)p(\ell)},
\end{aligned}\label{eq:prop1}
\end{equation}
where $\mathcal{Q}_t$ has been defined in Section \ref{sec:model} and $\log q(\by, \btau,\br \mid d,\bw,\balpha, \Phi,\Upsilon)$ is defined as
\begin{equation}
\begin{aligned}
    & \!\sum_{t=1}^T \sum_{(i,j)\in\mathcal{Q}_t} \log q(y_{ij,t}, \btau_{ij,t}, \br_{ij,t} \mid d, w_{ij,t}, \widehat{\bx}_{i:,t},\widehat{\bx}_{j:,t},\alpha_i,\alpha_j) \\
    & + \sum_{t=1}^T \log p\left( \widehat{\bx}_{::,t} \mid \widehat{\bx}_{::,t-1}, d, \Phi,\Upsilon \right) + \log p\left( \widehat{\bx}_{::,0} \mid  d \right)- \frac{d(T+1)N}{2}\log(Q^*),
\end{aligned}
\end{equation}
with $\widehat{\bx}_{i:,t}$ being a point of high posterior mass and $Q^* = \sum_{t=1}^T |\mathcal{Q}_t| + \sum_{(i,j)\in\mathcal{Q}_t} \I(y_{ij,t}>0)$.
\end{proposition}

Proposition~\ref{proposition:joint_posterior_dx} requires using a point of high posterior mass for the latent features, $\widehat{\bx}$, to perform a Laplace approximation. Since the latent features exhibit temporal dependence, we propose to use the output of the Kalman smoother as values for $\widehat{\bx}_{i:,t}$, $i=1,\dots,N$ and $t=1,\dots,T$.
Computationally, this approach would require running $\underline{d}$ Kalman smoothers at each iteration of the Gibbs sampler, one for each candidate value of the latent space dimension.
Once a new value of $d$ is drawn from eq.~\eqref{eq:prop1}, we sample $\bx_{:\ell,:}$, $\ell=1,\dots,d$, from the smoothed distribution (i.e., the full conditional posterior in the static case). Unfortunately, for large networks, both single-move MH and forward-filtering backward-sampling approaches \citep{fruhwirth1994data} are computationally too expensive, preventing the use of dynamic latent space models. We propose instead to draw trajectories of each node's latent features, all without a loop \citep[AWOL,][]{mccausland2011simulation}. This strategy significantly reduces the computing time and improves the mixing. 
We show that this sampling step can incorporate different prior dependence structures (\textit{node-wise} and \textit{feature-wise}), exploiting the assumption of joint Gaussian distribution and conditioning. This result is stated in the following proposition.

\begin{proposition}
\label{proposition:sampling_x}
Assume the latent space dimension, $d$, is given, and the quantities $\overline{\bz}_{i:,:}$, $\overline{\bx}_{i:,:}$, $K_i$, $\overline{K}_i$, $ G_i$, and $\widetilde\Sigma_i$ as in the Supplement. Then, one obtains the following:
\begin{enumerate}[label=(\roman*)]
    \item Under the \emph{node-wise} prior specification in Eq.~\eqref{eq:model_xt_node} and assuming the initial value for the process to be $\bx_{i:,0} \sim \mathcal{N}_d(\bx_{i:,0} \mid \underline{\bx}_{i:,0}, \Omega_0)$, the posterior (smoothed) distribution for $\bx_{i:,:}$ is $\bx_{i:,:} \mid d, \btau, \br, \balpha, \bx_{-i:,:}, \bx_{i:,0}, \Phi, \Upsilon \overset{ind}{\sim} \mathcal{N}_{dT}\big( \bx_{i:,:} \mid \widehat{\bx}_{i:,:}, \widehat{K}_i^{-1} \big)$, where $\widehat{K}_i = K_i + G_i' \widetilde\Sigma_i^{-1} G_i$ and $\widehat\bx_{i:,:} = \widehat{K}_i \big( K_i \overline{\bx}_{i:,:} + G_i' \widetilde\Sigma_i^{-1} \overline{\bz}_{i:,:} \big)$.
    
    \item Under the \emph{feature-wise} prior specification in Eq.~\eqref{eq:model_xt_feature} and assuming the initial value for the process to be $\bx_{:\ell,0} \sim \mathcal{N}_N(\bx_{:\ell,0} \mid \underline{\bx}_{:\ell,0}^*, \Omega_0^*)$, the posterior (smoothed) distribution for $\bx_{i:,:}$ is $\bx_{i:,:} \mid d, \btau, \br, \balpha, \bx_{-i:,:}, \bx_{i:,0}, \widetilde\Phi,\widetilde\Upsilon \overset{ind}{\sim} \mathcal{N}_{dT}\big( \bx_{i:,:} \mid \widehat{\bx}_{i:,:}, \widehat{K}_i^{-1} \big)$, where $\widehat{K}_i = \big(\overline{K}_i + G_i' \widetilde\Sigma_i^{-1} G_i \big)^{-1}$, $\widehat{\bx}_{i:,:} = \widehat{K}_i \big(\overline{K}_i \overline{\bx}_{i:,:} + G_i' \widetilde\Sigma_i^{-1} \overline{\bz}_i \big)$.
\end{enumerate}
\end{proposition}

Thanks to the sparse structure of $\widehat{K}_i$ and $\widehat{K}_i^*$, we can efficiently sample exactly $\bx_{i:,:}$, $i=1,\dots,N$, from their posterior (smoothed) distribution distribution $p(\bx, \bx_{::, 0} \mid d,\btau,\br,\bw,\balpha,\Phi,\Upsilon)$ , avoiding time loops.

Summarising, by Propositions~\ref{proposition:joint_posterior_dx} and~\ref{proposition:sampling_x}, sampling from $p(d \mid\by,\btau,\br,\bw,\balpha,\Phi,\Upsilon)$ and $p(\bx, \bx_{::, 0} \mid d,\btau,\br,\bw,\balpha,\Phi,\Upsilon)$ used in A) is performed by drawing $d$ and $\bx_{i:,:}$ with probability 
\begin{align}
    \frac{q(\by, \btau,\br \mid d, \bw, \balpha,\Phi,\Upsilon)p(d)}{\sum_{\ell=1}^{\underline{d}} q(\by, \btau,\br \mid \ell, \bw, \balpha,\Phi,\Upsilon)p(\ell)}, 
    \quad
    \mathcal{N}_{dT}\big( \bx_{i:,:} \mid \widehat{\bx}_{i:,:}, \widehat{K}_i^{-1} \big).
\label{MCMC:sample_d}
\end{align}
and $ \mathcal{N}_{Nd}\big( \bx_{::,0} \mid \widehat{\bx}_{::,0}, \widehat{K}_0^{-1} \big)$.
These steps must be performed in this order since $\bx$ is integrated out from the (marginal) posterior of $d$, and together these two steps represent a single draw from the joint full conditional posterior $p(d,\bx \mid \cdot)$.

The remaining parameters in step B) are sampled exactly from their full conditional distributions. Specifically, the distributions of the node fixed effects $p(\balpha \mid \btau,\br, \bw,\bx)$ and of the latent autoregressive coefficient matrix under the two specifications, $p(\bphi \mid \bx,\bx_{::,0}, \Upsilon)$ and $p(\widetilde\bphi \mid \bx,\bx_{::,0}, \widetilde\Upsilon)$ are:
\begin{equation}
    \mathcal{N}_N( \balpha \mid \overline{\bmu}_\alpha, \overline{\Sigma}_{\alpha}), \quad
    \mathcal{N}_{d^2}\big(\bphi \mid \overline{\bmu}_{\Phi}, \overline{\Sigma}_{\Phi} \big), \quad\text{and}\quad
    \mathcal{N}_{N^2}\big(\widetilde\bphi \mid \overline{\bmu}_{\widetilde\Phi}, \overline{\Sigma}_{\widetilde\Phi} \big).
\label{eq:full_phi}
\end{equation}
Note that the sampling scheme of the node fixed effects $\balpha$ is performed jointly instead of element-wise, leading to an improved mixing compared to the standard LSM literature (see Section A.5 of the Supplement for further details).

Under the \textit{feature-wise} specification and the graphical horseshoe prior assumption, the precision matrix $\widetilde\Omega = \widetilde\Upsilon^{-1}$ has a full conditional distribution proportional to
\begin{equation}
    \big|\widetilde\Omega\big|^{\frac{Nd}{2}} \exp\bigg( \!\! -\operatorname{tr}\Big(\frac{1}{2} S \widetilde\Omega \Big) \!\bigg) \prod_{i<j} \exp\bigg( \!\!-\frac{\omega_{ij}^2}{2 \varpi_{ij}^2 \rho_{\widetilde\Omega}^2} \bigg) \I(\widetilde\Omega \in\Sp^N)
\label{eq:post_gh_1}
\end{equation}
and it is updated one column/row at a time. The full conditional distributions $p(\varpi_{ij} \mid \eta_{ij}^\varpi, \omega_{ij}, \rho_{\widetilde\Omega})$ and $p(\eta_{ij}^\varpi \mid \varpi_{ij})$ of the prior hyperparameters are:
\begin{align}
    \mathcal{IG}\bigg( \varpi_{ij} \mid 1, \frac{1}{\eta_{ij}^\varpi} + \frac{\omega_{ij}^2}{2\rho_{\widetilde\Omega}^2} \bigg), \quad\text{and}\quad
    \mathcal{IG}\bigg( \eta_{ij}^\varpi \mid 1, 1 + \frac{1}{\varpi_{ij}^2} \bigg),
\label{eq:post_gh_2}
\end{align}
respectively, and those for $\rho_{\widetilde\Omega}^2$ and $\eta^\rho$ are obtained in an analogous way. Under the \textit{node-wise} specification and the conjugate inverse Wishart prior assumption, the full conditional of the scale matrix $\Upsilon$ is $\mathcal{IW}_{d}(\Upsilon \mid \overline{\Psi}_\Upsilon, \overline{\nu}_\Upsilon)$. If $\Upsilon$ is a diagonal matrix and $\Upsilon = \operatorname{diag}(\bupsilon)$ and $\underline{\Psi}_\Upsilon=\operatorname{diag}(\underline{\mathbf{b}})$ in the inverse Wishart prior, as commonly assumed in many applications, one obtains $\upsilon_\ell \sim \mathcal{IG}(\upsilon_\ell \mid \underline{a}_\upsilon, \underline{b}_\upsilon)$, $\ell=1,\dots,d$, and $p(\upsilon_\ell \mid \bx,\bx_0,\Phi) \propto \mathcal{IG}(\upsilon_\ell \mid  \overline{a}_\upsilon, \overline{b}_{\upsilon,\ell}),$ where $\overline{a}_\upsilon = \underline{a}_\upsilon + \frac{TN}{2}$ and $\overline{b}_{\upsilon,\ell} = \underline{b}_\upsilon + \frac{\overline{\Psi}_{\Upsilon,\ell\ell}}{2}$, with $\overline{\Psi}_{\Upsilon,\ell\ell}$ the $\ell$th diagonal element of the matrix $\overline{\Psi}_\Upsilon$ previously defined.

The full conditional $p(w_{ij,t} \mid y_{ij,t}, \bx_{i:,t},\bx_{j:,t}, \alpha_i,\alpha_j, z_{ij,t})$ of the zero-inflated mixture allocation variables is the Bernoulli distribution $\mathcal{B}ern(p_{ij,t}^*)$ with:
\begin{align}    p_{ij,t}^* = \frac{p_{ij,t} \mathcal{P}oi(0 \mid \lambda_{ij,t})}{(1-p_{ij,t}) + p_{ij,t} \mathcal{P}oi(0 \mid \lambda_{ij,t})}.
\label{eq:full_w}
\end{align}
The full conditionals of the latent utility $z_{ij,t}$ and of its coefficients $\beta_i$ are:
\begin{align}
    \!\!\!&z_{ij,t} = \mu_{ij,t}+ F_u^{-1}\big( w_{ij,t} + \kappa_{ijt}(1-w_{ij,t} -F_u(\mu_{ij,t}) \big), \quad\text{and}\quad 
    \mathcal{N}\big( \beta_i \mid \overline{m}_{\beta,i},\overline{v}_{\beta,i} \big),
\label{eq:full_z}
\end{align}
where $\kappa_{ij,t} \sim \mathcal{U}(0,1)$. Moreover, to improve the mixing of the latent utility sampler, we introduce the double expansion suggested by \citet{zens2024ultimate}, as detailed in the Supplement.

The auxiliary variables full conditional  distribution $p\big( \btau_{ij, t},\br_{ij,t} \mid y_{ij,t},w_{ij,t},\bx,\balpha \big)$ is
\begin{align}
    p\big( \tau_{ij, 1t} \mid \tau_{ij, 2t}, y_{ij,t},w_{ij,t}, \bx,\balpha \big)  p\big( \tau_{ij, 2t} \mid y_{ij,t},w_{ij,t} \big)\prod_{k=1}^{\min\{y_{ij,t}+1,2\}} \!  p\big( r_{ij, kt} \mid \tau_{ij, kt}, \bx,\balpha \big),
\label{eq:full_tau_r}
\end{align}
which requires to sample in the order $\tau_{ij, 2t}$, $\tau_{ij, 1t}$, then $(r_{ij, 1t}, r_{ij, 2t})$. From their conditional distributions $p(\tau_{ij,2 t} \mid y_{ij,t},w_{ij,t},\bx,\balpha)$ and $p(\xi_{ij, t} \mid \bx,\balpha)$ given by
\begin{align}
     & \mathcal{B}e\big( \tau_{ij, 2t} \mid y_{ij,t}, 1 \big), \quad
     \mathcal{E}xp\big( \xi_{ij,t} \mid \lambda_{ij,t} \big), \quad
     \tau_{ij, 1t}  = (1 + \xi_{ij,t}) - \tau_{ij, 2t}\mathbb{I}(y_{ij,t} > 0)
\label{eq:full_tau_r1} 
\end{align}
and $p(r_{ij,\ell t}=k \mid y_{ij,t},w_{ij,t},\bx,\balpha, \btau_{ij,t})$, $\ell=1,2$, proportional to
\begin{align}
     & c_k(1) \mathcal{N}\big( -\log(\tau_{ij, 1t}) -\log(\lambda_{ij,t}) \mid \mu_k(1), \sigma_k^2(1) \big),\\
     & c_k(y_{ij,t}) \mathcal{N}\big( -\tau_{ij, 2t} -\log(\lambda_{ij,t}) \mid \mu_k(y_{ij,t}), \sigma_k^2(y_{ij,t}) \big),
\label{eq:full_tau_r5}
\end{align}
respectively, the hyperparameters can be tabulated outside the main loop of the MCMC, speeding up the computations.

\begin{remark}
The proposed model and MCMC algorithm can be easily modified to handle other data types, such as binary and categorical, when the observation model is a logit, binomial, multinomial logit, and negative binomial. The data augmentation strategy in \citet{fruhwirth2009improved} yields a conditionally linear Gaussian state-space model. Positive real-valued data can be handled by introducing a Gaussian data augmentation, as in \citet{sewell2016latent}. The conditional distributions of $\btau,\br$ change according to the observation distribution, whereas the full conditionals of the latent process are Gaussian. 
\end{remark}
%%%%%%%%%%%%%%%%%%%%%%%%%%%%%%%%%%%%%%%%%%%%%%%%%%%%%%%%%%%

We conduct simulation experiments to assess the sampler's performance under the prior specification detailed in Section C.2 of the Supplement. A set of experiments demonstrates the accuracy of our sampler in recovering the ground-truth parameters under the true data-generating process in the presence of model misspecification. The second set of experiments shows the accuracy of our sampler for different sample sizes and proportions of zeros. A comparison between the IAMS-based sampler and a well-established alternative in the literature shows that the former exhibits better mixing performance, as measured by effective sample size. See Section C.3 and Section C.4 in the Supplement for further details.

%%%%%%%%%%%%%%%%%%%%%%%%%%%%%%%%%%%%%%%%%%%%%%%%%%%%%%%%%%%
\section{Real-data Applications}
\label{sec:application}
This section presents applications of the proposed method to three temporal network datasets: UN co-voting, international trade, and brain connectivity. The illustration demonstrates our model's ability to provide new evidence on latent space dimension node feature dynamics and structural zeros.

\subsection{UN Co-voting Networks}
We apply our model to a dynamic co-voting network constructed from the United Nations General Assembly roll-call voting dataset \citep{votingdata}. We consider a time series of $T = 21$ networks from 2004 to 2024, as there is minimal variation in country names over this timeframe. Each node represents one of the top-50 countries in terms of GDP, and each edge weight represents the occurrence of aligned votes (``yes''--``yes'', ``no''--``no'', ``abstain''--``abstain'') between country $i$ and country $j$ in year $t$. Voting networks are of interest because they provide a measure of bilateral relations, enabling both historical analysis and forecasting \citep[see][]{lauderdale2010unpredictable,kim2023dynamic}.

The posterior distribution of $d$ is highly concentrated at $\widehat{d} = 2$, and the evidence supports non-trivial network topologies with time-varying communities and islands (panel (a) in Figure~D.11 of the Supplement). The latent feature in circular projections in 2014 (left) and 2024 (right), (see also Section D.1 in the Supplement), shown in Fig. \ref{fig:plane_time_polMain}, exhibits two distinct clusters: the first comprises \emph{NATO} and \emph{EU} countries, and the second comprises \emph{BRICS} countries. The cluster composition is changing slowly, with the US progressively realigning with the rest of \emph{NATO} and Argentina changing its orientation in 2023-2024, as its foreign policy realigned towards the US and \emph{NATO} (red circles and green dots). The posterior estimate of the cross-country correlation (right) provides evidence of further blocking structure in the nodes that is finer than the \emph{NATO}-\emph{BRICS} bi-polarity. For instance, the position alignment of the China–Iran–Russia block is tighter than the alignment of other \emph{BRICS} members. There are other distinct clusters of dynamics: the EU and the Latin American–Asian group.

\begin{figure}[t]
\centering
\begin{tabular}{ccc}
\footnotesize 2014 & \footnotesize 2024  & \\
\includegraphics[width=0.3\textwidth]{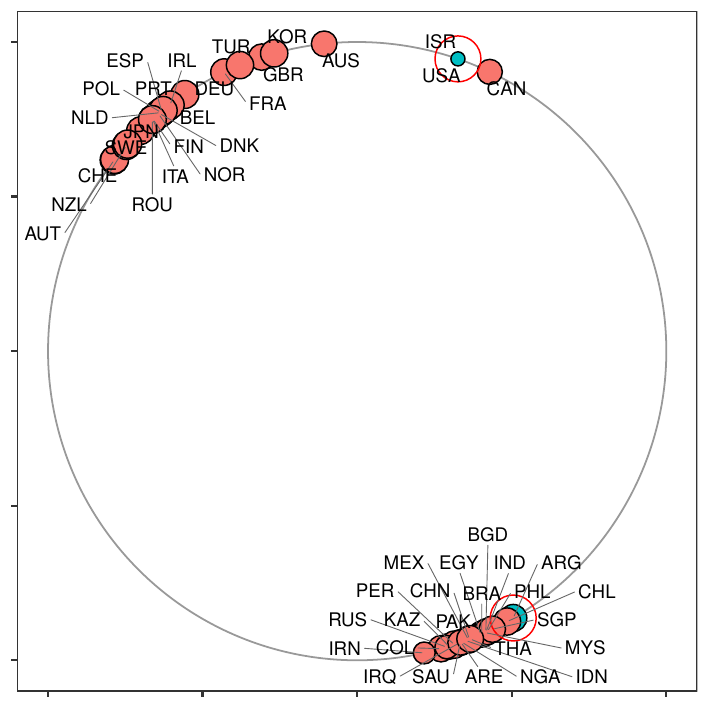} &
\includegraphics[width=0.3\textwidth]{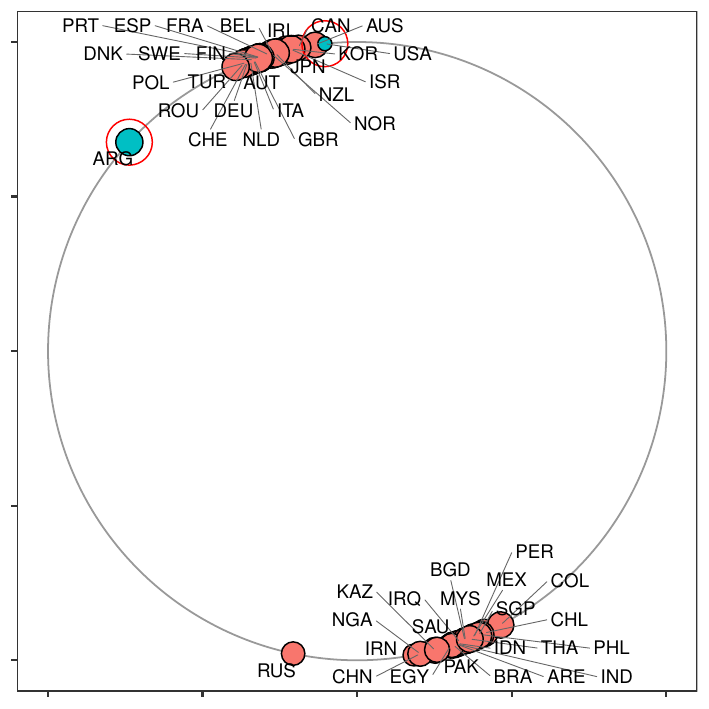}
&\includegraphics[width=0.3\textwidth]{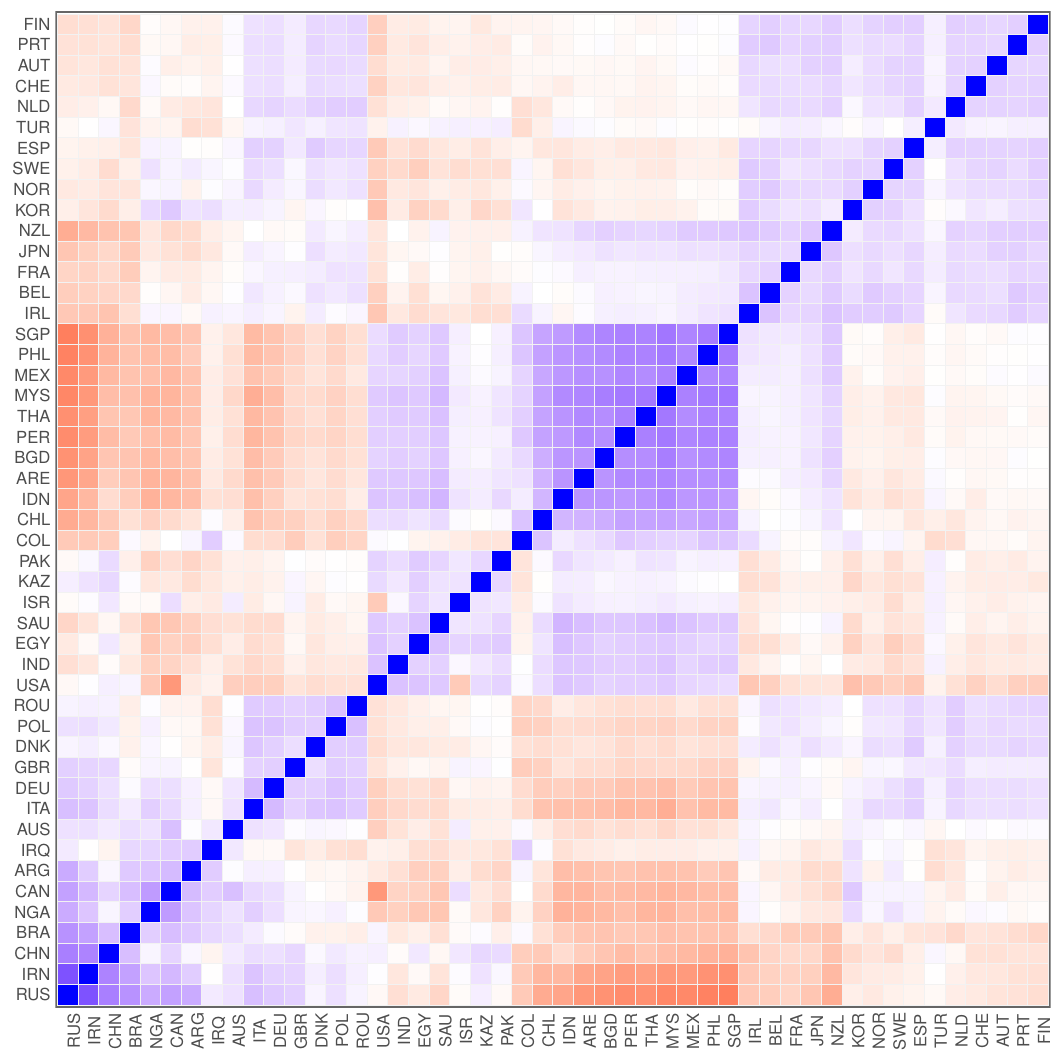}
\end{tabular}
\caption{UN General Assembly co-voting network. Circular projection of the latent space representation for the years 2014 and 2024 (left and middle), and node correlation posterior mean (right).}
\label{fig:plane_time_polMain}
\end{figure}

%%%%%%%%%%%%%%%%%%%%%%%%%%%%%%%%%%%%
\subsection{Trade Networks}
In this section, we apply our zero-inflated latent space model to a dynamic trade network constructed from the \textit{CEPII--BACI} dataset of international trade flows \citep{de2014network}. We consider a time series of $T = 21$ trade networks from 2004 to 2024. In these networks, each node represents one of the top-50 countries by GDP, and each edge weight represents the number of distinct HS2 product categories exchanged between countries $i$ and $j$ in year $t$. Considering trade variety rather than its economic value is of interest because it offers a different perspective on bilateral relations between countries. This measure can be considered a proxy for the extensive margin of trade \citep{hummels2005variety}, reflecting the diversity and strength of bilateral exchange relationships in the global trade system \citep{gemmetto2016multiplexity}.

\begin{figure}[t]
\centering
\setlength{\tabcolsep}{1pt}   % less horizontal 
\begin{tabular}{cc}
\includegraphics[width=0.45\textwidth]{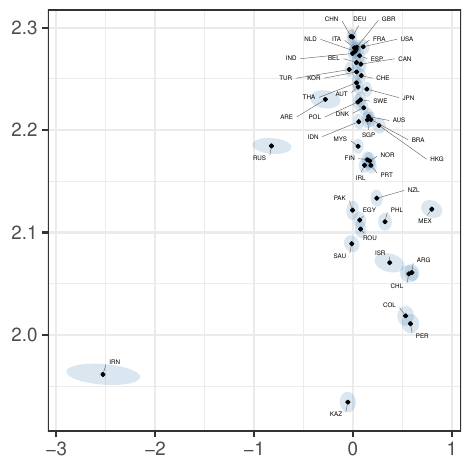}&
\includegraphics[width=0.45\textwidth]{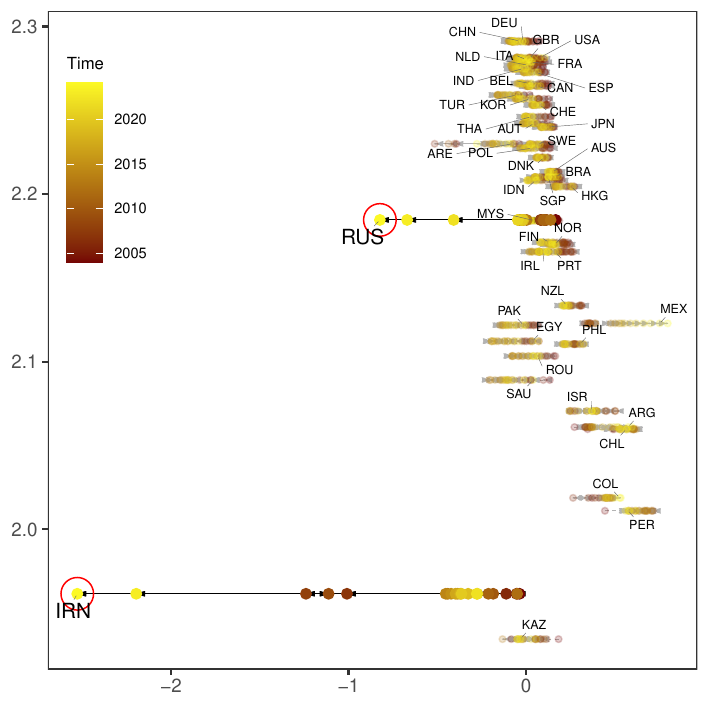}
\end{tabular}
\caption{Trade network. Latent space representation in the $(\bx_{:1,t},\balpha)$-plane (left) in 2024 with 95\% credible ellipses which are reported in blue, and position temporal evolution (right), with lighter node colours denoting more recent years.}
\label{fig:plane_timeMain}
\end{figure}

The model estimates a latent space dimension $\widehat{d} = 1$. For this reason, we represent the countries on the $(\bx_{:1,t}, \balpha)$-plane. The left panel of Figure \ref{fig:plane_timeMain} reports the $(\bx_{:1,t}, \balpha)$-plane representation for the year 2024. Two countries have features pointing in the same horizontal direction if they trade many goods with one another relative to their total number of goods exchanged, and/or if their trade with other countries is similar. Countries with higher $\alpha_i$ are globally more connected. This enables the identification of ``core'' countries, with many edges, and ``peripheral'' countries, more sparsely connected. The core, composed of European countries, the United States, and a few other advanced economies, is clearly identified and stable (top side of the plot). Other countries discount some degree of isolation in world trade and have peripheral positions in the plane. Other clusters emerge, such as the one related to Latin American countries (Argentina, Chile, Colombia, Mexico, and Peru).  There is also evidence of heterogeneity in the node trajectories. For example, from 2022 onward, Russia exhibits a displacement away from the core group, possibly reflecting trade disruptions and geopolitical fragmentation related to the conflict in Ukraine, while Iran recently moved far apart after some alternating periods, also in this case discounting geopolitical tensions (red circles in the right panel of Figure \ref{fig:plane_timeMain}).

%\FloatBarrier
%%%%%%%%%%%%%%%%%%%%%%%%%%%%%%%%

\subsection{Brain Networks}
We illustrate the effectiveness of our dynamic zero-inflated latent space model in capturing the structural connectivity patterns of the human brain. Some works focused on fMRI connectivity data (e.g., see \citealp{wang2025establishing}). We consider networks extracted from Diffusion Tensor Imaging (DTI) data. Latent Space models have previously shown their potential in modelling DTI-inferred network structure. \citet{durante2017nonparametric} provide a cross-sectional binary LSM to infer a common latent structure from a population of subjects, while \citet{aliverti2019spatial} proposes a cross-sectional brain-region LSM clustering. Instead, we consider a single individual and perform information pooling for the latent coordinates by leveraging the temporal dimension.

Our analysis is based on diffusion MRI data provided by the NeuroData repository (\url{https://neurodata.io/mri/}). Specifically, we focus on one anonymized subject, \texttt{0025982}, from the control group of the study by \citet{lu2011focal}. The subject underwent five diffusion MRI scanning sessions, each approximately 6 months apart, yielding a sequence of brain networks. The brain is segmented using the Schaefer 200 atlas \citep{schaefer2018local}, which partitions the cerebral cortex into 200 anatomical regions. Each region is a node in the brain network, while edges represent the strength of white-matter connectivity between pairs of regions.

Connectivity weights correspond to the number of white-matter fibers between pairs of regions. This representation yields a weighted, undirected adjacency matrix with count weights, in which structural zeros indicate the absence of detected fiber tracts. Such data naturally motivate the use of a zero-inflated modelling framework that can distinguish between true anatomical absences and sampling zeros arising from limited tractography sensitivity.

\begin{figure}[t]
\centering
\captionsetup{width=0.95\linewidth}
\setlength{\abovecaptionskip}{-15pt}
\begin{tabular}{cc}
    \multicolumn{2}{c}{
    \begin{tabular}{ccc}
    \centering
    \includegraphics[width=0.25\textwidth]{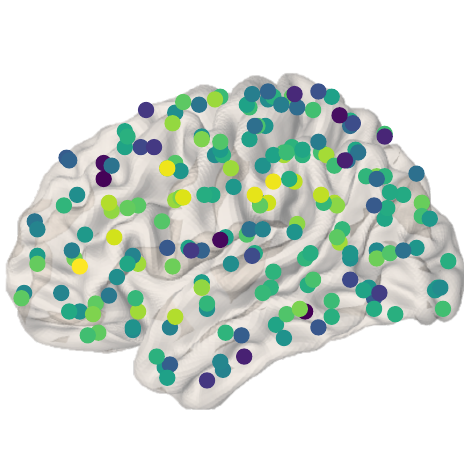} &
    \includegraphics[width=0.25\textwidth]{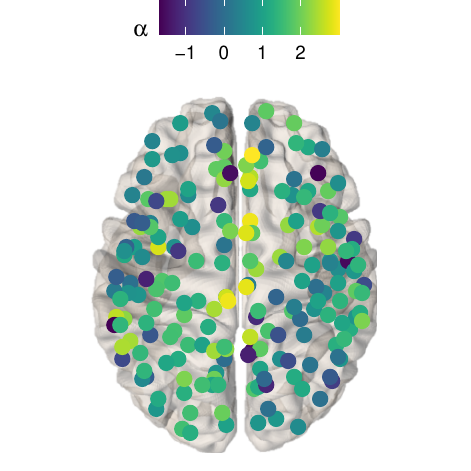}&
    \includegraphics[width=0.25\textwidth]{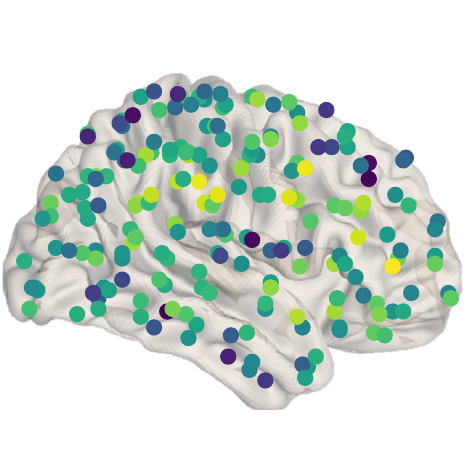}
    \end{tabular}
    }
    \\
    \includegraphics[width=0.4\textwidth, trim = 0cm -2cm 0cm 0cm, clip]{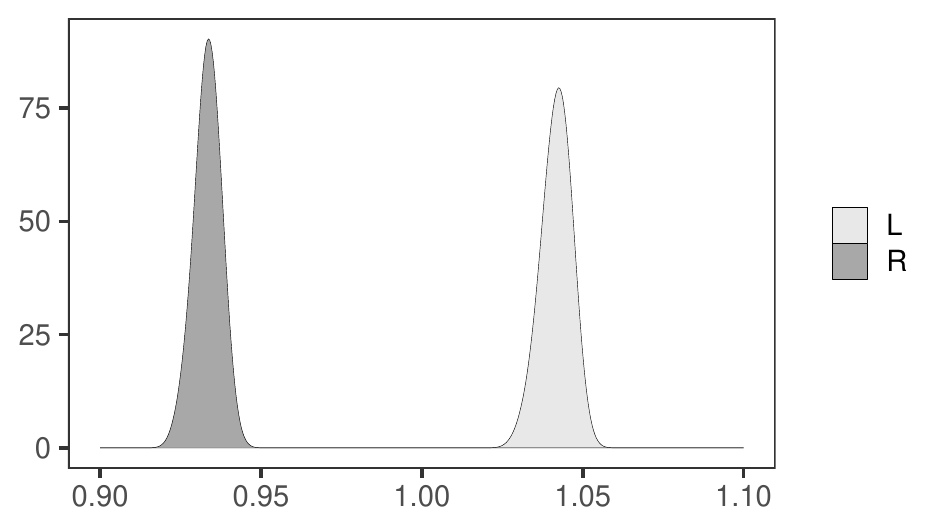} & \hspace*{-20pt}
    \includegraphics[width=0.4\textwidth, trim = 2cm 1cm 0cm 2cm, clip]{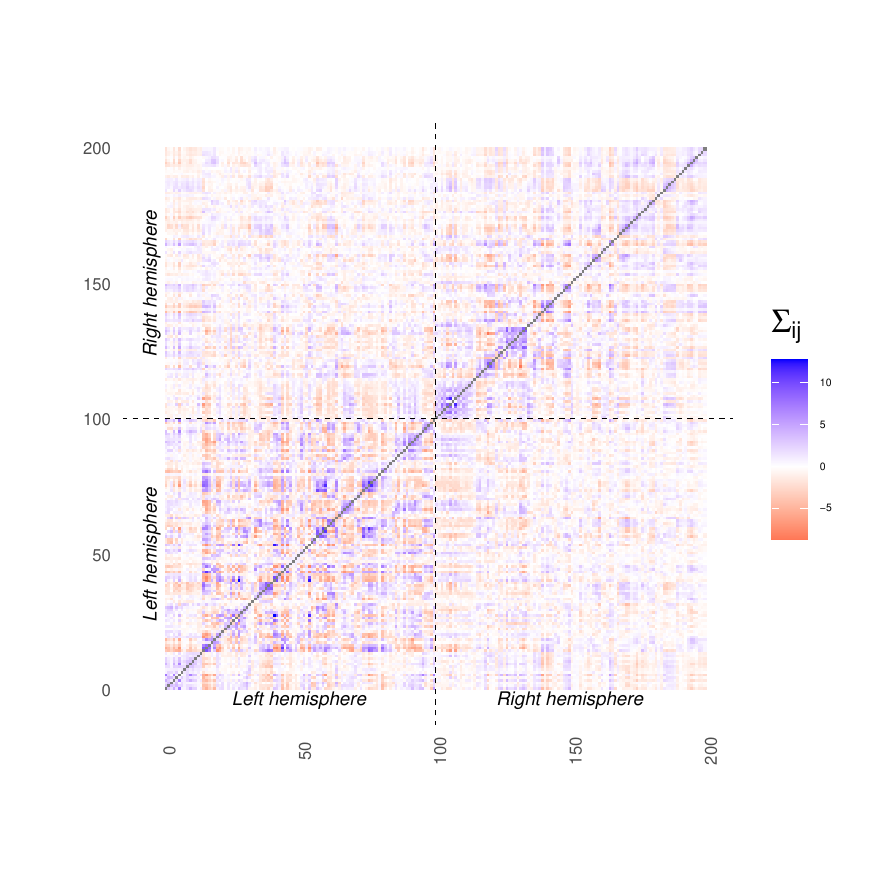}
\end{tabular}
\caption{Spatial and distributional characteristics of the parameter estimates.
Top: Brain regions (Schaefer-200 atlas) coloured by node-specific posterior mean $\hat{\alpha}_i$.
Bottom left: Posterior distribution of the average effects over left ($\sum_{i=1}^{100}\alpha_i/100$, light gray) and right hemisphere nodes ($\sum_{i=101}^{200}\alpha_i/100$, dark gray).
Bottom right: $\hat{\Sigma}$, estimated covariance matrix among $\bx_{:\ell, t}$, with dashed lines separating hemispheres.
}
\label{fig:alpha_brain_combined}
\end{figure}
Figure~\ref{fig:alpha_brain_combined} (top panel) reports a graphical representation of the Schaefer 200 brain regions coloured according to the posterior mean of the corresponding individual effect. 

\begin{figure}[t]
\centering

\captionsetup{width=0.95\linewidth}
\setlength{\tabcolsep}{2pt} 
\renewcommand{\arraystretch}{0.1} 
    \resizebox{0.9\textwidth}{!}{
    \begin{tabular}{ccc}
    \multicolumn{2}{c}{\scriptsize (a) Latent Coordinates} & \scriptsize (b) Structural Zeros\\
\includegraphics[width=0.35\textwidth]{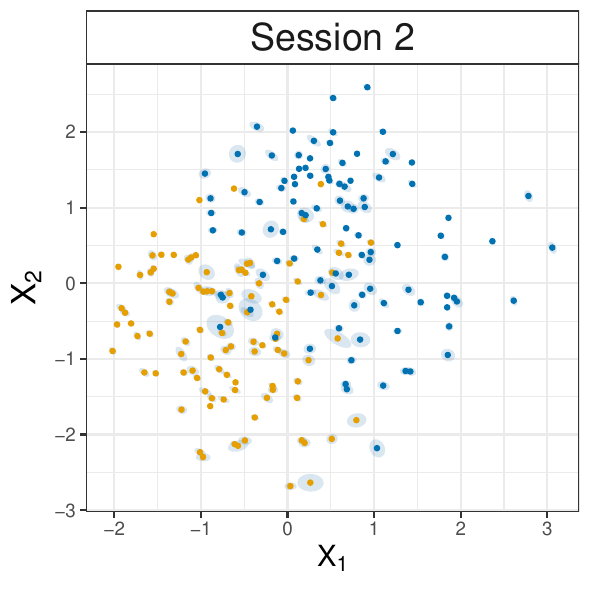} &
\includegraphics[width=0.35\textwidth]{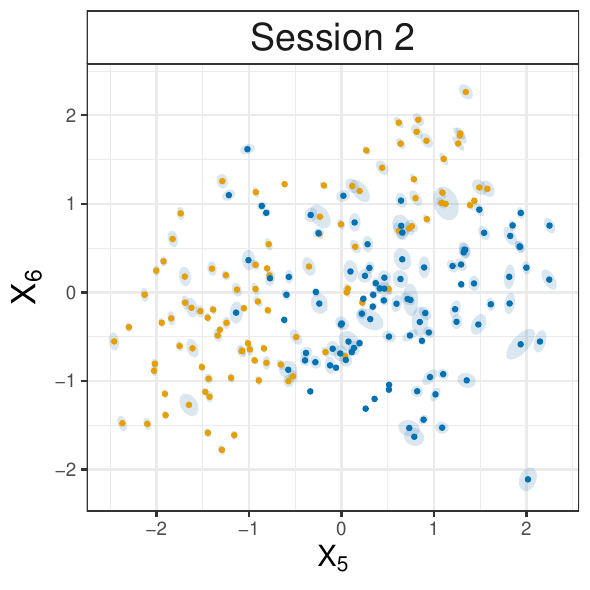} &
\includegraphics[width=0.35\textwidth]{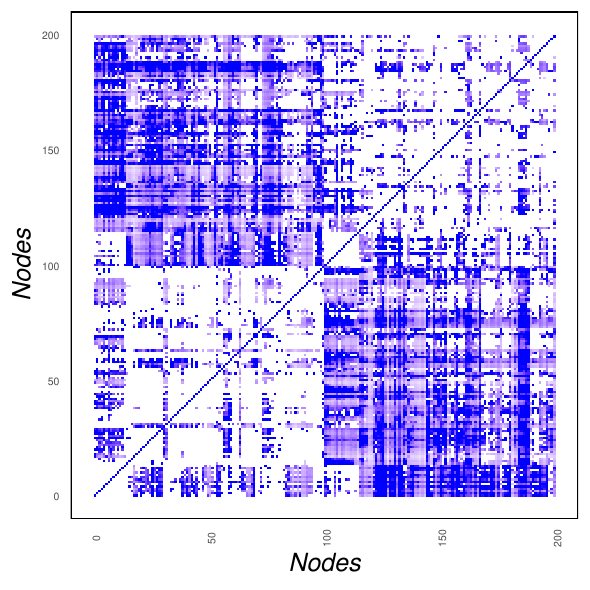} \\
\includegraphics[width=0.35\textwidth]{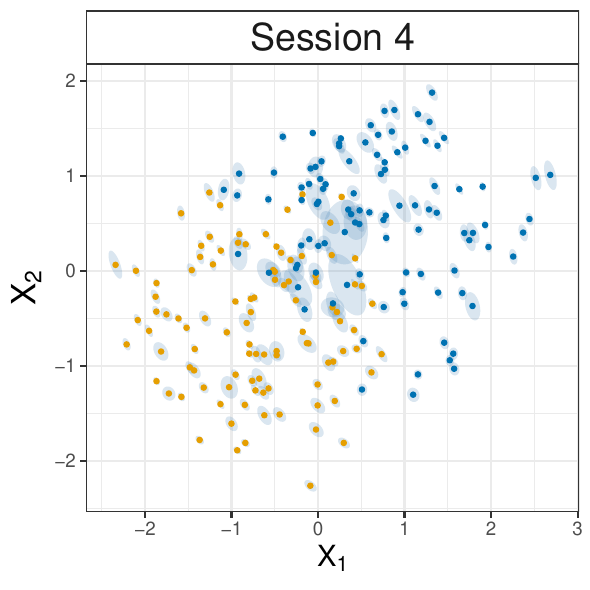} &
\includegraphics[width=0.35\textwidth]{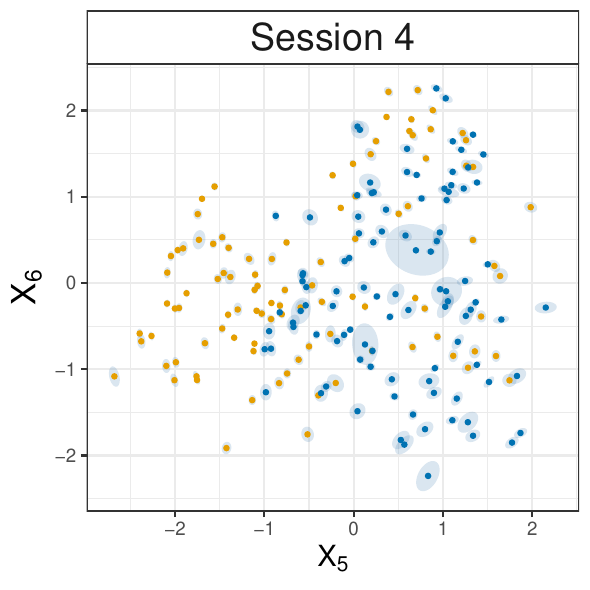} &
\includegraphics[width=0.35\textwidth]{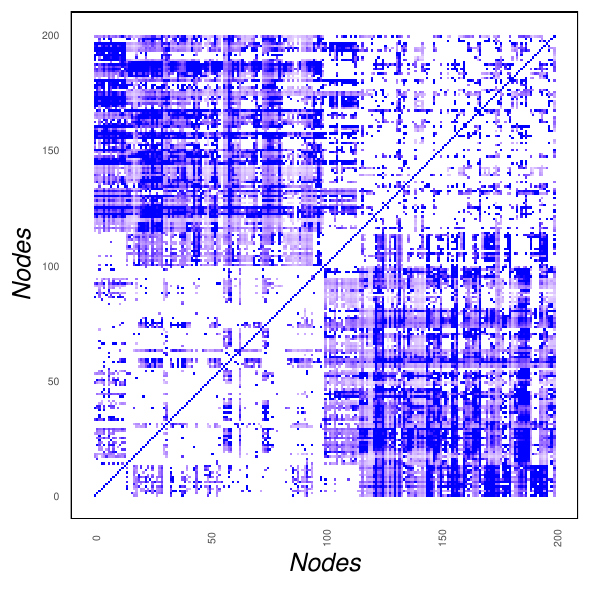} \\
\end{tabular}}
    \caption{Results across five MRI sessions (rows). 
    (a) Latent space representations. Estimated coordinate pairs $(\hat{x}_{i,1,t},\hat{x}_{i,2,t})$ and $(\hat{x}_{i,5,t},\hat{x}_{i,6,t})$  (left and middle columns) for a subsample $t=2,4$. Each point is a brain region as defined by Schaefer's 200 atlas: left hemisphere (orange) and right hemisphere (blue).
    (b) Posterior probability of structural zeroes, $\P(z_{ij,t}\leq 0\mid\mathbf{y})$. A darker colour denotes a higher probability of a structural zero.}
    \label{fig:LS_brain}
\end{figure}

Functional brain areas are characterized by high connectivity, i.e., large $\alpha_i$ values, where each $\alpha_i$ can be interpreted as the relative centrality of the brain regions within the network. The area with the highest connectivity in the middle panel corresponds to the motor cortices. The bottom-left panel reports the posterior distribution of the average of the individual effects for each hemisphere, i.e., $\overline{\alpha}_L = \frac{1}{100} \sum_{i=1}^{100} \alpha_i$ and $\overline{\alpha}_R = \frac{1}{100} \sum_{i=101}^{200} \alpha_i$. Regions in the left hemisphere are more central $\overline{\alpha}_L > \overline{\alpha}_R$, which is consistent with existing evidence on hemispheric network asymmetries \citep{iturria2011brain}. Our model allows us to study fluctuations in hemispheric asymmetry for each individual, as discussed below, thereby extending the evidence from the static framework. The estimated covariance matrix $\hat{\Sigma}$ (bottom-right) indicates that fluctuations across regions are dependent, with clustering effects (block-wise patterns), stronger correlations within hemispheres (red  and blue), and weaker correlations between hemispheres (white).  Panel (a) of Figure \ref{fig:LS_brain} reports the latent space representation along 4 of the 6 latent coordinates estimated over 5 scanning sessions. In each session, the separation between the left and right hemispheres (orange and blue, respectively) aligns with the data, with more closely connected brain regions showing higher white fiber counts. Our model also captured significant temporal fluctuations while filtering out noise due to measurement errors, which is appealing for monitoring the evolution of lesioned circuits and for studying patient recovery.

Different information on fluctuations in hemispheric connectivity is obtained from the structural zeros (panel b). The vast majority of zeros in the observed connectivity matrices are structural, reflecting genuine anatomical separations between the two hemispheres, whereas a relatively small fraction is attributed to random fluctuations (Poisson component), suggesting that these arise from measurement variability or noise. 

%\FloatBarrier
%%%%%%%%%%%%%%%%%%%%%%%%%%%%%%%%%%%%%%%%%%%%%%%%%%%%%%%%%%%
\section{Conclusion}
\label{sec:conclusion}

We proposed a new dynamic eigenmodel for count-valued temporal networks. The model accounts for excess zeros and dynamic latent features via alternative contemporaneous covariance structures. Our proposal leverages a data-augmentation scheme for the Poisson likelihood to obtain a conditionally Gaussian state-space representation. We developed a Bayesian inference procedure for the latent space dimension using a partially collapsed Gibbs sampler, which avoids trans-dimensional moves. The approach relies on a Laplace approximation of the marginal likelihood and performs well in terms of accuracy and computing time. Real-data applications demonstrate the methods' effectiveness in capturing dynamic features of the network connectivity.
Extensions of the proposed model that are worth investigating include taking into account homophily (a matrix $\Xi$ with real diagonal elements), overdispersion (negative binomial distribution), incomplete network (missing edge forecasting in space/time), and mixed-frequency data.

%%%%%%%%%%%%%%%%%%%%%%%%%%%%%%%%%
\bibliographystyle{chicago}   % chicago
\bibliography{biblio}

@article{kaur2024latent,
  title={A latent space model for multivariate count data time series analysis},
  author={Kaur, Hardeep and Rastelli, Riccardo},
  journal={arXiv preprint arXiv:2408.13162},
  year={2024}
}

@article{zens2024ultimate,
  title={Ultimate {P{\'o}lya Gamma} samplers -- {Efficient} {MCMC} for possibly imbalanced binary and categorical data},
  author={Zens, Gregor and Fr{\"u}hwirth-Schnatter, Sylvia and Wagner, Helga},
  journal={Journal of the American Statistical Association},
  volume={119},
  number={548},
  pages={2548--2559},
  year={2024},
  publisher={Taylor \& Francis}
}

@article{hoff2005bilinear,
  title={Bilinear mixed-effects models for dyadic data},
  author={Hoff, Peter D},
  journal={Journal of the American Statistical Association},
  volume={100},
  number={469},
  pages={286--295},
  year={2005},
  publisher={Taylor \& Francis}
}

@article{li2019graphical,
  title={The graphical horseshoe estimator for inverse covariance matrices},
  author={Li, Yunfan and Craig, Bruce A and Bhadra, Anindya},
  journal={Journal of Computational and Graphical Statistics},
  volume={28},
  number={3},
  pages={747--757},
  year={2019},
  publisher={Taylor \& Francis}
}

@article{pavone2025phylogenetic,
  title={Phylogenetic latent space models for network data},
  author={Pavone, Federico and Durante, Daniele and Ryder, Robin J},
  journal={arXiv preprint arXiv:2502.11868},
  year={2025}
}

@article{durante2014nonparametric,
  title={Nonparametric {Bayes} dynamic modelling of relational data},
  author={Durante, Daniele and Dunson, David B},
  journal={Biometrika},
  volume={101},
  number={4},
  pages={883--898},
  year={2014},
  publisher={Oxford University Press}
}

@article{sewell2015analysis,
  title={Analysis of the formation of the structure of social networks by using latent space models for ranked dynamic networks},
  author={Sewell, Daniel K and Chen, Yuguo},
  journal={Journal of the Royal Statistical Society Series C: Applied Statistics},
  volume={64},
  number={4},
  pages={611--633},
  year={2015},
  publisher={Oxford University Press}
}

@article{fruhwirth1994data,
  title={Data augmentation and dynamic linear models},
  author={Fr{\"u}hwirth-Schnatter, Sylvia},
  journal={Journal of Time Series Analysis},
  volume={15},
  number={2},
  pages={183--202},
  year={1994},
  publisher={Wiley Online Library}
}

@article{van2008partially,
  title={Partially collapsed {Gibbs} samplers: {Theory} and methods},
  author={Van Dyk, David A and Park, Taeyoung},
  journal={Journal of the American Statistical Association},
  volume={103},
  number={482},
  pages={790--796},
  year={2008},
  publisher={Taylor \& Francis}
}

@article{turnbull2023sequential,
  title={Sequential estimation of temporally evolving latent space network models},
  author={Turnbull, Kathryn and Nemeth, Christopher and Nunes, Matthew and McCormick, Tyler},
  journal={Computational Statistics \& Data Analysis},
  volume={179},
  pages={107627},
  year={2023},
  publisher={Elsevier}
}

@article{sewell2016latent,
  title={Latent space models for dynamic networks with weighted edges},
  author={Sewell, Daniel K and Chen, Yuguo},
  journal={Social Networks},
  volume={44},
  pages={105--116},
  year={2016},
  publisher={Elsevier}
}

@article{sewell2015latent,
  title={Latent space models for dynamic networks},
  author={Sewell, Daniel K and Chen, Yuguo},
  journal={Journal of the American Statistical Association},
  volume={110},
  number={512},
  pages={1646--1657},
  year={2015},
  publisher={Taylor \& Francis}
}

@article{sarkar2005dynamic,
  title={Dynamic social network analysis using latent space models},
  author={Sarkar, Purnamrita and Moore, Andrew W},
  journal={Acm Sigkdd Explorations Newsletter},
  volume={7},
  number={2},
  pages={31--40},
  year={2005},
  publisher={ACM New York, NY, USA}
}

@article{artico2023dynamic,
  title={Dynamic latent space relational event model},
  author={Artico, Igor and Wit, Ernst C},
  journal={Journal of the Royal Statistical Society Series A: Statistics in Society},
  volume={186},
  number={3},
  pages={508--529},
  year={2023},
  publisher={Oxford University Press US}
}

@article{hoff2002latent,
  title={Latent space approaches to social network analysis},
  author={Hoff, Peter D and Raftery, Adrian E and Handcock, Mark S},
  journal={Journal of the American Statistical Association},
  volume={97},
  number={460},
  pages={1090--1098},
  year={2002},
  publisher={Taylor \& Francis}
}

@article{hoff2008modeling,
  title={Modeling homophily and stochastic equivalence in symmetric relational data},
  author={Hoff, Peter D.},
  journal={Advances in Neural Information Processing Systems},
  volume={1},
  year={2008},
  pages={00--00}    
}

@book{bishop2006pattern,
  title={Pattern recognition and machine learning},
  author={Bishop, Christopher M and Nasrabadi, Nasser M},
  volume={4},
  year={2006},
  publisher={Springer},
  address={New York}
}

@article{loyal2025spike,
  title={A spike-and-slab prior for dimension selection in generalized linear network eigenmodels},
  author={Loyal, Joshua D and Chen, Yuguo},
  journal={Biometrika},
  volume={112},
  number={3},
  pages={asaf014},
  year={2025},
  publisher={Oxford University Press}
}

@article{li2023statistical,
  title={Statistical inference on latent space models for network data},
  author={Li, Jinming and Xu, Gongjun and Zhu, Ji},
  journal={arXiv preprint arXiv:2312.06605},
  year={2023}
}

@article{macdonald2022latent,
  title={Latent space models for multiplex networks with shared structure},
  author={MacDonald, Peter W and Levina, Elizaveta and Zhu, Ji},
  journal={Biometrika},
  volume={109},
  number={3},
  pages={683--706},
  year={2022},
  publisher={Oxford University Press}
}

@article{loyal2023eigenmodel,
  title={An eigenmodel for dynamic multilayer networks},
  author={Loyal, Joshua Daniel and Chen, Yuguo},
  journal={Journal of Machine Learning Research},
  volume={24},
  number={128},
  pages={1--69},
  year={2023}
}

@article{chan2009efficient,
  title = {Efficient Simulation and Integrated Likelihood Estimation in State Space Models},
  author = {Chan, Joshua C.C. and Jeliazkov, Ivan},
  year = {2009},
  journal = {International Journal of Mathematical Modelling and Numerical Optimisation},
  volume = {1},
  number = {1-2},
  pages = {101--120},
  publisher = {Inderscience Publishers},
  doi = {10.1504/IJMMNO.2009.03009},
}

@article{mccausland2011simulation,
  title = {Simulation Smoothing for State--Space Models: {A} Computational Efficiency Analysis},
  author = {McCausland, William J. and Miller, Shirley and Pelletier, Denis},
  year = {2011},
  journal = {Computational Statistics \& Data Analysis},
  volume = {55},
  number = {1},
  pages = {199--212},
  doi = {10.1016/j.csda.2010.07.009},
}

@article{king2025warped,
  title={Warped Dynamic Linear Models for Time Series of Counts},
  author={King, Brian and Kowal, Daniel R},
  journal={Bayesian Analysis},
  volume={20},
  number={1},
  pages={1331--1356},
  year={2025},
  publisher={International Society for Bayesian Analysis}
}

@article{fruhwirth2009improved,
  title={Improved Auxiliary mixture sampling for hierarchical models of non-{Gaussian} data},
  author={Fr{\"u}hwirth-Schnatter, Sylvia and Fr{\"u}hwirth, Rudolf and Held, Leonhard and Rue, H{\aa}vard},
  journal={Statistics and Computing},
  volume={19},
  pages={479--492},
  year={2009},
  publisher={Springer}
}

@article{fruhwirth2006auxiliary,
  title={Auxiliary mixture sampling for parameter-driven models of time series of counts with applications to state space modelling},
  author={Fr{\"u}hwirth-Schnatter, Sylvia and Wagner, Helga},
  journal={Biometrika},
  volume={93},
  number={4},
  pages={827--841},
  year={2006},
  publisher={Oxford University Press}
}

@article{d2023efficient,
  title={Efficient posterior sampling for {Bayesian Poisson} regression},
  author={D’Angelo, Laura and Canale, Antonio},
  journal={Journal of Computational and Graphical Statistics},
  volume={32},
  number={3},
  pages={917--926},
  year={2023},
  publisher={Taylor \& Francis}
}

@article{casarin2025media,
  title={Media bias and polarization through the lens of a {M}arkov switching latent space network model},
  author={Casarin, Roberto and Peruzzi, Antonio and Steel, Mark FJ},
  journal={The Annals of Applied Statistics},
  volume={19},
  number={4},
  pages={3416--3437},
  year={2025},
  publisher={Institute of Mathematical Statistics}
}

@article{ParkBA1147,
author = {Jong Hee Park and Yunkyu Sohn},
title = {Detecting Structural Changes in Longitudinal Network Data},
volume = {15},
journal = {Bayesian Analysis},
number = {1},
publisher = {International Society for Bayesian Analysis},
pages = {133--157},
year = {2020},
doi = {10.1214/19-BA1147},
URL = {https://doi.org/10.1214/19-BA1147}
}

@article{yu2021spatial,
  title={Spatial Voting Models in Circular Spaces: {A} Case Study of the {US} House of Representatives},
  author={Yu, Xingchen and Rodriguez, Abel},
  journal={The Annals of Applied Statistics},
  volume={15},
  number={4},
  pages={1897--1922},
  year={2021},
  publisher={Institute of Mathematical Statistics}
}

@article{durante2017nonparametric,
  title={Nonparametric {Bayes} Modeling of Populations of Networks},
  author={Durante, Daniele and Dunson, David B and Vogelstein, Joshua T},
  journal={Journal of the American Statistical Association},
  volume={112},
  number={520},
  pages={1516--1530},
  year={2017},
  publisher={Taylor \& Francis}
}

@article{wang2023joint,
  title={Joint Latent Space Model for Social Networks with Multivariate Attributes},
  author={Wang, Selena and Paul, Subhadeep and De Boeck, Paul},
  journal={Psychometrika},
  volume={88},
  number={4},
  pages={1197--1227},
  year={2023},
  publisher={Springer}
}

@article{lu2025zeroPois,
  title={A zero-inflated {Poisson} latent position cluster model},
  author={Lu, Chaoyi and Rastelli, Riccardo and Friel, Nial},
  journal={arXiv preprint arXiv:2502.13790},
  year={2025}
}

@article{handcock2007model,
  title={Model-based clustering for social networks},
  author={Handcock, Mark S and Raftery, Adrian E and Tantrum, Jeremy M},
  journal={Journal of the Royal Statistical Society Series A: Statistics in Society},
  volume={170},
  number={2},
  pages={301--354},
  year={2007},
  publisher={Oxford University Press}
}

@article{gwee2025latent,
  title={A latent shrinkage position model for binary and count network data},
  author={Gwee, Xian Yao and Gormley, Isobel Claire and Fop, Michael},
  journal={Bayesian Analysis},
  volume={20},
  number={2},
  pages={405--433},
  year={2025},
  publisher={International Society for Bayesian Analysis}
}

@article{albert1993bayesian,
  title={Bayesian analysis of binary and polychotomous response data},
  author={Albert, James H and Chib, Siddhartha},
  journal={Journal of the American Statistical Association},
  volume={88},
  number={422},
  pages={669--679},
  year={1993},
  publisher={Taylor \& Francis}
}

@article{gollini2016joint,
  title={Joint modeling of multiple network views},
  author={Gollini, Isabella and Murphy, Thomas Brendan},
  journal={Journal of Computational and Graphical Statistics},
  volume={25},
  number={1},
  pages={246--265},
  year={2016},
  publisher={Taylor \& Francis}
}

@incollection{geweke1991evaluating,
  title={Evaluating the accuracy of sampling-based approaches to the calculation of posterior moments},
  author={Geweke, John F},
booktitle={Bayesian Statistics 4},
  year={1992},
  editor={Berger, J.O. and Bernardo, J.M. and Dawid, A.P. and Smith, A.F.M.},
publisher={Oxford: Oxford University Press},
pages={169-194}
}

@article{de2014network,
  title={Network analysis of world trade using the {BACI-CEPII} dataset},
  author={De Benedictis, Luca and Nenci, Silvia and Santoni, Gianluca and Tajoli, Lucia and Vicarelli, Claudio},
  journal={Global Economy Journal},
  volume={14},
  number={03n04},
  pages={287--343},
  year={2014},
  publisher={World Scientific}
}

@article{hummels2005variety,
  title={The variety and quality of a nation's exports},
  author={Hummels, David and Klenow, Peter J},
  journal={American Economic Review},
  volume={95},
  number={3},
  pages={704--723},
  year={2005},
  publisher={American Economic Association}
}

@article{gemmetto2016multiplexity,
  title={Multiplexity and multireciprocity in directed multiplexes},
  author={Gemmetto, Valerio and Squartini, Tiziano and Picciolo, Francesco and Ruzzenenti, Franco and Garlaschelli, Diego},
  journal={Physical Review E},
  volume={94},
  number={4},
  pages={042316},
  year={2016},
  publisher={APS}
}

@article{lu2011focal,
  title={Focal Pontine Lesions Provide Evidence that Intrinsic Functional Connectivity Reflects Polysynaptic Anatomical Pathways},
  author={Lu, Jie and Liu, Hesheng and Zhang, Miao and Wang, Danhong and Cao, Yanxiang and Ma, Qingfeng and Rong, Dongdong and Wang, Xiaoyi and Buckner, Randy L and Li, Kuncheng},
  journal={Journal of Neuroscience},
  volume={31},
  number={42},
  pages={15065--15071},
  year={2011},
  publisher={Society for Neuroscience}
}

@article{aliverti2019spatial,
  title={Spatial modeling of brain connectivity data via latent distance models with nodes clustering},
  author={Aliverti, Emanuele and Durante, Daniele},
  journal={Statistical Analysis and Data Mining: The ASA Data Science Journal},
  volume={12},
  number={3},
  pages={185--196},
  year={2019},
  publisher={Wiley Online Library}
}

@article{wang2025establishing,
  title={Establishing group-level brain structural connectivity incorporating anatomical knowledge under latent space modeling},
  author={Wang, Selena and Wang, Yiting and Xu, Frederick H and Shen, Li and Zhao, Yize and Alzheimer’s Disease Neuroimaging Initiative and others},
  journal={Medical Image Analysis},
  volume={99},
  pages={103309},
  year={2025},
  publisher={Elsevier}
}

@article{schaefer2018local,
  title={Local-global parcellation of the human cerebral cortex from intrinsic functional connectivity {MRI}},
  author={Schaefer, Alexander and Kong, Ru and Gordon, Evan M and Laumann, Timothy O and Zuo, Xi-Nian and Holmes, Avram J and Eickhoff, Simon B and Yeo, BT Thomas},
  journal={Cerebral Cortex},
  volume={28},
  number={9},
  pages={3095--3114},
  year={2018},
  publisher={Oxford University Press}
}

@article{iturria2011brain,
  title={Brain hemispheric structural efficiency and interconnectivity rightward asymmetry in human and nonhuman primates},
  author={Iturria-Medina, Yasser and P{\'e}rez Fern{\'a}ndez, Alejandro and Morris, David M and Canales-Rodr{\'\i}guez, Erick J and Haroon, Hamied A and Garc{\'\i}a Pent{\'o}n, Lorna and Augath, Mark and Gal{\'a}n Garc{\'\i}a, L{\'\i}dice and Logothetis, Nikos and Parker, Geoffrey JM and others},
  journal={Cerebral Cortex},
  volume={21},
  number={1},
  pages={56--67},
  year={2011},
  publisher={Oxford University Press}
}

@article{chen2021autoregressive,
  title={Autoregressive models for matrix-valued time series},
  author={Chen, Rong and Xiao, Han and Yang, Dan},
  journal={Journal of Econometrics},
  volume={222},
  number={1},
  pages={539--560},
  year={2021},
  publisher={Elsevier}
}

@article{handcock2008fitting,
  title={Fitting latent cluster models for networks with latentnet},
  author={Handcock, Mark S and Krivitsky, Pavel N},
  journal={Journal of Statistical Software},
  volume={24},
  number={5},
  year={2008},
  pages={1--23},
  publisher={American Statistical Association}
}

@article{sosa2021review,
  title={A Review of Latent Space Models for Social Networks},
  author={Sosa, Juan and Buitrago, Lina},
  journal={Revista Colombiana de Estad{\'\i}stica},
  volume={44},
  number={1},
  pages={171--200},
  year={2021},
  publisher={Universidad Nacional de Colombia}
}

@article{ma2020universal,
  title={Universal latent space model fitting for large networks with edge covariates},
  author={Ma, Zhuang and Ma, Zongming and Yuan, Hongsong},
  journal={Journal of Machine Learning Research},
  volume={21},
  number={4},
  pages={1--67},
  year={2020}
}

@article{kim2023dynamic,
  title={A dynamic additive and multiplicative effects network model with application to the United Nations voting behaviors},
  author={Kim, Bomin and Niu, Xiaoyue and Hunter, David and Cao, Xun},
  journal={The Annals of Applied Statistics},
  volume={17},
  number={4},
  pages={3283},
  year={2023}
}

@article{lauderdale2010unpredictable,
  title={Unpredictable voters in ideal point estimation},
  author={Lauderdale, Benjamin E},
  journal={Political Analysis},
  volume={18},
  number={2},
  pages={151--171},
  year={2010},
  publisher={Cambridge University Press}
}

@misc{votingdata,
  author       = {{United Nations}},
  title        = {United Nations General Assembly Voting Data: Resolutions 1 (11 December 1946) to 79/328 (5 September 2025)},
  year         = {2025},
  publisher    = {United Nations},
  address      = {New York},
  url          = {http://digitallibrary.un.org/record/4060887},
  note         = {Accessed: 24 February 2026}
}

@article{gardini2026note,
  title={A note on auxiliary mixture sampling for {B}ayesian {P}oisson models.},
  author={Gardini, Aldo and Greco, Fedele and Trivisano, Carlo},
  journal={Statistics and Computing},
  volume={36},
  number={1},
  pages={27},
  year={2026},
  publisher={Springer}
}

\clearpage

\appendix
\onehalfspacing

%%%%%%%%%%%%%%%%%%%%%%%%%%%%%%%%%%%%%%%%%%%%%%%%%%%%%%%%
{\centering\section*{{\LARGE Supplementary Materials}}}

Appendix \ref{sec:posterior_details} of this document contains the 
% supplementary material for the article ``A Dynamic Latent Space Model for Weighted Networks''. The
proofs. Appendix \ref{sec:MCMC_algos} reports a description of the algorithmic implementation.
Appendix \ref{sec:simulation_further} presents the simulation results, while Appendix \ref{sec:applicaiton_further} provides further results related to the empirical applications.

%%%%%%%%%%%%%%%%%%%%%%%%%%%%%%%%%%%%%%%%%%%%%%%%%%%%%%%%%%%

\numberwithin{equation}{section}

\renewcommand\thesection {\Alph{section}}
\renewcommand\theequation{\Alph{section}.\arabic{equation}}
\renewcommand\thefigure  {\Alph{section}.\arabic{figure}}
\renewcommand\thetable   {\Alph{section}.\arabic{table}}

\section{Proof of the Results}
\label{sec:posterior_details}

\subsection{Matrix form representation of the model}

The proposed model can be written in matrix form as
\begin{subequations}
\begin{align}
    Y_{t} \mid \Lambda_{t}, Z_{t} & \overset{d}{=} P_{t}(Z_{t}) \odot N_t + (\biota_N\biota_N' -P_{t}(Z_{t})) \odot D_t\delta_{\{0\}}(y_{ij,t}) \\
    \log(\Lambda_t) & = \balpha \biota_N' + \biota_N \balpha' + X_t \Xi X_t' \\
    X_{t} & = \widetilde\Phi X_{t-1}\Phi' + H_{t} \qquad H_t \overset{iid}{\sim} \mathcal{MN}_{N, d}(H_{t}\mid\mathbf{O}_{N, d}, \widetilde\Upsilon, \Upsilon) \\
    Z_t & = V_{t}\bbeta \biota_N' + \biota_N \bbeta' V_{t}' + U_{t}, \qquad 
    U_{t} \overset{iid}{\sim} \mathcal{MN}_{N,N}(U_{t} \mid \mathbf{0}_{N,N},\Id_N,\Id_N),
\end{align}
\end{subequations}
where $N_t \in \mathbb{N}^{N\times N}$, $D_t \in \{0,1\}^{N\times N}$, and $P_t(Z_t) \in (0,1)^{N\times N}$, with $ij$th element given by $N_{ij,t} \overset{iid}{\sim} \mathcal{P}oi\big( y_{ij,t} \mid \lambda_{ij,t} \big)$, $D_{ij,t} \overset{iid}{\sim} \delta_{\{0\}}(y_{ij,t})$, and $p_{ij,t}(z_{ij,t})$, respectively.
Moreover, $\biota_N$ is a vector of 1s of length $N$, $\odot$ represents the Hadamard product and the expression $\log(\Lambda_t)$ is meant to apply the natural logarithm to each element of $\Lambda_t$ separately.

\subsection{Proof of Lemma \ref{lemma:auxiliary_mixture}}

Without loss of generality, we omit the indices $ij$ to ease the notation.
The first step of data augmentation allows to obtain a linear non-Gaussian equation by recreating a Poisson process for each observation $y_t$ via the introduction of two latent variables for each nonzero observation: the arrival time $\tau_{2t}$ of the $y_t$th jump and the interarrival time $\tau_{1t}$ between the $y_t$th jump and the next one.
Therefore, if $y_t>0$ jumps occurred in the unit time interval, then $\tau_{2t}$ is the arrival time of the last jump before time $u=1$, whereas $\tau_{2t} +\tau_{1t}$ is the arrival time of the first jump after $u=1$.
% This result will be used to sample from the posterior distribution of $\tau_{i1}$ and $\tau_{i2}$ given $y_i$.
%
By the definition of Poisson process, the interarrival time $\tau_{1t} \sim \mathcal{E}xp(\tau_{1t} \mid \lambda_t)$ with rate $\lambda_t$, whereas the arrival time of the $y_t$th jump, $\tau_{2t}$, being obtained as the sum of $y_t$ exponentially distributed times, follows a Gamma distribution $\tau_{2t} \sim \mathcal{G}a(\tau_{2t} \mid y_t, \lambda_t)$.

Therefore, for $y_t = 0$, one has
\begin{align*}
    \P(Y_t = 0 \mid \lambda_t) & = \P(\tau_{1t} > 1 \mid \lambda_t) = \int_{1}^{\infty} \lambda_t \exp(-\lambda_t \tau_{1t}) \: \mathrm{d}\tau_{1t} = \exp(-\lambda_t)
\end{align*}
Instead, for $y_t > 0$, one has 
\begin{align*}
    \P(Y_t = y_t \mid \lambda_t) & = \P(\tau_{2t} < 1 \mid \lambda_t, y_t) \P(\tau_{1t} > 1-\tau_{2t} \mid \lambda_t) \\
    & = \int_0^1 \frac{\lambda_t^{y_t}}{\Gamma(y_t)} \tau_{2t}^{y_t-1} \exp(-\lambda_t \tau_{2t}) \int_{1-\tau_{2t}}^{\infty} \lambda_t \exp(-\lambda_t \tau_{1t}) \: \mathrm{d}\tau_{1t} \: \mathrm{d}\tau_{2t} \\
    % & = \int_0^1 \frac{\lambda_t^{y_t}}{\Gamma(y_t)} \tau_{2t}^{y_t-1} \exp(-\lambda_t \tau_{2t}) \big( -\exp(-\lambda_t \tau_{1t}) \big) \Big|_{1-\tau_{2t}}^{\infty} \: \mathrm{d}\tau_{2t} \\
    & = \int_0^1 \frac{\lambda_t^{y_t}}{\Gamma(y_t)} \tau_{2t}^{y_t-1} \exp(-\lambda_t \tau_{2t}) \exp(-\lambda_t(1-\tau_{2t})) \: \mathrm{d}\tau_{2t} \\
    & = \int_0^1 \frac{\lambda_t^{y_t}}{(y_t-1)!} \tau_{2t}^{y_t-1} \exp(-\lambda_t) \: \mathrm{d}\tau_{2t} \\
    & = \frac{\lambda_t^{y_t} \exp(-\lambda_t)}{(y_t-1)!} \int_0^1 \tau_{2t}^{y_t-1} \: \mathrm{d}\tau_{2t} = \frac{\lambda_t^{y_t} \exp(-\lambda_t)}{y_t!}.
\end{align*}
Summarising, one can leverage this relationship to obtain the joint distribution
\begin{equation}
    p(y_t, \tau_{1t}, \tau_{2t} \mid \lambda_t) = \begin{cases}
        p(\tau_{1t} \mid \lambda_t) \I(\tau_{1t} > 1) & \text{ if } y_t = 0 \\
        p(\tau_{2t} \mid \lambda_t, y_t) p(\tau_{1t} \mid \tau_{2t}, \lambda_t) \I(\tau_{2t} < 1 < \tau_{1t}+\tau_{2t}) & \text{ if } y_t > 0
    \end{cases}
\end{equation}
In short, the distributions of $\tau_{1t}$ and $\tau_{2t}$ can be equivalently expressed as
\begin{align}
    % \tau_{i1} & = \frac{\xi_{i1}}{\lambda_i}, 
    % & \xi_{i1} & \sim \mathcal{E}xp(1) \\
    % \tau_{i2} & = \frac{\xi_{i2}}{\lambda_i}, 
    % & \xi_{i2} & \sim \mathcal{G}a(y_i, 1).
    \tau_{1t} & = \frac{\xi_{1t}}{\lambda_t}, \quad
    \xi_{1t}  \sim \mathcal{E}xp(\xi_{1t}  \mid 1), 
    & \tau_{2t} & = \frac{\xi_{2t}}{\lambda_t}, \quad
    \xi_{2t}  \sim \mathcal{G}a(\xi_{2t}  \mid y_t, 1).
\label{eq:prior_tau}
\end{align}
Taking logarithms, one obtains a conditionally linear non-Gaussian model for the logarithm of the intensity of the original Poisson as
\begin{align}
    -\log(\tau_{1t}) & = \log(\lambda_t) + \varepsilon_{1t},
    & -\log(\tau_{2t}) & = \log(\lambda_t) + \varepsilon_{2t},
\end{align}
where $\varepsilon_{lt} = -\log(\xi_{lt}) \in\R$, $l=1,2$, have a negative log-Gamma distribution with integer shape parameter $\nu$ equal to $1$ and $y_t$, respectively, and density
\begin{equation}
	p_\varepsilon(\varepsilon \mid \nu) = \frac{\exp(-\nu \varepsilon -\exp(-\varepsilon))}{\Gamma(\nu)}.
\end{equation}
%
% Therefore, to obtain a conditionally linear Gaussian model \cite{fruhwirth2009improved} proposed to approximate the negative log-Gamma distribution with a finite mixture of Gaussian components, that is
% \begin{equation*}
% 	p_\varepsilon(\varepsilon \mid \nu) \approx \sum_{k=1}^{R(\nu)} w_k(\nu) \phi\big(\varepsilon \mid \mu_k(\nu), \sigma_k^2(\nu) \big),
% \end{equation*}
% where $\phi(x \mid \mu,\sigma^2)$ denotes the density of a Gaussian distribution with mean $\mu$ and variance $\sigma^2$. We remark that the location, scale and number of components in the mixture varies as a function of the (integer) shape parameter of the negative log-Gamma to be approximated.

% In a second data augmentation step, we introduce latent variables $r_{lt}$, $l=1,2$, that allocate each $\tau_{lt}$ to a component of the mixture, resulting in the conditionally linear Gaussian model:
% \begin{align}
% 	-\log(\tau_{lt}) = \lambda_t + \varepsilon_{lt}, \qquad \varepsilon_{lt} \mid r_{lt} \sim \mathcal{N}(\varepsilon_{lt} \mid \mu_{r_{lt}}, \sigma^2_{r_{lt}}).
% \label{eq:model_log_tau_conditional}
% \end{align}
% We shall remark that in the presence of a zero observation, $y_t=0$, the data augmentation consists of only two variables, $(\tau_{1t},r_{1t})$.

The second step of data augmentation is used to obtain a conditionally Gaussian distribution by approximating the densities of $\varepsilon_{1t}$ and $\varepsilon_{2t}$ by a finite mixture of Gaussian then introducing the latent component indicators $r_t = (r_{1t}, r_{2t})$.
As a result, one obtains
\begin{align}
    p_{\varepsilon}(\varepsilon \mid \nu) & 
    % = \frac{\exp \left(-\nu \varepsilon -e^{-\varepsilon}\right)}{\Gamma(\nu)} 
    \approx \sum_{k=1}^{R(\nu)} c_k(\nu) \phi(\varepsilon \mid \mu_k(\nu), \sigma_k^2(\nu)),
\end{align}
where $\phi(\cdot \mid \mu,\sigma^2)$ denotes a Gaussian density with mean $\mu$ and variance $\sigma^2$ and $R(\nu)$ is the number of components of the mixture. 
Therefore, conditionally on the latent variables $\btau$ and $\br$, the nonlinear non-Gaussian model is transformed into a linear Gaussian model where the mean is a linear function of the parameters (if $y_t=0$, we consider only the first equation):
\begin{equation}
\begin{aligned}
    -\log(\tau_{1t}) & = \log(\lambda_t) + \mu_{r_{1t}}(1) + \varepsilon_{1t} 
    & \quad \varepsilon_{1t} \mid r_{1t} & \sim \mathcal{N}\big(\varepsilon_{1t}  \mid 0, \sigma_{r_{1t}}^2(1) \big), \\
    -\log(\tau_{2t}) & = \log(\lambda_i) + \mu_{r_{2t}}(y_t) + \varepsilon_{2t} 
    & \quad \varepsilon_{2t} \mid r_{2t} & \sim \mathcal{N}\big( \varepsilon_{2t}  \mid 0, \sigma_{r_{2t}}^2(y_t) \big).
\end{aligned}
\end{equation}
We shall remark that in the presence of a zero, $y_t=0$, the data augmentation consists of only two variables, $(\tau_{1t},r_{1t})$.

%%%%%%%%%%%%%%%%%%%%%%%%%%%%%%%%%%%%%%%%%%%%%%%%%%%%
\subsection{Proof of Proposition \ref{proposition:joint_posterior_dx}}

We derive the decomposition of the joint distribution of $(d, \bx)$, conditionally on all the other parameters, as $p(d,\bx \mid \cdot) = p(\bx \mid d, \cdot) p(d \mid \cdot)$ and then use Laplace approximation. 

% Under the \textit{node-wise} specification, the posterior distribution of the latent features' trajectory for each node $i=1,\dots,N$, conditionally on $d$ and all other nodes' trajectories, is obtained following Lemma~\ref{lemma:AWOL}.
The marginal posterior for $d$ is obtained after integrating out the latent features $\bx$, that is
\begin{align}
    p(d \mid \by,\btau,\br,\bw, \balpha,\Phi,\Upsilon) & = \frac{p(\by, \btau,\br \mid  d, \bw, \balpha,\Phi,\Upsilon)p(d)}{\sum_{\ell=1}^{\underline{d}} p(\by, \btau,\br \mid  \ell, \bw, \balpha,\Phi,\Upsilon)p(\ell)},  \quad  d\in\mathcal{D}.
\label{eq:posterior_d}
\end{align}
Let $\mathcal{Q}_t = \{(i,j): i\in\{1,\dots,N\}, \: j\in\{i+1,\dots,N\}, w_{ij,t} = 1\}$ be the set of edges in $Y_t$ allocated to the Poisson component, with cardinality $Q_t = |\mathcal{Q}_t|$. Also, let us introduce the function
\begin{align*}
    \varphi_{ij,t}(\bx_{i:,t},\bx_{j:,t}) & =  p(y_{ij,t}, \btau_{ij,t}, \br_{ij,t} \mid d, w_{ij,t}, \bx_{i:,t},\bx_{j:,t}, \alpha_i,\alpha_j) w_{ijt} + \delta_{\{0\}}(y_{ij,t})(1-w_{ijt}),
\end{align*}
where $p(y_{ij,t}, \btau_{ij,t}, \br_{ij,t} \mid d, w_{ij,t}, \bx_{i:,t},\bx_{j:,t}, \alpha_i,\alpha_j)$ is defined in eq.~\eqref{eq:complete_likelihood_yijt}.
The joint density $p(\by, \btau, \br \mid d, \bw, \balpha)$ is defined as the integral
\begin{align}
    \notag
    & p(\by, \btau, \br \mid d, \bw, \balpha, \Phi,\Upsilon)  = \int_{\R^{dTN}}  \prod_{t=1}^T \big(\prod_{i=1}^N \prod_{j=i+1}^N  \varphi_{ijt}(\bx_{i:,t},\bx_{j:,t})\big)  p(\bx_{::,t} \mid \bx_{::,t-1},d, \Phi,\Upsilon) \: \mathrm{d}\bx \\
    \notag
    & = \big(\prod_{t=1}^T \prod_{(i,j)\notin\mathcal{Q}_t} \!\! \varphi_{ijt}(\bx_{:i,t},\bx_{:j,t})\big) \int_{\R^{dTN}}  \prod_{t=1}^T \prod_{(i,j)\in\mathcal{Q}_t}  \varphi_{ijt}(\bx_{i:,t},\bx_{j:,t}) p(\bx_{::,t} \mid \bx_{::,t-1},d, \Phi,\Upsilon) \: \mathrm{d}\bx \\
    \notag
    & = \int_{\R^{dTN}} \prod_{t=1}^T \prod_{(i,j)\in\mathcal{Q}_t} \!\! p(y_{ij,t}, \btau_{ij,t}, \br_{ij,t} \mid d, w_{ij,t}, \bx_{i:,t},\bx_{j:,t}, \alpha_i,\alpha_j) p(\bx_{::,t} \mid \bx_{::,t-1},d, \Phi,\Upsilon) \: \mathrm{d}\bx \\
    & \approx \int_{\R^{dTN}} \prod_{t=1}^T \prod_{(i,j)\in\mathcal{Q}_t} \!\! q(y_{ij,t}, \btau_{ij,t}, \br_{ij,t} \mid d, w_{ij,t}, \bx_{i:,t},\bx_{j:,t}, \alpha_i,\alpha_j) p(\bx_{t} \mid \bx_{::,t-1},d, \Phi,\Upsilon) \: \mathrm{d}\bx,
\label{eq:marginal_like_d}
\end{align}
where $p(\bx_{::,t} \mid \bx_{::,t-1}, d, \Phi,\Upsilon)$ defined in eq.~\eqref{eq:joint_xt} and the last line follows from the  approximation in Lemma \ref{lemma:auxiliary_mixture}.
Since the integral in \eqref{eq:marginal_like_d} is not available in closed form, we use a Laplace method \citep[e.g., see][]{bishop2006pattern} to get the approximation\footnote{We have assumed that the Hessian matrix of the negative log posterior has full rank \citep[]{bishop2006pattern}.}
\begin{align}
\label{eq:Laplace_approx_d}
    \notag
    \log p(\by, \btau, \br & \mid d,\bw,\balpha, \Phi,\Upsilon)  \approx \log q(\by, \btau,\br \mid d,\bw,\balpha, \Phi,\Upsilon) \\
    \notag
    & = \sum_{t=1}^T \sum_{(i,j)\in\mathcal{Q}_t} \!\log q\big( y_{ij,t}, \btau_{ij, t}, \br_{ij,t} \mid d,w_{ij,t},\widehat{\bx}_{i:,t},\widehat{\bx}_{j:,t}, \alpha_i,\alpha_j \big) \\ 
    & \quad + \sum_{t=1}^T \log p\big( \widehat{\bx}_{::,t} \mid \widehat{\bx}_{::,t-1}, d, \Phi,\Upsilon \big)- \frac{dTN}{2}\log(Q^*),
    %  &\quad{\color{blue} + \frac{NDT}{2}\log(2\pi) - \frac{NDT}{2}\log(QTN(N-1)/2) + \frac{NDT}{2}\log(|H|)},
    % -\frac{TN(N-1)Q/2}{2}\log(TN),
\end{align}
% where $\log q\big( \btau_{ij, t}, \br_{ij,t} \mid \alpha_i,\alpha_j,\omega_{ij,t} \big) = \log p\big( \btau_{ij, t}, \br_{ij, t} \mid \widehat{\bx}_{i,t}, \widehat{\bx}_{j,t}, \alpha_i,\alpha_j, \omega_{ij,t} \big)$, 
where $\widehat{\bx}_{i:,t}$ is a point of high posterior mass (e.g., the MAP) and $Q^* = \sum_{t=1}^T Q_t + \sum_{(i,j)\in\mathcal{Q}_t} \mathbb{I}(y_{ijt} > 0)$.

Finally, using the approximation \eqref{eq:Laplace_approx_d} in eq.~\eqref{eq:posterior_d} one obtains
\begin{align}
    p(d \mid \by, \btau,\br, \bw, \balpha,\Phi,\Upsilon) & \approx \frac{q(\by, \btau,\br \mid d, \bw, \balpha,\Phi,\Upsilon)p(d)}{\sum_{\ell=1}^{\underline{d}} q(\by, \btau,\br \mid \ell, \bw, \balpha,\Phi,\Upsilon)p(\ell)},  \quad  d\in\mathcal{D}.
\label{eq:posterior_d_approx}
\end{align}

%%%%%%%%%%%%%%%%%%%%%%%%%%%%%%%%%%%%%%%%%%%%%%%%%%%%
\subsection{Proof of Proposition  \ref{proposition:sampling_x}}

%\subsection{Sampling $\bx_i$ with ``node-specific'' prior}
%\label{sec:app_nodewise}

Starting from eq.~\eqref{eq:complete_likelihood_yijt}, we exploit independence across $q=1,2$ to compactly rewrite the model for $\btau_{ij,t}$ conditionally on $\br_{ij,t}$ as
\begin{align*}
    p\big( \btau_{ij,t} \mid \cdot \big) & = \mathcal{N}_2\big( \widetilde\btau_{ij,t} \mid \widetilde\bmu_{ij,t}, \widetilde\Sigma_{ij,t} \big),
\end{align*}
where
\begin{align*}
    \widetilde\btau_{ij,t} & = \begin{pmatrix}
        -\log\big( \tau_{ij, 1t} \big) \\ -\log\big( \tau_{ij, 2t} \big)
    \end{pmatrix}, &  
    \widetilde\Sigma_{ij,t} & = \operatorname{diag}\big( \sigma_{r_{ij, 1t}}^2, \sigma_{r_{ij, 2t}}^2 \big), &
    \widetilde\bmu_{ij,t} & = \begin{pmatrix}
         \log(\lambda_{ij,t}) + \mu_{r_{ij, 1t}} \\
         \log(\lambda_{ij,t}) + \mu_{r_{ij, 2t}}
    \end{pmatrix}.
\end{align*}
For convenience, we rewrite the vector $\widetilde\bmu_{ij,t}$ as follows
\begin{align*}
    \widetilde\bmu_{ij,t} & = \begin{pmatrix}
         \log(\lambda_{ij,t}) + \mu_{r_{ij, 1t}} \\
         \log(\lambda_{ij,t}) + \mu_{r_{ij, 2t}}
    \end{pmatrix} = \begin{pmatrix}
         \alpha_i + \alpha_j + \bx_{i:,t}'\bx_{j:,t} + \mu_{r_{ij, 1t}} \\
         \alpha_i + \alpha_j + \bx_{i:,t}'\bx_{j:,t} + \mu_{r_{ij, 2t}}
    \end{pmatrix} \\
    & = \underbrace{\begin{pmatrix}
         \alpha_i + \alpha_j + \mu_{r_{ij, 1t}} \\
         \alpha_i + \alpha_j + \mu_{r_{ij, 2t}}
    \end{pmatrix}}_{\overline\bmu_{ij,t}} + \bx_{i:,t}'\bx_{j:,t} \biota_2 = \overline\bmu_{ij,t} + \bx_{i:,t}'\bx_{j:,t} \biota_2.
\end{align*}
Let us define the vector of partial residuals as $\overline{\bz}_{ij,t} = \widetilde\btau_{ij,t} - \overline\bmu_{ij,t}$, thus obtaining
\begin{align}
    p\big( \overline{\bz}_{ij,t} \mid \br_{ij,t},\bx_{i:,t},\bx_{j:,t}, \alpha_i,\alpha_j,\omega_{ij,t} \big) = \mathcal{N}_2\big( \overline{\bz}_{ij,t} \mid \biota_2 \bx_{j:,t}'\bx_{i:,t}, \, \widetilde\Sigma_{ij,t} \big).
\end{align}
% Combining the latter with the transition for the state vector yields the state space representation
% \begin{align}
%     p\big( \bz_{ij,t} \mid \br_{ij,t},\bx_{i:,t},\bx_{j:,t},\alpha_i,\alpha_j,\omega_{ij,t} \big) & = \mathcal{N}_2\big( \bz_{ij,t} \mid \bx_{j:,t}'\bx_{i:,t}, \, \widetilde\Sigma_{ij,t} \big), \\
%     p(\bx_{i:,t} \mid \bx_{i:,t-1}, \Phi, \Upsilon) & = \mathcal{N}_d(\bx_{i:,t} \mid \Phi \bx_{i:,t-1}, \Upsilon).
% \end{align}
%
By stacking vertically all the vectors $\overline{\bz}_{i:,t} = \big( \overline{\bz}_{i1,t}^{\prime},\dots, \overline{\bz}_{i,i-1,t}^{\prime}, \overline{\bz}_{i,i+1,t}^{\prime}, \dots,\overline{\bz}_{iN,t}^{\prime})' \in\R^{2(N-1)}$ and $\br_{i:,t} = \big( \br_{i1,t}^{\prime},\dots, \br_{i,i-1,t}^{\prime}, \br_{i,i+1,t}^{\prime}, \dots,\br_{iN,t}^{\prime})' \in\R^{2(N-1)}$ one obtains
\begin{align}
    p\big( \overline{\bz}_{i:,t} \mid \br_{i:,t},\bx_{i:,t},\bx_{j:,t}, \alpha_i,\alpha_j \big) & = \mathcal{N}_{2(N-1)}\big( \overline{\bz}_{i:,t} \mid G_{i,t} \bx_{i:,t}, \, \widetilde\Sigma_{i,t} \big),
\end{align}
where, denoting with $\operatorname{blkdiag}(A,B)$ a block-diagonal matrix with diagonal blocks $A$ and $B$, we defined $G_{i,t} = \big( (\biota_2\bx_{1:,t}')',\dots, (\biota_2\bx_{i-1:,t}')', (\biota_2\bx_{i+1:,t}')',\dots, (\biota_2\bx_{N:,t}')' \big)' \in\R^{2(N-1)\times d}$ and $\widetilde\Sigma_{i,t} = \operatorname{blkdiag}\big( \widetilde\Sigma_{i1,t},\dots, \widetilde\Sigma_{i,i-1,t}, \widetilde\Sigma_{i,i+1,t}, \dots,\widetilde\Sigma_{i,N,t} \big) \in\Sp^{2(N-1)}$, that is
\begin{align*}
    G_{i,t} & = \begin{pmatrix}
        \biota_2 \bx_{1:,t}' \\
        \vdots \\
        \biota_2 \bx_{i-1:,t}' \\
        \biota_2 \bx_{i+1:,t}' \\
        \vdots \\
        \biota_2 \bx_{N:,t}'
    \end{pmatrix}, \quad
    \widetilde\Sigma_{i,t} = \begin{pmatrix}
\widetilde\Sigma_{i1,t} & \widetilde\Sigma_{i2,t} & \mathbf{O}_{2,2}  & \cdots & \mathbf{O}_{2,2} \\
\widetilde\Sigma_{i1,t} & \widetilde\Sigma_{i2,t} & \widetilde\Sigma_{i3,t} & \ddots & \vdots \\
\mathbf{O}_{2,2} & \ddots & \ddots & \ddots & \mathbf{O}_{2,2} \\
\vdots & \ddots & \widetilde\Sigma_{i,i-1,t} & \widetilde\Sigma_{ii,t} & \widetilde\Sigma_{i,i+1,t} \\
\mathbf{O}_{2,2} & \cdots & \mathbf{O}_{2,2} & \widetilde\Sigma_{i,N-1,t} & \widetilde\Sigma_{iN,t}
\end{pmatrix}.
\end{align*}
Finally, we stack all vectors over the time dimension and define $G_i = \operatorname{blkdiag}(G_{i,1},\dots,G_{i, T})\in\R^{2(N-1)T \times dT}$, $\widetilde\Sigma_i = \operatorname{blkdiag}(\widetilde\Sigma_{i, 1},\dots,\widetilde\Sigma_{i,T}) \in\Sp^{2(N-1)T}$, $\overline{\bz}_{i:,:} = (\overline{\bz}_{i,1}^{\prime},\dots,\overline{\bz}_{i,T}^{\prime})' \in\R^{2(N-1)T}$, $\bx_{i:,:} = (\bx_{i:,1}',\dots,\bx_{i:,T}')' \in\R^{dT}$.
%, $\overline{\Psi} = (\Id_T \otimes \Phi)$, and $\overline{\Upsilon} = (\Id_T \otimes \Upsilon)$.

\bigskip
%%%%%%%%%%%%%%%%%%%%%%%%%%%%%%%%%%%%%%%%%%%%%%%%%%%%%%%%%
% \textit{node-wise} specification
\paragraph{(i) Posterior of $\bx$, node-wise.}
Regarding the \textit{node-wise} specification, the entire path of the node-$i$ latent dynamic features, $\bx_{i:,:} = (\bx_{i:,1}',\dots,\bx_{i:,T}')'$ is jointly sampled all without a loop \citep[e.g, see][]{chan2009efficient,mccausland2011simulation}.
To obtain the posterior (smoothed) distribution of $\bx_{i:,:}$ following a procedure similar to \cite{chan2009efficient}, start by rewriting the data-augmented model in a state space form. 

We assume the initial value for the process to be $\bx_{i:,0} \sim \mathcal{N}_d(\bx_{i:,0} \mid \underline{\bx}_{i:,0}, \Omega_0)$.
Combining the transition and measurement equations yields the state space representation
\begin{align}
    p\big( \overline{\bz}_{i:,:} \mid \br,\bx_{i:,:},\bx_{-i,::},\balpha \big) & = \mathcal{N}_{2(N-1)T}\big( \overline{\bz}_{i:,:} \mid G_i \bx_{i:,:}, \, \widetilde\Sigma_i \big), \\
    p(\bx_{i:,:} \mid \bx_{i:, 0}, \Phi, \Upsilon) & = \mathcal{N}_{dT}(\bx_{i:,:} \mid \overline{\bx}_{i:,:}, K_i^{-1}),
\end{align}
where
\begin{align*}
    K_i = H_i' S_i^{-1} H_i, \qquad \overline{\bx}_{i:,:} = H_i^{-1} \begin{pmatrix}
        \underline{\bx}_{i:,0} \\ \bO_d \\ \vdots \\ \bO_d
    \end{pmatrix} 
\end{align*}
and we defined $S_i = \operatorname{blkdiag}(\Omega_0, \Upsilon,\dots,\Upsilon)$ and
\begin{align*}
    H_i = \begin{pmatrix}
        % \Id_d & O_{d}&O_{d} & \cdots&  \cdots& O_{d}\\
        % -\Phi & \Id_d & O_{d}& \cdots&  \cdots& O_{d}\\
        % O_{d}& -\Phi & \Id_d & \cdots&\cdots& O_{d}  \\
        % \vdots& & \ddots & \ddots & & \vdots\\
        % & & & -\Phi & \Id_d &O_{d}\\
        % O_{d}& \cdots & \cdots& O_{d}& -\Phi & \Id_d
I_d      & \mathbf{O}_{d,d}      & \mathbf{O}_{d,d}      & \cdots & \mathbf{O}_{d,d} \\
-\Phi  & I_d      & \mathbf{O}_{d,d}      & \cdots & \mathbf{O}_{d,d} \\
\mathbf{O}_{d,d}      & -\Phi  & I_d      & \ddots & \vdots \\
\vdots   & \ddots   & \ddots   & \ddots & \mathbf{O}_{d,d} \\
\mathbf{O}_{d,d}      & \cdots   & \mathbf{O}_{d,d}      & -\Phi & I_d
\end{pmatrix}.
\end{align*}
As a result, the posterior (smoothed) distribution for $\bx_i$ is obtained as
\begin{align}
    \bx_{i:,:} \mid \overline{\bz}_{i:,:}, \br,\bx_{-i:,:},\balpha, \Phi, \Upsilon, \bx_{i:,0} \sim \mathcal{N}_{dT}\big( \bx_{i:,:} \mid \widehat{\bx}_{i:,:} \widehat{K}_i^{-1} \big),
\end{align}
where the mean and precision are
\begin{align}
    \widehat{K}_i & = K_i + G_i' \widetilde\Sigma_i^{-1} G_i, \qquad
    \widehat{\bx}_{i:,:} = \widehat{K}_i \big( K_i \overline{\bx}_{i:,:} + G_i' \widetilde\Sigma_i^{-1} \overline{\bz}_{i:,:} \big).
\end{align}

Finally, define $\bx_{::,0} = (\bx_{1:,0}',\dots,\bx_{N:,0}')' \in \R^{dN}$, $F = \Phi \otimes \Id_N$, and $Q = \Upsilon^{-1} \otimes \Id_N$, and define $Q_0 = \Omega_0 \otimes \Id_N$ and $\underline{\bx}_{::,0} = (\underline{\bx}_{1:,0}',\dots,\underline{\bx}_{N:,0}')' \in \R^{dN}$. Then, we obtain
\begin{align*}
    p( \bx_{::,0} \mid \bx_{::,1}) 
    & \propto \exp\Big( -\frac{1}{2} \big(\bx_{::,1} -F\bx_{::,0} \big)' Q \big(\bx_{::,1} -F\bx_{::,0} \big) \Big) \\
    & \quad \times \exp\Big( -\frac{1}{2} (\bx_{::,0} - \underline{\bx}_{::,0})' Q_0 (\bx_{::,0} - \underline{\bx}_{::,0}) \Big) \\
    & \propto \exp\big( -\frac{1}{2} \big( \bx_{::,0}'\big(F'QF + Q_0\big) \bx_{::,0} - 2\bx_{::,0}'F'Q\bx_{::,1} -2\bx_{::,0}'Q_0 \underline{\bx}_{::,0} \big) \Big).
\end{align*}
Equivalently, $\bx_{::,0} \mid \bx_{::,1} \sim \mathcal{N}_{Nd}(\bx_{::,0} \mid \overline{\bx}_0, (\overline{\Omega}_0)^{-1})$, where
\begin{equation}
\overline{\Omega}_0 = F'QF + Q_0, \qquad
\overline{\bx}_0 = \overline{\Omega}_0^{-1}\big( F'Q\bx_{::,1} + Q_0 \underline{\bx}_{::,0} \big).
\end{equation}

\bigskip
%%%%%%%%%%%%%%%%%%%%%%%%%%%%%%%%%%%%%%%%%%%%%%%%%%%%%%%%%
% \textit{feature-wise} specification
\paragraph{(ii) Posterior of $\bx$, feature-wise.}
Regarding the \textit{feature-wise} specification, the ``dimension-specific'' prior postulates a first-order vector autoregressive process for each coordinate $\bx_{:\ell,:}$:
\begin{equation*}
    \bx_{:\ell,0} \sim \mathcal{N}_{N}(\bx_{:\ell,0}  \mid \underline{\bx}_{:\ell,0}^*, \Omega_0^*), \qquad
    \bx_{:\ell,t} \mid \bx_{:\ell,t-1} \sim \mathcal{N}_{N}(\bx_{:\ell,t} \mid \widetilde\Phi \bx_{:\ell,t-1}, \widetilde\Upsilon),
\end{equation*}
for $t=1,\dots,T$, where $\underline{\bx}_{:\ell,0}^* \in\R^N$ is a fixed prior mean and $\Omega_0^* \in \Sp^{N}$ is a fixed prior covariance for the initial state. Our goal is to obtain a joint prior distribution for the vector $\bx$, then derive the conditional of $\bx_{i:,:} \mid \bx_{-i:,:}$ along the same line as in \textbf{(i)}.

Recall that $\bx_{::,t} = \operatorname{vec}(X_t) = (\bx_{:1,t}',\dots,\bx_{:d,t}')' \in \R^{dN}$ the vector of all latent variables -- across nodes and features -- at time $t$, and assume the joint distribution of $\bx_{::,t}$ (conditionally on $\bx_{::,t-1}$) is Gaussian with zero cross-dimension covariance:
\begin{align}
    % \bx_t \mid \bx_{t-1} \sim \mathcal{N}_{Nd}\!\left( (\Id_d \otimes \widetilde\Phi)\, \bx_{t-1}, \ \operatorname{blkdiag}(\Upsilon_1,\dots,\Upsilon_d) \right).
    \bx_{::,t} \mid \bx_{::,t-1} \sim \mathcal{N}_{Nd}\!\big(\bx_{::,t} \mid (\Id_d \otimes \widetilde\Phi)\, \bx_{::,t-1}, \ \Id_d \otimes \widetilde\Upsilon \big). \label{eq:joint_xt}
\end{align}
Therefore, the joint distribution of $\bx = (\bx_{::,1}^{\prime},\dots,\bx_{::;T}^{\prime})' \in \R^{NdT}$ and $\bx_{::,0}$ factors as
\begin{align*}
    p(\bx, \bx_{::,0}) = p(\bx_{::,0}) \prod_{t=1}^T p(\bx_{::,t} \mid \bx_{::,t-1}).
\end{align*}

% Denoting with $F = \Id_d \otimes \Phi$, $\Upsilon = \operatorname{blkdiag}(\Upsilon_1,\dots,\Upsilon_d)$ and $Q = \Upsilon^{-1} = \operatorname{blkdiag}(\Upsilon_1^{-1},\dots,\Upsilon_d^{-1})$, we obtain
\noindent Defining $F = \Id_d \otimes \widetilde\Phi$ and $Q  = \Id_d \otimes \widetilde\Upsilon^{-1}$, we obtain
\begin{align*}
    p(\bx\mid \bx_{::,0}) & = (2\pi)^{-\frac{NdT}{2}} |Q|^{\frac{T}{2}} \exp\Big( -\frac{1}{2} \sum_{t=1}^T \big(\bx_{::,t} -F\bx_{::,t-1} \big)' Q \big(\bx_{::,t} -F\bx_{::,t-1} \big) \Big).
\end{align*}
Equivalently, $\bx \mid \widetilde\bx_{::,0} \sim \mathcal{N}_{NdT}(\bx \mid \bmu, \Psi^{-1})$, with
\begin{equation*}
    \bmu = \begin{pmatrix} F \bx_{::,0}\\ F^2 \bx_{::,0}\\ \vdots\\ F^T \bx_{::,0} \end{pmatrix}, 
    \qquad
    \Psi = \begin{pmatrix}
    Q+F'QF & -F' Q & \mathbf{O}_{dN,dN}  & \cdots & \mathbf{O}_{dN,dN} \\
    -QF & Q+F'QF & -F' Q & \ddots & \vdots \\
    \mathbf{O}_{dN,dN} & \ddots & \ddots & \ddots & \mathbf{O}_{dN,dN} \\
    \vdots & \ddots & -QF & Q+F'QF & -F'Q \\
    \mathbf{O}_{dN,dN} & \cdots & \mathbf{O}_{dN,dN} & -QF & \ \,Q
    \end{pmatrix}.
\end{equation*}
By directly stacking all $\bx_{:\ell,t}$ defined in the \textit{feature-wise} specification, we end up with a vector $\widetilde\bx = (\bx_{:1,:}', \dots, \bx_{:d,:}')'$ having the same length as $\bx = (\bx'_{::,1}, \dots, \bx'_{::,T})'$, but with entries in different position, as a consequence of the different ordering in which we stacked the elements $x_{ij,t}$. In the following, we let $\bx_{-i:,:}$ denote all remaining elements of $\bx$ after removing $\bx_{i:,:}$ from the collection.

To move from $\widetilde\bx$ to $\bx$ and reconcile the two vectors, first, we need to identify the indices of the elements in $\widetilde\bx$ that correspond to each element in $\bx$, that is the permutation sequence $\mathtt{s}_i = \{ s_i \}_{i=1}^{NdT}$, with $s_i \in \{1,\dots,NdT\}$, such that the $s_i$th element of $\widetilde\bx$ coincides with the $i$th element of $\bx$.
Second, to obtain the conditional distribution of $\bx_{i:,:} \mid \bx_{-i:,:}$ for each $i=1,\dots,d$, we need to select the appropriate elements of $\bmu$ and $\Psi$, according to $\mathtt{s}_i$.
For $\Psi$, we denote with $\Psi_{ii}$, the sub-matrix of $\Psi$ consisting of the rows/columns whose index is in $\mathtt{s}_i$, with $\Psi_{i,-i}$ the sub-matrix with row index in $\mathtt{s}_i$ and column index not in $\mathtt{s}_i$, and with $\Psi_{-i,-i}$ the sub-matrix whose rows/columns have no index in $\mathtt{s}_i$. Notice that $\Psi_{-i,i} = \Psi_{i,-i}'$.
Then, with a slight abuse of notation, one has
\begin{align*}
    \bmu_i & = \{ \mu_k \in \bmu : k \in \mathtt{s}_i \}, \quad
    \bmu_{-i} = \{ \mu_k \in \bmu : k \notin \mathtt{s}_i \}, \quad
    \Psi = \begin{pmatrix}
    \Psi_{ii} & \Psi_{i,-i} \\
    \Psi_{-i,i} & \Psi_{-i,-i}
    \end{pmatrix}.
\end{align*}
Then, by Gaussian conditioning, one obtains
\begin{equation}
    \bx_{i:,:} \mid \bx_{-i:,:} \sim \mathcal{N}_{dT}\big(\bx_{i:,:}\mid \overline{\bx}_{i:,:}, \overline{K}_i^{-1} \big),
\end{equation}
where
\begin{equation}
    \overline{K}_i^{-1} = \Psi_{ii}^{-1}, 
    \qquad
    \overline{\bx}_{i:,:} = \bmu_i - \Psi_{ii}^{-1} \Psi_{i,-i} (\bx_{-i:,:}-\bmu_{-i}),
\end{equation}
% with $\bmu_i$ and $\bmu_{-i}$ denoting the corresponding subvectors of $\bmu$ and $\Psi$ block-partitioned as
% \begin{equation*}
%     \Psi = \begin{pmatrix}
%     \Psi_{ii} & \Psi_{i,-i} \\
%     \Psi_{-i,i} & \Psi_{-i,-i}
%     \end{pmatrix}.
% \end{equation*}

Combining the transition and measurement equations in compact form yields the state space representation
\begin{align}
    p\big( \overline{\bz}_{i:,:} \mid \br,\bx_{i,:,:},\bx_{-i:,:},\balpha \big) & = \mathcal{N}_{2(N-1)T}\big( \overline{\bz}_{i:,:} \mid G_i \bx_{i:,:}, \, \widetilde\Sigma_i \big), \\
    p(\bx_{i:,:} \mid  \bx_{-i:,:}, \widetilde\Phi, \widetilde\Upsilon) & = \mathcal{N}_{dT}(\bx_{i:,:} \mid \overline{\bx}_{i:,:}, \overline{K}_i^{-1}).
\end{align}
As a result, the posterior (smoothed) distribution for $\bx_{i:,:}$ is obtained as
\begin{align}
    \bx_{i:,:} \mid  \bx_{-i:,:},\bx_{i:,0}, \overline{\bz}_{i:,:}, \br,\balpha, \widetilde\Phi,\widetilde\Upsilon \sim \mathcal{N}_{dT}\big(\bx_{i:,:} \mid  \widehat{\bx}_{i:,:}, \widehat{K}_i^{-1} \big),
\end{align}
where the covariance and mean are
\begin{align}
    \widehat{K}_i & = \overline{K}_i + G_i' \widetilde\Sigma_i^{-1} G_i,
    \qquad
    \widehat{\bx}_{i:,:} = \widehat{K}_i \big( \overline{K}_i \overline{\bx}_{i:,:} + G_i' \widetilde\Sigma_i^{-1} \overline{\bz}_{i:,:} \big).
\end{align}

Recall that $\bx_{::,0} = (\bx_{:1,0}',\dots,\bx_{:d,0}')' \in \R^{dN}$ and $\widetilde\bx_{::,0} = (\bx_{:1,0}',\dots,\bx_{:d,0}')' \in \R^{dN}$ are obtained from different vectorisations of $X_t$, therefore they have elements ordered differently.
To match the order of the elements of the two vectors, we introduce the commutation matrix $U^{(N,N)}$ such that the order of the elements of $U^{(N,N)} \bx_{::,1}$ matches that of $\widetilde\bx_{::,0}$.
Then, recall the definitions of $\widetilde{F} = \Id_d \otimes \widetilde\Phi$, and $\widetilde{Q} = \Id_d \otimes \widetilde\Upsilon^{-1}$, and define $\widetilde{Q}_0 = \Id_d \otimes \Omega_0^*$ and $\underline{\bx}_{::,0}^* = (\underline{\bx}_{:1,0}^{*\prime},\dots,\underline{\bx}_{:d,0}^{*\prime})' \in \R^{dN}$. Then, we obtain
\begin{align*}
    p( \widetilde\bx_{::,0} \mid \bx_{::,1}) 
    & \propto \exp\Big( -\frac{1}{2} \big(U^{(N,N)}\bx_{::,1} -\widetilde{F} \widetilde\bx_{::,0} \big)' \widetilde{Q} \big(U^{(N,N)}\bx_{::,1} -\widetilde{F} \widetilde\bx_{::,0} \big) \Big) \\
    & \quad \times \exp\Big( -\frac{1}{2} (\widetilde\bx_{::,0} - \underline{\bx}_{::,0}^*)' \widetilde{Q}_0 (\widetilde\bx_{::,0} - \underline{\bx}_{::,0}^*) \Big) \\
    & \propto \exp\big( -\frac{1}{2} \big( \widetilde\bx_{::,0}' \big(\widetilde{F}'\widetilde{Q}\widetilde{F} + \widetilde{Q}_0\big) \widetilde\bx_{::,0} - 2\widetilde\bx_{::,0}'\widetilde{F}'\widetilde{Q} U^{(N,N)}\bx_{::,1} -2\widetilde\bx_{::,0}'\widetilde{Q}_0 \underline{\bx}_{::,0}^* \big) \Big).
\end{align*}
Equivalently, $\widetilde\bx_{::,0} \mid \bx_{::,1} \sim \mathcal{N}_{Nd}(\widetilde\bx_{::,0} \mid \overline{\bx}_0^*, (\overline{\Omega}_0^*)^{-1})$, where
\begin{equation}
\overline{\Omega}_0^* = \widetilde{F}'\widetilde{Q}\widetilde{F} + \widetilde{Q}_0, \qquad
\overline{\bx}_0^* = (\overline{\Omega}_0^*)^{-1}\big( \widetilde{F}'\widetilde{Q} U^{(N,N)}\bx_{::,1} + \widetilde{Q}_0 \underline{\bx}_{::,0}^* \big).
\end{equation}

%%%%%%%%%%%%%%%%%%%%%%%%%%%%%%%%%%%%%%%%%%%%%%%%%%%%
\subsection{Proof of the Results in eq. \eqref{eq:full_phi}}
\label{sec:posterior_alpha_phi}

\paragraph{(i) Posterior of $\balpha$.} To derive the posterior full conditional distribution of the vector of node fixed effects, $\balpha$, to ease the notation, we use $\log(\Lambda_t)$ and $f(\bx_t)$ to denote the $N\times N$ matrices whose $ij$th entries are $\log(\lambda_{ij,t})$ and $\bx_{i:,t}' \bx_{j:,t}$, respectively.
Let us define the following quantities:
\begin{align*}
    \widetilde\Lambda_t & = \log(\Lambda_t) - f(\bx_t) = \balpha \biota_N' + \biota_N \balpha' \\
    \operatorname{vec}(\widetilde\Lambda_t) & = (\biota_N \otimes \Id_N) \balpha + (\Id_N \otimes \biota_N) \balpha = \widetilde{D} \balpha
\end{align*}
where $\widetilde{D} \coloneqq (\biota_N \otimes \Id_N) + (\Id_N \otimes \biota_N) \in \{0,1\}^{N^2 \times N}$.

We need to elaborate the likelihood term at time $t$ as a function of $\operatorname{vec}(\widetilde\Lambda_t)$.
First, let us consider the case where all entries are positive counts, which implies two series of auxiliary variables associated to each observation.
Let us define the following matrices:
\begin{equation}
\begin{aligned}
    \btau_t^* & \coloneqq
    \left(\begin{array}{c}
         \big\{ -\log(\tau_{ij,1t}) \big\}_{ij} \\[4pt]
         \big\{ -\log(\tau_{ij,2t}) \big\}_{ij}
    \end{array}\right) \in\R^{2N\times N} \qquad
    \bmu_t^* \coloneqq
    \left(\begin{array}{c}
         \big\{ \mu_{ij,1t} \big\}_{ij} \\[4pt]
         \big\{ \mu_{ij,2t} \big\}_{ij}
    \end{array}\right) \in\R^{2N\times N} \\
    \Lambda_t^* & \coloneqq \biota_2 \otimes \widetilde\Lambda_t =
    \left(\begin{array}{c}
         \widetilde\Lambda_t \\[4pt]
         \widetilde\Lambda_t
    \end{array}\right) \in\R^{2N\times N}
\end{aligned}
\end{equation}
Then, the covariance matrix of the vector $\operatorname{vec}(\btau_t^*)$ is 
\begin{equation}
\begin{aligned}
    \Sigma_t^* & \coloneqq \mathbb{V}\mathrm{ar}\big(\operatorname{vec}(\btau_t^*)\big) = \operatorname{blkdiag}(\Sigma_{1t}^*, \Sigma_{2t}^*) =
    \left(\begin{array}{cc}
       \Sigma_{1t}^*  &  \\
         & \Sigma_{2t}^*
    \end{array}\right) \in\Sp^{2N^2} \\
    \Sigma_{kt}^* & \coloneqq \operatorname{diag}\big( \sigma^2_{r_{11,kt}},\dots,\sigma^2_{r_{1N,kt}},\dots,
    \sigma^2_{r_{N1,kt}},\dots,\sigma^2_{r_{NN,kt}} \big), \qquad k=1,2.
\end{aligned}
\end{equation}
Therefore, we have that
\begin{equation}
    \operatorname{vec}(\btau_t^*) \mid \balpha, \cdot \sim \mathcal{N}_{2N^2}\Big( \operatorname{vec}(\btau_t^*) \mid \operatorname{vec}(\bmu_t^*) + \operatorname{vec}(\widetilde\Lambda_t^*), \: \Sigma_t^* \Big).
\end{equation}
We now define the quantity
\begin{equation}
    \bzeta_t^* \coloneqq \operatorname{vec}(\btau_t^*) - \operatorname{vec}(\bmu_t^*).
\end{equation}
We are going to show that $\operatorname{vec}(\widetilde\Lambda_t^*)$ is linear in $\balpha$. Recalling the definition of $\widetilde\Lambda_t^*$ and the property of the vectorization and Kronecker product, one gets
\begin{align*}
    \operatorname{vec}(\widetilde\Lambda_t^*) & = \operatorname{vec}\big( \biota_2 \otimes \widetilde\Lambda_t \big) \\
    & = D^* \Big( \operatorname{vec}(\biota_2) \otimes \operatorname{vec}\big( \widetilde\Lambda_t \big) \Big) 
     = D^* \Big( \biota_2 \otimes \big( \widetilde{D}\balpha \big) \Big) \\
    & = D^* \Big( \big( \biota_2 \cdot 1 \big) \otimes \big( \widetilde{D} \balpha \big) \Big) = D^* \Big( \big( \biota_2 \otimes \widetilde{D} \big) \big( 1 \otimes \balpha \big) \Big)  \\
    & = D^* \big( \biota_2 \otimes \widetilde{D} \big) \balpha,
\end{align*}
where $D^* \coloneqq (1 \otimes K_{(N,2)} \otimes \Id_N) \in\{0,1\}^{2N^2\times 2N^2}$ and $K_{(N,2)} \in\{0,1\}^{2N\times 2N}$ is a commutation matrix.
One can exploit this result to derive the conditional distribution of $\bzeta_t^*$, which, coupled with a Gaussian prior for $\balpha$, results in
\begin{equation}
\begin{aligned}
    \bzeta_t^* \mid \balpha, \br & \sim \mathcal{N}_{2N^2}\Big( \bzeta_t^* \mid D^* \big( \biota_2 \otimes \widetilde{D} \big) \balpha, \: \Sigma_t^* \Big) \\
    \balpha & \sim \mathcal{N}_N( \balpha \mid \underline{\bmu}_\alpha, \underline{\Sigma}_{\alpha}).
\end{aligned}
\end{equation}
Collecting all time points, one obtains the posterior full conditional of the vector of node fixed effects as
\begin{equation}
    \balpha \mid \btau,\br,\bw,\bx \sim \mathcal{N}_N( \balpha \mid \overline{\bmu}_\alpha, \overline{\Sigma}_{\alpha}),
\end{equation}
where
\begin{align*}
    \overline{\Sigma}_\alpha & = \Big( \underline{\Sigma}_\alpha^{-1} + \big( \biota_2' \otimes \widetilde{D}' \big) D^{*\prime} \Big( \sum_{t=1}^T (\Sigma_t^*)^{-1} \Big) D^* \big( \biota_2 \otimes \widetilde{D} \big) \Big)^{-1}
    \\
    \overline{\bmu}_\alpha & = \overline{\Sigma}_\alpha \Big( \underline{\Sigma}_\alpha^{-1} \underline{\bmu}_\alpha + \big( \biota_2' \otimes \widetilde{D}' \big) D^{*\prime} \Big( \sum_{t=1}^T (\Sigma_t^*)^{-1} \bzeta_t^* \Big) \Big).
\end{align*}
We remark that the matrices $\widetilde{D}$ and $D^*$ are binary and very sparse, while $\Sigma_t^*$ are diagonal, therefore all the posterior quantities can be computed very fast.

Recall that a zero observation allocated to the Poisson component implies the augmentation of the likelihood via a single collection of auxiliary variables, $\tau_{ij,1t},r_{ij,1t}$. In this case, the result above shall be changed to account for this reduced likelihood contribution.
Therefore, let $\kappa_t$ denote the number of zero elements in $Y_t$ and let $S_t \in\{ 0,1 \}^{(2N^2-\kappa_t)\times 2N^2}$ denote a selection matrix that removes from $\bzeta_t^*$ those entries corresponding the zero elements in $Y_t$. Then, define the reduced vector
\begin{equation}
    \bzeta_t^\dag \coloneqq S_t \bzeta_t^* \in\R^{2N^2-\kappa_t}
\end{equation}
whose distribution is
\begin{equation}
    \bzeta_t^\dag \mid \balpha, \br  \sim \mathcal{N}_{2N^2-\kappa_t}\Big( \bzeta_t^\dag \mid S_t D^* \big( \biota_2 \otimes \widetilde{D} \big) \balpha, \: S_t \Sigma_t^* S_t' \Big).
\end{equation}
The resulting posterior for $\balpha$ is analogous to the previous case, with parameters
\begin{align*}
    \overline{\Sigma}_\alpha & = \Big( \underline{\Sigma}_\alpha^{-1} + \big( \biota_2' \otimes \widetilde{D}' \big) D^{*\prime} \Big( \sum_{t=1}^T S_t' (S_t \Sigma_t^* S_t')^{-1} S_t \Big) D^* \big( \biota_2 \otimes \widetilde{D} \big) \Big)^{-1}
    \\
    \overline{\bmu}_\alpha & = \overline{\Sigma}_\alpha \Big( \underline{\Sigma}_\alpha^{-1} \underline{\bmu}_\alpha + \big( \biota_2' \otimes \widetilde{D}' \big) D^{*\prime} \sum_{t=1}^T S_t' (S_t \Sigma_t^* S_t')^{-1} \bzeta_t^\dag \Big).
\end{align*}

\medskip

\begin{remark}[Alternative sampler for $\alpha_i$]
Alternatively, one can obtain the full conditional of $\alpha_i$ as follows. First, let us define the quantities
\begin{align*}
    \widetilde{\boldeta}_{ij,t} & = \begin{pmatrix}
         \alpha_j + \bx_{i:,t}'\bx_{j:,t} + \mu_{r_{ij, 1t}} \\
         \alpha_j + \bx_{i:,t}'\bx_{j:,t} + \mu_{r_{ij, 2t}}
    \end{pmatrix} \\
    \widetilde{\bz}_{ij,t} & = \widetilde\btau_{ij,t} - \widetilde\boldeta_{ij,t}
\end{align*}
therefore
\begin{align}
    p\big( \widetilde\bz_{ij,t} \mid \btau_{ij,t},\br_{ij,t},y_{ij,t}, \bx_{i:,t},\bx_{j:,t}, \alpha_i,\alpha_j, w_{ij,t} \big) = \mathcal{N}_2\big( \widetilde\bz_{ij,t} \mid \alpha_i, \, \widetilde\Sigma_{ij,t} \big).
\end{align}
By stacking all the vectors $\widetilde\bz_{i:,t} = \big( \widetilde\bz_{i1,t}^{\prime},\dots, \widetilde\bz_{i,i-1,t}^{\prime}, \widetilde\bz_{i,i+1,t}^{\prime}, \dots, \widetilde\bz_{iN,t}^{\prime})' \in\R^{2(N-1)}$ one obtains
\begin{align*}
    p\big( \widetilde\bz_{i:,t} \mid \btau_{ij,t},\br_{ij,t},y_{ij,t}, \bx_{i:,t},\bx_{j:,t}, \alpha_i,\alpha_j, w_{ij,t} \big) & = \mathcal{N}_{2(N-1)}\big( \widetilde\bz_{i:,t} \mid \biota_{2(N-1)} \alpha_i, \, \widetilde\Sigma_{i,t} \big),
\end{align*}
where we defined $\widetilde\Sigma_{i,t} = \operatorname{blkdiag}\big( \widetilde\Sigma_{i1,t},\dots, \widetilde\Sigma_{i,i-1, t},$ $\widetilde\Sigma_{i,i+1, t}, \dots,\widetilde\Sigma_{iN,t} \big)$.
Finally, we stack all vectors over the time dimension and define $\widetilde\Sigma_i = \operatorname{blkdiag}(\widetilde\Sigma_{i,1},\dots,\widetilde\Sigma_{i,T})$, $\widetilde\bz_{i:,:} = (\widetilde\bz_{i:,1}^{\prime}, \dots, \widetilde\bz_{i:,T}^{\prime})'$.
Combining the latter result with a Gaussian prior yields
\begin{align*}
    p(\alpha_i \mid \btau_{i:,:},\br_{i:,:},\by_{i:,:}, \bx,\balpha_{-i},\bw_{i:,:}) = \mathcal{N}(\alpha_i \mid \overline{m}_{\alpha,i}, \overline{v}_{\alpha,i}),
\end{align*}
with
\begin{align*}
    \overline{v}_{\alpha,i} = \big( \underline{v}_\alpha^{-1} + \biota_{2T(N-1)}' \widetilde\Sigma_i^{-1} \biota_{2T(N-1)} \big)^{-1}, \qquad
    \overline{m}_{\alpha,i} = \overline{v}_{\alpha,i} \big( \underline{v}_\alpha^{-1}\underline{m}_\alpha + \biota_{2T(N-1)}' \widetilde\Sigma_i^{-1} \widetilde\bz_{i:,:} \big).
\end{align*}
\end{remark}

\medskip
\paragraph{(ii) Posterior of $\Phi$.}
Regarding the full conditional of $\Phi$, let $\bphi = \operatorname{vec}(\Phi)$. Conditionally on the paths of all the latent features, $\bx = (\bx_{::,1}',\dots,\bx_{::,T}')' \in\R^{NdT}$ with $\bx_{::,t} \in\R^{Nd}$, we sample the matrix $\Phi$ from their full conditional distributions.
The vectorised coefficient matrix is drawn from a multivariate Gaussian distribution
\begin{align*}
    p(\bphi \mid \bx,\Upsilon) & \propto \exp\Big( -\frac{1}{2} \Big( \bphi' \underline{\Sigma}_{\Phi}^{-1} \bphi + \sum_{t=1}^T\sum_{i=1}^N  
    -2(\bx'_{i:,t-1}\otimes\bx'_{i:,t} \Upsilon^{-1})\bphi 
    +\bphi'((\bx_{i:,t-1}\bx'_{i:,t-1})\otimes\Upsilon^{-1})\bphi \Big) \Big) \\
    & \propto \mathcal{N}_{d^2}(\bphi \mid \overline{\bmu}_{\Phi}, \overline{\Sigma}_{\Phi}),
\end{align*}
 where
\begin{align*}
    \overline{\Sigma}_{\Phi} = \Big( \underline{\Sigma}_{\Phi}^{-1} + \sum_{t=1}^T \sum_{i=1}^N 
    ((\bx_{i:,t-1}\bx'_{i:,t-1})\otimes\Upsilon^{-1})
    \Big)^{-1}, \qquad
    \overline{\bmu}_{\Phi} = \overline{\Sigma}_{\Phi} \Big( \sum_{t=1}^T  \sum_{i=1}^N (\bx'_{i:,t-1}\otimes\bx'_{i:,t} \Upsilon^{-1}) \Big).
\end{align*}

If $\Phi$ is a diagonal matrix, $\Phi = \operatorname{diag}(\bphi_{*})$, the vector of coefficients is drawn from a multivariate Gaussian distribution
\begin{align*}
    p(\bphi_{*} \mid \bx,\Upsilon) & \propto \mathcal{N}_{d}(\bphi_{*} \mid \overline{\bmu}_{\Phi_{*}}, \overline{\Sigma}_{\Phi_{*}}),
\end{align*}
with $\widetilde{\bX}_{i, t} = \operatorname{diag}(\bx_{i:, t})$ and 
\begin{align*}
    \overline{\Sigma}_{\Phi_{*}} = \Big( \underline{\Sigma}_{\Phi_{*}}^{-1} + \sum_{t=1}^T\sum_{i=1}^N \widetilde{\bX}_{i, t-1}'\Upsilon^{-1}\widetilde{\bX}_{i, t-1} \Big)^{-1}, \qquad
    \overline{\bmu}_{\Phi_{*}} = \overline{\Sigma}_{\Phi_{*}} \Big( \sum_{t=1}^T\sum_{i=1}^N \widetilde{\bX}_{i,t-1}'\Upsilon^{-1}_n\bx_{i:,t} \Big).
\end{align*}

\medskip
\paragraph{(ii) Posterior of $\widetilde\Phi$.}
Regarding the full conditional of  $\widetilde\Phi$ we derive the distribution of $\widetilde\bphi = \operatorname{vec}(\widetilde\Phi)$ conditionally given the paths of all the latent features, $\bx = (\bx_{::,1}',\dots,\bx_{::,T}')' \in\R^{NdT}$ and obtain a multivariate Gaussian distribution:
\begin{align*}
    p(\widetilde\bphi \mid \bx,\widetilde\Upsilon) & \propto \exp\Big( -\frac{1}{2} \Big( \widetilde\bphi' \underline{\Sigma}_{\widetilde\Phi}^{-1} \widetilde\bphi + 
    \sum_{t=1}^T\sum_{\ell=1}^d  
    -2(\bx'_{:\ell,t-1}\otimes\bx'_{:\ell,t} \widetilde\Upsilon^{-1})\widetilde\bphi
    +\widetilde\bphi'( (\bx_{:\ell,t-1}\bx'_{:\ell,t-1}) \otimes \widetilde\Upsilon^{-1})\widetilde\bphi \Big) \Big) \\
    & \propto \mathcal{N}_{N^2}(\widetilde\bphi \mid \overline{\bmu}_{\widetilde\Phi}, \overline{\Sigma}_{\widetilde\Phi}),
\end{align*}
where
\begin{align*}
    \overline{\Sigma}_{\widetilde\Phi} = \Big( \underline{\Sigma}_{\widetilde\Phi}^{-1} + \sum_{t=1}^T \sum_{\ell=1}^d 
    ((\bx_{:\ell,t-1}\bx'_{:\ell,t-1})\otimes\widetilde\Upsilon^{-1})
    \Big)^{-1}, \qquad
    \overline{\bmu}_{\widetilde\Phi} = \overline{\Sigma}_{\widetilde\Phi} \Big( \sum_{t=1}^T  \sum_{\ell=1}^d (\bx'_{:\ell,t-1}\otimes\bx'_{:\ell,t} \widetilde\Upsilon^{-1}) \Big).
\end{align*}

If $\widetilde\Phi$ is a diagonal matrix, $ \widetilde\Phi = \operatorname{diag}(\widetilde\bphi_{*})$, the vector of coefficients is drawn from a multivariate Gaussian distribution
\begin{align*}
    p(\widetilde\bphi_{*} \mid \bx,\widetilde\Upsilon) & \propto \mathcal{N}_{N}(\widetilde\bphi_{*} \mid \overline{\bmu}_{\widetilde\Phi_{*}}, \overline{\Sigma}_{\widetilde\Phi_{*}}),
\end{align*}
where  $\bX_{\ell, t, *} = \operatorname{diag}(\widetilde\bx_{:\ell, t})$
% assuming $\underline{\Sigma}^{-1}_{\Phi_{*}}$ is a $ND$ diagonal matrix
\begin{align*}
    \overline{\Sigma}_{\widetilde\Phi_{*}} = \Big( \underline{\Sigma}^{-1}_{\widetilde\Phi_{*}} + \sum_{t=1}^T  \sum_{\ell=1}^d \bX'_{\ell, t, *}\widetilde\Upsilon^{-1}\bX_{\ell, t, *}
    \Big)^{-1}, \qquad
    \overline{\bmu}_{\widetilde\Phi_{*}} = \overline{\Sigma}_{\widetilde\Phi_{*}} \Big( \sum_{t=1}^T \sum_{\ell=1}^d \bX_{\ell,t-1,*}\widetilde\Upsilon^{-1}\widetilde\bx_{:\ell, t}\Big).
\end{align*}

%%%%%%%%%%%%%%%%%%%%%%%%%%%%%%%%%%%%%%%%%%%%%%%%%%%%
\subsection{Proof of the Results in eq. \eqref{eq:post_gh_1} and  eq. \eqref{eq:post_gh_2}}

Let $\boldsymbol{\varpi} = \{\varpi_{ij}| i,j = 1, \dots, N\}$, a graphical Horseshoe prior \citep{li2019graphical} for the precision matrix $\widetilde\Upsilon^{-1} = \widetilde\Omega = \{\omega_{ij}\}_{ij = 1}^{N}$ assumes the following specification:
\begin{align}
    p(\widetilde\Omega, \rho_\Omega, \boldsymbol{\varpi}) & \propto \mathcal{C}^+(\rho_\Omega \mid 0,1)\prod_{i<j} \mathcal{N}(\omega_{ij} \mid \varpi_{ij}^2, \rho_\Omega^2) \mathcal{C}^+(\varpi_{ij} \mid 0,1) \times \I(\widetilde\Omega  \in\Sp^N),
\end{align}
where $\mathcal{C}^+$ denotes the half Cauchy distribution and $p(\omega_{ii}) \propto 1$ for $i=1,\dots,N$.  Combining the likelihood and prior, one obtains:
\begin{align*}
    p(\widetilde\Omega, \rho_\Omega, \boldsymbol{\varpi} \mid \bx,\widetilde\Phi) & \;\propto\; \mathcal{C}^+(\rho_\Omega \mid 0,1)\prod_{i<j} \mathcal{N}\!\big(\omega_{ij}\mid 0, \tau^2\varpi_{ij}^2\big)\mathcal{C}^+(\varpi_{ij} \mid 0,1) \\
     & \quad \times \abs{\widetilde\Omega}^{Td/2} \exp\!\Big(-\frac{1}{2}\operatorname{tr}(S\widetilde\Omega)\Big) \times \I(\widetilde\Omega \in \Sp^N),
\end{align*}
with $S = \sum_{t = 1}^T \sum_{\ell = 1}^d(\widetilde\bx_{\ell, t} - \widetilde\Phi\widetilde\bx_{\ell, t-1} )(\widetilde\bx_{\ell, t} - \widetilde\Phi\widetilde\bx_{\ell, t-1})'$.
Therefore, the posterior distribution is proportional to:
\begin{equation*}
    p(\widetilde\Omega \mid \bx,\widetilde\Phi,\rho_\Omega,\boldsymbol{\varpi}) \propto \big| \widetilde\Omega \big|^{\frac{Nd}{2}} \exp\big( -\operatorname{tr}\big(\frac{1}{2} S \widetilde\Omega\big)\big) \prod_{i<j} \exp\bigg( -\frac{\omega_{i j}^2}{2 \varpi_{i j}^2 \rho_\Omega^2}\bigg) \I(\widetilde\Omega \in\Sp^N).
\end{equation*}

Sampling from this joint distribution is made one column/row at a time \citep{li2019graphical}, where $\bomega_i = (\omega_{i1},\dots,\omega_{iN})'$ denotes the $i$th row of $\widetilde\Omega$ (a similar notation is used for other matrices). Without loss of generality, consider sampling the last column and the last row.
First, letting $(-N)$ denote the set of all indices $\{1,\dots,N\}$ except for $N$, we partition the matrices
\begin{equation*}
\begin{aligned}
    \widetilde\Omega & = \begin{pmatrix}
    \Omega_{(-N)(-N)} & \bomega_{(-N) N} \\
    \bomega_{(-N) N}' & \omega_{N N}
    \end{pmatrix}, \,\,
    S = \begin{pmatrix}
    S_{(-N)(-N)} & \mathbf{s}_{(-N) N} \\
    \mathbf{s}_{(-N) N}' & s_{N N}
    \end{pmatrix}, \,\,
    \boldsymbol{\varpi} = \begin{pmatrix}
    \boldsymbol{\varpi}_{(-N)(-N)} & \bvarpi_{(-N) N} \\
    \bvarpi_{(-N) N}' & 1
    \end{pmatrix}
\end{aligned}
\end{equation*}
where $\boldsymbol{\varpi}_{(-N)(-N)}$ and $\bvarpi_{(-N) N}$ have entries $\varpi_{ij}^2$, whereas the diagonal elements of $\boldsymbol{\varpi}_{(-N)(-N)}$ are set to 1.
Then, the full conditional of the last column of $\widetilde\Omega$ is
\begin{equation*}
    \begin{aligned}
    p(\bomega_{(-N) N}, \omega_{NN} & \mid \Omega_{(-N)(-N)}, \bx, \widetilde\Phi, \boldsymbol{\varpi}, \rho) \propto (\omega_{N N} -\bomega_{(-N) N}' \Omega_{(-N)(-N)}^{-1} \bomega_{(-N) N})^{n/2} \\
    & \quad \times \exp\big(-\mathbf{s}_{(-N) N}' \bomega_{(-N) N} -s_{N N} \omega_{N N} / 2-\bomega_{(-N) N}' \big(\boldsymbol{\varpi}^* \rho^2\big)^{-1} \bomega_{(-N) N} / 2\big)
    \end{aligned}
\end{equation*}
where $\boldsymbol{\varpi}^*$ is a diagonal matrix with $\bvarpi_{(-N) N}$ in the diagonal.
Second, we make a variable change $(\bomega_{(-N)N}, \omega_{NN}) \to (\bbeta_\Omega,\gamma)$ via transformation $\bbeta_\Omega = \bomega_{(-N) N}$ and $\gamma = \omega_{N N}- \bomega_{(-N) N}' \Omega_{(-N)(-N)}^{-1} \bomega_{(-N) N}$, which has constant Jacobian.
Therefore, the full conditional distribution of $\bbeta$ and $\gamma$ is
\begin{equation*}
    \begin{aligned}
    p(\bbeta_\Omega, \gamma \mid \Omega_{(-N)(-N)}, \bx, \widetilde\Phi, \rho_\Omega, \boldsymbol{\varpi}) & \propto \gamma^{n / 2} \exp\left(-\frac{1}{2}\left(s_{NN} \gamma + \bbeta_\Omega' s_{NN} \Omega_{(-N)(-N)}^{-1} \bbeta_\Omega \right.\right. \\
    & \left.\left.\quad + \bbeta_\Omega' (\boldsymbol{\varpi}^* \rho_\Omega^2)^{-1} \bbeta_\Omega +2 \mathbf{s}_{(-N)N}'\bbeta_\Omega \right)\right) \\
    & \propto \mathcal{G}a(\gamma \mid n/2+1, s_{NN} / 2) \mathcal{N}_{N-1}(\bbeta \mid -C\mathbf{s}_{(-N)N}, C),
    \end{aligned}
\end{equation*}
where $C = \big(s_{NN} \Omega_{(-N)(-N)}^{-1} + (\boldsymbol{\varpi}^* \rho_\Omega^2)^{-1} \big)^{-1}$. Finally, by inverting the transformation, one obtains a draw from $\bomega_N$. Applying this approach recursively to each column/row yields an update to $\widetilde\Omega$.

Concerning the local and global shrinkage hyperparameters, $\varpi_{ij}$ and $\rho_\Omega$, we exploit the stochastic representation of a half-Cauchy as a mixture of inverse gamma distributions.
Thus, introducing the latent variables $\eta_{ij}^\varpi$ and $\eta^\rho$, one obtains the full conditional posterior distribution of $\varpi_{ij}$ and $\eta_{ij}^\varpi$ obtained as
\begin{equation*}
    \varpi_{ij} \mid \eta_{ij}^\varpi, \omega_{ij}, \rho_\Omega \sim \mathcal{IG}\bigg( \varpi_{ij}\mid 1, \frac{1}{\eta_{ij}^\varpi} + \frac{\omega_{ij}^2}{2\rho_\Omega^2} \bigg), \qquad
    \eta_{ij}^\varpi \mid \varpi_{ij} \sim \mathcal{IG}\bigg( \eta_{ij}^\varpi \mid 1, 1 + \frac{1}{\varpi_{ij}^2} \bigg).
\end{equation*}
The full conditional posteriors for $\rho_\Omega^2$ and the auxiliary variable $\eta^\rho$ are obtained in an analogous way.

If instead we assume a node-wise specification, a conjugate inverse Wishart prior for the covariance matrix $\Upsilon$ results in the following inverse Wishart full conditional distribution
\begin{align*}
    p(\Upsilon \mid \bx,\Phi) & \propto \abs{\Upsilon}^{-\frac{\underline{\nu}_\Upsilon}{2}} \exp\Big( -\frac{1}{2} \operatorname{tr}(\underline{\Psi}_\Upsilon \Upsilon^{-1}) \Big) \\
    & \quad \times \abs{\Upsilon}^{-\frac{TN}{2}} \exp\Big( -\frac{1}{2} \sum_{t=1}^T\sum_{i=1}^N \operatorname{tr}\big( (\bx_{it}-\Phi\bx_{i:,t-1}) (\bx_{it}-\Phi\bx_{i:,t-1})' \Upsilon^{-1} \big) \Big) \\
    & \propto \mathcal{IW}_{d}( \Upsilon \mid \overline{\Psi}_\Upsilon, \overline{\nu}_\Upsilon),
\end{align*}
with
\begin{align*}
    \overline{\Psi}_\Upsilon = \underline{\Psi}_\Upsilon + \sum_{t=1}^T\sum_{i=1}^N (\bx_{i:,t}-\Phi\bx_{i:,t-1}) (\bx_{i:,t}-\Phi\bx_{i:,t-1})', \qquad
    \overline{\nu}_\Upsilon = \underline{\nu}_\Upsilon + TN.
\end{align*}

%%%%%%%%%%%%%%%%%%%%%%%%%%%%%%%%%%%%%%%%%%%%%%%%%%%%
\subsection{Proof of the Results in eq. \eqref{eq:full_w} - eq. \eqref{eq:full_z}}
\label{sec:posterior_ZI_details}
%Posterior Sampling for Zero-Inflated Poisson
Before deriving the full conditional, note that in the Zero-Inflated Poisson, 
for each pair of nodes $i,j$ and time $t$, the count-valued observed edge is distributed as
\begin{equation}
    y_{ij,t} \mid \lambda_{ij,t}, z_{ij,t} \sim p_{ij,t} \mathcal{P}oi(y_{ij,t} \mid \lambda_{ij,t}) + (1-p_{ij,t}) \delta_{\{0\}}(y_{ij,t}),
\end{equation}
where $p_{ij,t} = \P(z_{ij,t} > 0)$ and
    $z_{ij,t} = \bbeta_i'\bv_{i,t} + \bbeta_j'\bv_{j,t} + \varepsilon_{ij,t}, \qquad 
    \upsilon_{ij,t} \sim \mathcal{N}( \upsilon_{ij,t} \mid 0,1)$. Introducing an auxiliary allocation variable, $w_{ij,t}$ to resolve the mixture, one obtains
\begin{equation}
    w_{ij,t} = \I(z_{ij,t} > 0), \qquad
    p(w_{ij,t} =1) = p_{ij,t} = \P(z_{ij,t} > 0).
\end{equation}
Therefore, the complete data likelihood for $y_{ij,t}$ is obtained as
\begin{align*}
p(y_{ij,t} \mid \lambda_{ij,t}, w_{ij,t}) = \Bigg( \frac{\lambda_{ij,t}^{\,y_{ij,t}} e^{-\lambda_{ij,t}}}{y_{ij,t}!} \Bigg)^{\I(w_{ij,t} = 1)} \times \Big( \delta_{\{0\}}(y_{ij,t}) \Big)^{\I(w_{ij,t} = 0)}.
\end{align*}

Notice that an equivalent formulation is obtained by specifying
\begin{equation}
    y_{ij,t} = w_{ij,t} \widetilde{y}_{ij,t},
\end{equation}
where $w_{ij,t}$ is defined as before and $\widetilde{y}_{ij,t}$ is a partially observed count variable (i.e., it is observed only when $w_{ij,t} = 1$).

\medskip
\paragraph{(i) Posterior of $w_{ij,t}$.}
The posterior full conditional distribution for the allocation variable $w_{ij,t}$ given in Eq. \ref{eq:full_w} is a Bernoulli and is obtained as follows. By construction of the ZIP, $y_{ij,t} >0$ is necessarily associated with $w_{ij,t}=1$, thereby restricting the update of $w_{ij,t}$ to the edges such that $y_{ij,t} = 0$.
Therefore one has
\begin{align*}
    p(w_{ij,t} = 1 \mid y_{ij,t} > 0, z_{ij,t},\lambda_{ij,t}) & = 1 \\
    p(w_{ij,t} = 0 \mid y_{ij,t} > 0, z_{ij,t},\lambda_{ij,t}) & = 0 \\
    p(w_{ij,t} = 1 \mid y_{ij,t} = 0, z_{ij,t},\lambda_{ij,t}) & \propto p_{ij,t} \mathcal{P}oi( 0 \mid \lambda_{ij,t}) \\
    p(w_{ij,t} = 0 \mid y_{ij,t} = 0, z_{ij,t},\lambda_{ij,t}) & \propto (1-p_{ij,t}),
\end{align*}
which imply
\begin{align}
    w_{ij,t} \mid z_{ij,t},\lambda_{ij,t} & \sim \mathcal{B}ern( w_{ij,t}\mid p_{ij,t}^*), \qquad
    p_{ij,t}^*  = \frac{p_{ij,t} \mathcal{P}oi( 0 \mid \lambda_{ij,t})}{(1-p_{ij,t}) + p_{ij,t} \mathcal{P}oi( 0 \mid \lambda_{ij,t})}.
\end{align}

\medskip
\paragraph{(ii) Posterior of $\bbeta_i$.}
Regarding the full conditional of $\bbeta_i$, it can be derived following a strategy analogous to the one used for $\alpha_i$ in the non-inflated case. Define
\begin{align*}
    \widehat{z}_{ij,t} & = \bbeta_j'\bv_{j,t}, 
    & \widehat{u}_{ij,t} & = z_{ij,t} - \widehat{z}_{ij,t}.
\end{align*}
It follows that
\begin{align}
    p\big(\widehat{u}_{ij,t} \mid \bbeta_i, \bbeta_j, \bv_{i,t}, \bv_{j,t} \big) = \mathcal{N}\!\big( \widehat{u}_{ij,t} \mid \bbeta_i' \bv_{i,t}, \, 1 \big).
\end{align}
By stacking all values for $j\neq i$ at time $t$, let us define
\begin{equation*}
    \widehat{\mathbf{u}}_{i,t} = \big( \widehat{u}_{i1,t},\dots,\widehat{u}_{i,i-1,t}, \widehat{u}_{i,i+1,t},\dots,\widehat{u}_{iN,t} \big)' \in \mathbb{R}^{N-1},
\end{equation*}
thus obtaining
\begin{align*}
    p\big( \widehat{\mathbf{u}}_{i,t} \mid \bbeta_i, \bbeta_{-i}, \bv_{i,t}, \bv_{-i,t} \big) = \mathcal{N}_{N-1}\big( \widehat{\mathbf{u}}_{i,t} \mid \biota_{N-1}\, \bbeta_i'\bv_{i,t}, \, \Id_{N-1} \big).
\end{align*}
Stacking over the time dimension, one obtains
\begin{equation*}
    \widehat{\mathbf{u}}_i = (\widehat{\mathbf{u}}_{i,1}',\dots,\widehat{\mathbf{u}}_{i,T}')' \in \mathbb{R}^{(N-1)T}, \qquad
    V_i = \begin{pmatrix}
        \biota_{N-1}\bv_{i,1}'\\
        \vdots\\
        \biota_{N-1}\bv_{i,T}'
        \end{pmatrix}
    \in \mathbb{R}^{(N-1)T \times L}.
\end{equation*}
Then
\begin{align*}
    p\big( \widehat{\mathbf{u}}_i \mid \bbeta_i, \bbeta_{-i}, \bv_i, \bv_{-i} \big) = \mathcal{N}_{(N-1)T}\big( \widehat{\mathbf{u}}_i \mid V_i \bbeta_i, \, \Id_{(N-1)T} \big).
\end{align*}
Combining this likelihood with the Gaussian prior
\begin{equation*}
    \bbeta_i \sim \mathcal{N}_L( \bbeta_i \mid \underline{\mathbf{m}}_{\beta}, \underline{\Sigma}_{\beta}),
\end{equation*}
yields the full conditional posterior distribution
\begin{align*}
    p(\bbeta_i \mid \bbeta_{-i}, \bv) = \mathcal{N}_L\big( \bbeta_i \mid \overline{\mathbf{m}}_{\beta,i}, \overline{\Sigma}_{\beta,i} \big),
\end{align*}
with
\begin{align*}
    \overline{\Sigma}_{\beta,i} & = \Big( \underline{\Sigma}_{\beta}^{-1} + V_i'V_i \Big)^{-1},
    & \overline{\mathbf{m}}_{\beta,i} & = \overline{\Sigma}_{\beta,i}
    \Big(\underline{\Sigma}_{\beta}^{-1}\underline{\mathbf{m}}_{\beta} + V_i'\widehat{\mathbf{u}}_i \Big).
\end{align*}

\medskip
\paragraph{(ii) Posterior of $z_{ij,t}$.}
The full conditional of the latent variables $z_{ij,t}$ for $i,j = 1, \dots, N$, $i>j$, and $t = 1, \dots, T$ is given in Eq. \ref{eq:full_z}. Given the collection of all $w_{ij,t} = \mathbb{I}(y_{ij,t}>0)$ and $\mu_{z,ij,t} = \bbeta_i'\bv_{i,t} + \bbeta_j'\bv_{j,t} $ for $i,j = 1, \dots, N$ with $j > i$ and $t = 1, \dots, T$, the latent variables are conditionally independent with conditional posterior
\begin{equation*}
    p(z_{ij,t} \mid w_{ij,t}, \bbeta_i,\bbeta_j) \propto \P(w_{ij,t}=1 \mid z_{ij,t}) f_u(z_{ijt}- \mu_{z,ij,t}).
\end{equation*}
The posterior of $z_{ij,t}$ is obtained as
\begin{align*}
    p(z_{ij,t} \mid w_{ij,t}, \bbeta_i,\bbeta_j) = \begin{cases}
        f_u(z_{ij,t}-\mu_{z,ij,t})\I(z_{ij,t} \leq 0) & \text{if } y_{ij,t} = 0 \\
        f_u(z_{ij,t}-\mu_{z,ij,t})\I(z_{ij,t} > 0) & \text{if } y_{ij,t} > 0
    \end{cases}
\end{align*}
therefore $z_{ij,t} = \mu_{z,ij,t} + u_{ij,t}$, where 
\begin{equation*}
    u_{ij,t} \sim \begin{cases}
        f_u(u_{ij,t})\I(u_{ij,t} \leq -\mu_{z,ij,t}) & \text{if } w_{ij,t} = 0 \\
        f_u(u_{ij,t})\I(u_{ij,t} > -\mu_{z,ij,t}) & \text{if } w_{ij,t} = 1
    \end{cases}
\end{equation*}
For the probit and logit specifications the quantile function $F_u^{-1}(p)$ is available in closed form ($F_u^{-1}(p) = \Phi^{-1}(p)$ and $F_u^{-1}(p) = \log(p) -\log(1-p))$, respectively) and in these cases $f_u(u)$ is symmetric around 0. Thus, we can sample from the posterior density $z_{ij,t}$ by drawing $\kappa_{ij,t} \sim \mathcal{U}(0,1)$ then computing
\begin{align}
    z_{ij,t} = \mu_{ij,t} + F_u^{-1}\big( w_{ij,t} + \kappa_{ijt}(1 -w_{ij,t} -F_u(\mu_{ij,t}) \big).
\label{eq:posterior_zt}
\end{align}

Let us define the vectors $\bbeta_{ij} = (\bbeta_i', \bbeta_j')' \in\R^{2L}$ and $\bv_{ijt} = (\bv_{i,t}', \bv_{j,t}')' \in\R^{2L}$, to obtain
\begin{equation*}
    z_{ij,t} = \bv_{ijt}' \bbeta_{ij} + \upsilon_{ij,t}, \qquad \upsilon_{ij,t} \sim \mathcal{N}(\upsilon_{ij,t}  \mid 0,1).
\end{equation*}

Following \cite{zens2024ultimate}, we introduce two auxiliary variables making a location and scale transformation of $z_{ij,t}$ to improve the mixing of the MCMC algorithm. 

% The location-based transform consists in the expanded model
% \begin{equation}
% \begin{aligned}
%     \widetilde{z}^L_{ij,t} & = z_{ij,t} + \gamma_{1t} \\
%     w_{ij,t} & = \I(\widetilde{z}^L_{ij,t} > \gamma_{1t}) \\
%     \widetilde{z}^L_{ij,t} & = \gamma_{1t} + \bv_{ij,t}'\bbeta_{ij} + \upsilon_{ij,t}.
% \end{aligned}
% \end{equation}
% Instead, the scale-based transformation is given by
% \begin{equation}
% \begin{aligned}
%     \widetilde{z}^S_{ij,t} & = \sqrt{\gamma_{2t}} z^L_{ij,t} \\
%     w_{ij,t} & = \I(\widetilde{z}^S_{ij,t} > 0) \\
%     \widetilde{z}^S_{ij,t} & = \sqrt{\gamma_{2t}} \bv_{ij,t}'\bbeta_{ij} + \sqrt{\gamma_{2t}}\upsilon_{ij,t}
% \end{aligned}
% \end{equation}

% \bigskip

A posteriori, after having drawn $z_{ij,t}$ from eq.~\eqref{eq:posterior_zt}, we perform the location-based parameter expansion as follows:
\begin{itemize}
    \item sample $\widetilde{\gamma}_{1t} \sim \mathcal{N}(\widetilde{\gamma}_{1t} \mid 0, \underline{G}_{\gamma,1t})$ and propose $\widetilde{z}_{ij,t} = z_{ij,t} + \widetilde{\gamma}_{1t}$
    \item sample $\gamma_{1t}^* \mid \widetilde{\gamma}_{1t}, \bz_t, \bw_t \sim T\mathcal{N}\big(\gamma_{1t}^*\mid \overline{\mu}_{\gamma,1t} , \overline{G}_{\gamma,1t} ; \big( L(\widetilde{\gamma}_{1t}), U(\widetilde{\gamma}_{1t}) \big) \big)$, a Gaussian truncated on the interval $\big( L(\widetilde{\gamma}_{1t}), U(\widetilde{\gamma}_{1t}) \big)$
    \item define $z_{ij,t}^L = \widetilde{z}_{ij,t} - \gamma_{1t}^*$
\end{itemize}
Let us define $\widetilde{z}^L_{ij,t} = z_{ij,t} + \widetilde{\gamma}_{1t}$ in the expanded model
\begin{equation*}
\begin{aligned}
    w_{ij,t} & = \I(\widetilde{z}_{ij,t} > \gamma_{1t}^*) \\
    \widetilde{z}_{ij,t} & = \gamma_{1t}^* + \bv_{ijt}' \bbeta_{ij} + \upsilon_{ij,t}, \qquad \upsilon_{ij,t} \sim \mathcal{N}(\upsilon_{ij,t} \mid 0,1).
\end{aligned}
\end{equation*}
Therefore, combining the Gaussian prior $\bbeta_{ij} \sim \mathcal{N}_{2L}(\bbeta_{ij} \mid \mathbf{0}_{2L}, \underline{\Sigma}_\beta)$ with the conditional distribution $\widetilde{z}_{ij,t} \mid \bbeta_{ij}, \gamma_{1t}^* \sim \mathcal{N}\big(\widetilde{z}_{ij,t} \mid \gamma_{1t}^* + \bv_{ijt}' \bbeta_{ij}, \, 1 \big)$ to obtain the full conditional posterior
\begin{equation}
    \bbeta_{ij} \mid \bz_{ij}, \gamma_{1t}^*, \widetilde{\gamma}_{1t} \sim \mathcal{N}_{2L}\big(\bbeta_{ij} \mid \overline{\bmu}_\beta, \overline{\Sigma}_\beta \big),
\end{equation}
with
\begin{align*}
    \overline{\Sigma}_\beta = \Bigg( \underline{\Sigma}_\beta^{-1} + \sum_{t=1}^T \bv_{ij,t}\bv_{ij,t}' \Bigg)^{-1} \qquad
    \overline{\bmu}_\beta = \overline{\Sigma}_\beta \Bigg( \sum_{t=1}^T (z_{ij,t}+\widetilde{\gamma}_{1t} -\gamma_{1t}^* ) \bv_{ij,t} \Bigg).
\end{align*}
The posterior distribution of $\gamma_{1t}^*$ is obtained conditionally on $\bz$ and $\widetilde{\gamma}_{1t}$, but marginally with respect to $\bbeta$, as follows:
\begin{align}
    p(\gamma_{1t}^{*} \mid \widetilde{\gamma}_{1t}, \bz) = p(\gamma_{1t}^{*}) \int \prod_{i=1}^N \prod_{j < i} p(\widetilde{z}^L_{ij,t} \mid \gamma_{1t}^{*}, \bbeta_{ij}) p(\bbeta_{ij})\mathbb{I}(L(\widetilde\gamma_{1t}) \leq \gamma_{1t}^{*} \leq U(\widetilde\gamma_{1t}))\: \mathrm{d}\bbeta_{ij}.
\end{align}
Using the previous result and making explicit the dependence on $\gamma_{1t}$, the integral can be evaluated as
{\footnotesize
\begin{align*}
    & \int  \prod_{i=1}^N \prod_{j < i} p(\widetilde{z}^L_{ij,t} \mid \gamma_{1t}^{*}, \bbeta_{ij}) p(\bbeta_{ij})\mathbb{I}(L(\widetilde\gamma_{1t}) \leq \gamma_{1t}^{*} \leq U(\widetilde\gamma_{1t})) \: \mathrm{d}\bbeta_{ij} = \\
    % & \propto \int  \prod_{i=1}^N \prod_{j < i} \exp\Bigg(\!\! -\frac{1}{2} \big( \bbeta_{ij}' \bv_{ijt}\bv_{ijt}' \bbeta_{ij} -2 (z_{ij,t}+\widetilde{\gamma}_{1t} -\gamma_{1t}) \bv_{ijt}' \bbeta_{ij} + (z_{ij,t}+\widetilde{\gamma}_{1t} -\gamma_{1t})^2 \big) \Bigg) \: \mathrm{d}\bbeta_{ij} \\
    & \propto \exp\Bigg(\!\! -\frac{1}{2} \sum_{i=1}^N \sum_{j < i} (z_{ij,t}+\widetilde{\gamma}_{1t} -\gamma_{1t}^{*})^2 \Bigg)\mathbb{I}(L(\widetilde\gamma_{1t}) \leq \gamma_{1t}^{*} \leq U(\widetilde\gamma_{1t})) \\
    & \quad \times \prod_{i=1}^N \prod_{j < i} \int \exp\Bigg(\!\! -\frac{1}{2} \big( \bbeta_{ij}' \overline{\Sigma}_\beta^{-1} \bbeta_{ij} -2 \overline{\bmu}_\beta' \overline{\Sigma}_\beta^{-1} \bbeta_{ij} + \overline{\bmu}_\beta' \overline{\Sigma}_\beta^{-1} \overline{\bmu}_\beta - \overline{\bmu}_\beta' \overline{\Sigma}_\beta^{-1} \overline{\bmu}_\beta \big) \Bigg) \: \mathrm{d}\bbeta_{ij} \\
    & \propto \exp\Bigg(\!\! -\frac{1}{2} \sum_{i=1}^N \sum_{j < i} \gamma_{1t}^{*2} \big(1 - \bv_{ij,t}' \overline{\Sigma}_\beta \bv_{ij,t} \big) -2\gamma_{1t} \big( (z_{ij,t}+\widetilde{\gamma}_{1t}) - \bv_{ij,t}'\overline{\Sigma}_\beta (z_{ij,t}+\widetilde{\gamma}_{1t})\bv_{ij,t} \big) \Bigg)\mathbb{I}(L(\widetilde\gamma_{1t}) \leq \gamma_{1t}^{*} \leq U(\widetilde\gamma_{1t})),
\end{align*}}

where $L(\widetilde\gamma_{1t}) = max_{ij: w_{ij,t} = 0} z_{ij,t} + \widetilde\gamma_{1t}$ and $U(\widetilde\gamma_{1t}) = min_{ij: w_{ij,t} = 1} z_{ij,t} + \widetilde\gamma_{1t}$.

Therefore, combined with a Gaussian prior, $\gamma_{1t}^{*} \sim \mathcal{N}(0, \underline{G}_{\gamma,1t})$, one obtains the conditional posterior %\textbf{M: with a truncation???}
\begin{equation}
    \gamma_{1t}^{*} \mid \widetilde{\gamma}_{1t}, \bz \sim \mathcal{N}( \gamma_{1t}^{*} \mid \overline{\mu}_{\gamma,1t}, \overline{G}_{\gamma,1t})\mathbb{I}(L(\widetilde\gamma_{1t}) \leq \gamma_{1t}^{*} \leq U(\widetilde\gamma_{1t}) ),
\end{equation}
where
\begin{align*}
    \overline{G}_{\gamma,1t} & = \Bigg( \underline{G}_{\gamma,1t}^{-1} + \sum_{i=1}^N \sum_{j < i} \big(1 - \bv_{ij,t}' \overline{\Sigma}_\beta \bv_{ij,t} \big) \Bigg)^{-1}, \\
    \overline{\mu}_{\gamma,1t} & = \underline{G}_{\gamma,1t} \Bigg( \sum_{i=1}^N \sum_{j < i} (z_{ij,t}+\widetilde{\gamma}_{1t}) - \bv_{ij,t}'\overline{\Sigma}_\beta (z_{ij,t}+\widetilde{\gamma}_{1t}) \bv_{ij,t}  \Bigg).
\end{align*}

\noindent Then, perform the scale-based parameter expansion as follows:
\begin{itemize}
    \item sample $\widetilde{\gamma}_{2t} \sim \mathcal{IG}(\widetilde{\gamma}_{2t} \mid \underline{a}_{\gamma,2}, \underline{b}_{\gamma,2})$ and propose $\widetilde{z}_{ij,t}^S = \sqrt{\widetilde{\gamma}_{2t}} z_{ij,t}^L$
    \item sample $\gamma_{2t}^* \mid \widetilde{\gamma}_{2t}, \bz_t^L, \bw_t \sim \mathcal{IG}(\gamma_{2t}^* \mid  \overline{a}_{\gamma,2t},    \overline{b}_{\gamma,2t})$
    \item define $z_{ij,t}^{LS} = \sqrt{\dfrac{\widetilde{\gamma}_{2t}}{\gamma_{2t}^*}} z_{ij,t}^L$
\end{itemize}
Let us define $\widetilde{z}_{ij,t}^S = \sqrt{\widetilde{\gamma}_2} z_{ij,t}^L$ in the expanded model
\begin{equation*}
\begin{aligned}
    w_{ij,t} & = \I(\widetilde{z}_{ij,t}^S > 0) \\
    \widetilde{z}_{ij,t}^S & = \sqrt{\gamma_{2t}^{*}}\bv_{ijt}' \bbeta_{ij} + \sqrt{\gamma_{2t}^{*}}\upsilon_{ij,t}, \qquad \upsilon_{ij,t} \sim \mathcal{N}(\upsilon_{ij,t} \mid 0,1).
\end{aligned}
\end{equation*}
Therefore, combining the Gaussian prior $\bbeta_{ij} \sim \mathcal{N}_{2L}(\bbeta_{ij} \mid \mathbf{0}_{2L}, \underline{\Sigma}_\beta)$ with the conditional distribution $\widetilde{z}_{ij,t}^S \mid \bbeta_{ij}, \gamma_{2t}^{*} \sim \mathcal{N}\big( \widetilde{z}_{ij,t}^S \mid \sqrt{\gamma_{2t}^{*}}\bv_{ijt}' \bbeta_{ij}, \, \gamma_{2t}^{*} \big)$ to obtain the full conditional posterior
\begin{equation}
    \bbeta_{ij} \mid \bz_{ij}, \gamma_{2t}^{*}, \widetilde{\gamma}_{2t} \sim \mathcal{N}_{2L}\big( \bbeta_{ij} \mid \overline{\bmu}_\beta, \overline{\Sigma}_\beta \big),
\end{equation}
with
\begin{align*}
    \overline{\Sigma}_\beta = \Bigg( \underline{\Sigma}_\beta^{-1} + \sum_{t=1}^T \bv_{ij,t}\bv_{ij,t}' \Bigg)^{-1} \qquad
    \overline{\bmu}_\beta = \overline{\Sigma}_\beta \Bigg( \sqrt{\frac{\widetilde{\gamma}_{2t}}{\gamma_{2t}^{*}}} \sum_{t=1}^T z_{ij,t}^L \bv_{ij,t} \Bigg).
\end{align*}
The posterior distribution of $\gamma_{2t}^{*}$ is obtained conditionally on $\bz$ and $\widetilde{\gamma}_{2t}$, but marginally with respect to $\bbeta$, as follows:
\begin{align}
    p(\gamma_{2t}^{*} \mid \widetilde{\gamma}_{2t}, \bz) = p(\gamma_{2t}^{*}) \int \prod_{i=1}^N \prod_{j < i} p(\widetilde{z}_{ij,t}^L \mid \gamma_{2t}^{*}, \bbeta_{ij}) p(\bbeta_{ij}) \: \mathrm{d}\bbeta_{ij}.
\end{align}
Using the previous result and making explicit the dependence on $\gamma_{2t}^{*}$, the integral can be evaluated as
\begin{align*}
    & \int \prod_{i=1}^N \prod_{j < i} p(\widetilde{z}_{ij,t}^S \mid \gamma_{2t}^{*}, \bbeta_{ij}) p(\bbeta_{ij}) \: \mathrm{d}\bbeta_{ij} \propto \\
    % & \propto \int \prod_{i=1}^N \prod_{j < i} (\gamma_{2t}^{*})^{-1/2} \exp\Bigg(\!\! -\frac{1}{2} \Bigg( \bbeta_{ij}' \bv_{ij,t}\bv_{ij,t}' \bbeta_{ij} -2 \sqrt{\frac{\widetilde{\gamma}_{2t}}{\gamma_{2t}^*}} z_{ij,t} \bv_{ij,t}' \bbeta_{ij} + \frac{\widetilde{\gamma}_{2t}}{\gamma_{2t}^{*}} z_{ij,t}^2 \Bigg) \Bigg) \: \mathrm{d}\bbeta_{ij} \\
    & \propto (\gamma_{2t}^{*})^{-\frac{N(N-1)}{4}} \exp\Bigg(\!\! -\frac{1}{2} \frac{\widetilde{\gamma}_{2t}}{\gamma_{2t}^{*}}  \sum_{i=1}^N \sum_{j < i} z_{ij,t}^2 \Bigg) \\
    & \quad \times \prod_{i=1}^N \prod_{j < i} \int \exp\Bigg(\!\! -\frac{1}{2} \Bigg( \bbeta_{ij}' \overline{\Sigma}_\beta^{-1} \bbeta_{ij} -2 \overline{\bmu}_\beta' \overline{\Sigma}_\beta^{-1} \bbeta_{ij} + \overline{\bmu}_\beta' \overline{\Sigma}_\beta^{-1} \overline{\bmu}_\beta - \overline{\bmu}_\beta' \overline{\Sigma}_\beta^{-1} \overline{\bmu}_\beta \Bigg) \Bigg) \: \mathrm{d}\bbeta_{ij} \\
    & \propto (\gamma_{2t}^{*})^{-\frac{N(N-1)}{4}} \exp\Bigg(\!\! -\frac{1}{2} \frac{\widetilde{\gamma}_{2t}}{\gamma_{2t}^{*}} \sum_{i=1}^N \sum_{j < i} z_{ij,t}^2 \Bigg) \\
    & \quad \times \prod_{i=1}^N \prod_{j < i} \exp\Bigg( \frac{1}{2} \big( \sqrt{\frac{\widetilde{\gamma}_{2t}}{\gamma_{2t}^{*}}} z_{ij,t} \bv_{ij,t}' \Bigg) \overline{\Sigma}_\beta \Bigg( \sqrt{\frac{\widetilde{\gamma}_{2t}}{\gamma_{2t}^{*}}}  z_{ij,t} \bv_{ij,t} \Bigg) \\
    & \propto (\gamma_{2t}^{*})^{-\frac{N(N-1)}{4}} \exp\Bigg(\!\! -\frac{1}{2} \frac{\widetilde{\gamma}_{2t}}{\gamma_{2t}^{*}}  \sum_{i=1}^N \sum_{j < i} z_{ij,t}^2 \Bigg)  \prod_{i=1}^N \prod_{j < i} \exp\Bigg( \frac{1}{2} \frac{\widetilde{\gamma}_{2t}}{\gamma_{2t}^{*}} \underbrace{(z_{ij,t} \bv_{ij,t})'}_{\widetilde{\bv}_{ij}'} \overline{\Sigma}_\beta \underbrace{(z_{ij,t} \bv_{ij,t})}_{\widetilde{\bv}_{ij}} \Bigg) \\
    % & \propto (\gamma_{2t}^{*})^{-\frac{N(N-1)}{4}} \exp\Bigg(\!\! -\frac{1}{2} \frac{\widetilde{\gamma}_{2t}}{\gamma_{2t}^{*}} \sum_{i=1}^N \sum_{j < i} z_{ij,t}^2 \Bigg)  \exp\Bigg( \frac{1}{2} \frac{\widetilde{\gamma}_{2t}}{\gamma_{2t}^{*}} \sum_{i=1}^N \sum_{j < i} \widetilde{\bv}_{ij}' \overline{\Sigma}_\beta \widetilde{\bv}_{ij} \Bigg) \\
    & \propto (\gamma_{2t}^{*})^{-\frac{N(N-1)}{4}} \exp\Bigg(\!\! -\frac{1}{\gamma_{2t}^{*}} \frac{\widetilde{\gamma}_{2t}}{2} \sum_{i=1}^N \sum_{j < i} \big( z_{ij,t}^2 - \widetilde{\bv}_{ij}' \overline{\Sigma}_\beta \widetilde{\bv}_{ij} \big) \Bigg).
\end{align*}
Therefore, combined with an inverse gamma prior, $\gamma_{2t}^{*} \sim \mathcal{IG}(\gamma_{2t}^{*} \mid \underline{a}_{\gamma,2}, \underline{b}_{\gamma,2})$, one obtains the conditional posterior
\begin{equation}
    \gamma_{2t}^{*} \mid \widetilde{\gamma}_{2t}, \bz \sim \mathcal{IG}( \gamma_{2t}^{*} \mid \overline{a}_{\gamma,2t}, \overline{b}_{\gamma,2t}),
\end{equation}
where
\begin{align*}
    \overline{a}_{\gamma,2t} = \underline{a}_{\gamma,2} + \frac{N(N-1)}{4}, \qquad
    \overline{b}_{\gamma,2t} = \underline{b}_{\gamma,2} + \frac{\widetilde{\gamma}_{2t}}{2} \Bigg( \sum_{i=1}^N \sum_{j < i} \big( z_{ij,t}^2 - \widetilde{\bv}_{ij}' \overline{\Sigma}_\beta \widetilde{\bv}_{ij} \big) \Bigg).
\end{align*}

\medskip
\paragraph{(iv) Posterior of $\btau,\br,\alpha_i$ and $\bx_i$.}
To derive the full conditional of $\btau,\br,\alpha_i$ and $\bx_i$, one can follow the same procedure as in the non-inflated case, except that we condition on $\{ w_{ij,t} \}_{ijt}$ and hence we use as observations only the $y_{ij, t}$ for which $w_{ij, t} =1$.

%%%%%%%%%%%%%%%%%%%%%%%%%%%%%%%%%%%%%%%%%%%%%%%%%%%%%%%%%%%

\subsection{Proof of the Results in eq. \eqref{eq:full_tau_r} - eq. \eqref{eq:full_tau_r5}}

Based on the Poisson process interpretation, knowing that $y_{ij,t} >0$ jumps occurred during the ``fictitious'' time interval $[0,1]$ implies that $\tau_{ij, 2t}$ is the arrival time of the last jump before $\ell =1$ and $\tau_{ij, 1t}+\tau_{ij, 2t}$ is the arrival time of the first jump after $\ell =1$.
This allows us to obtain the posterior distribution of $\btau_{ij,t} = (\tau_{ij, 1t},\tau_{ij, 2t})$ given $y_{ij,t}$ (and the parameters of the model) as follows.
Conditionally on $y_{ij,t} >0$, the arrival time $\tau_{ij, 2t}$ of the $y_{ij,t}$th jump is the maximum of $y_{ij,t}$ random variables with distribution $\mathcal{U}(0,1)$, and follows a Beta distribution with parameters $(y_{ij,t}, 1)$.
Instead, the waiting time until the first jump after $\ell = 1$ has an exponential distribution with mean $\lambda_{ij,t}$, which implies $\tau_{ij, 1t} = 1 + \xi_{ij, t} - \tau_{ij, 2t}$, with $\xi_{ij, t}$ exponentially distributed with mean $\lambda_{ij,t}$.

To summarise, we obtain a draw from the posterior full conditional distribution of $\btau_{ij,t} = (\tau_{ij, 1t},\tau_{ij, 2t})$ by first drawing
\begin{align*}
    \tau_{ij, 2t} \mid y_{ij,t} & \sim \mathcal{B}e\big( \tau_{ij, 2t} \mid y_{ij,t}, 1 \big), &
    \xi_{ij, t} \mid \bx,\balpha & \sim \mathcal{E}xp\big( \xi_{ij, t} \mid \lambda_{ij,t} \big),
\end{align*}
then setting $\tau_{ij, 1t}$ as:
\begin{align*}
    % \tau_{ij, 1t} \mid y_{ij,t}, \bx,\balpha = \begin{cases}
    \tau_{ij, 1t} = \begin{cases}
        1 + \xi_{ij, t} & \text{ if } y_{ij,t} = 0 \\
        1 + \xi_{ij, t} - \tau_{ij, 2t} & \text{ if } y_{ij,t} > 0 
    \end{cases}
\end{align*}

%\subsection{Sampling $\br$}
Each indicator $r_{ij,qt}$ is sampled independently from the others.
For $q=1$, we sample $r_{ij, 1t}$ from a discrete distribution on $D_1 = \{ 1,\dots,R(1) \}$, with unnormalised probabilities
\begin{align*}
    \P\big( r_{ij, 1t} = k \mid \tau_{ij, 1t}, \theta \big) \propto c_k(1) \mathcal{N}\big( -\log(\tau_{ij, 1t}) -\log(\lambda_{ij,t}) \mid \mu_k(1), \sigma_k^2(1) \big), \quad k\in D_1.
\end{align*}
Instead, the allocation $r_{ij, 2t}$ is sampled from a discrete distribution on $D_2 = \{ 1,\dots,R(y_{ij,t}) \}$, with unnormalised probabilities
\begin{align*}
    \P\big( r_{ij, 2t} = k \mid \tau_{ij, 2t}, \theta \big) \propto c_k(y_{ij,t}) \mathcal{N}\big( -\tau_{ij, 2t} -\log(\lambda_{ij,t}) \mid \mu_k(y_{ij,t}), \sigma_k^2(y_{ij,t}) \big), \quad k\in D_2.
\end{align*}
The number of mixture components and the mixture weights, location, and scale parameters vary depending on an integer parameter, $\nu$, which takes value $\nu = 1$ in the posterior for $r_{ij, 1t}$, and $\nu = y_{ij,t}$ in the posterior for $r_{ij, 2t}$.
For each possible value of $\nu$, \cite{fruhwirth2009improved} provides a method to compute the associated number of mixture components and the mixture weights, location, and scale parameters.

\clearpage
%%%%%%%%%%%%%%%%%%%%%%%%%%%%%%%%%%%%%%%%%%%%%%%%%%%%%%%%%%%
\section{MCMC algorithms}
\label{sec:MCMC_algos}

The structure of the proposed methodologies and samplers is modular, meaning that they can be simplified to obtain still novel models and algorithms for multivariate count data.
For instance, setting $w_{ij,t}=1$ and removing the steps for $(\bw,\bz,\bbeta)$ allows us to deal with non-zero-inflated counts, whereas a static model can be estimated by simplifying Step 1 in a simple draw from a multivariate Gaussian.

Notably, the proposed methodologies (\textit{node-wise} and \textit{feature-wise}) and MCMC algorithms (with fixed and random $d$) can be easily simplified to the static case, which represents an additional contribution of this article.

\begin{algorithm}[H]
\caption{PCG sampler for dynamic model with random dimension $d$}
\label{alg:MCMC_dynamic_random_d}
\begin{algorithmic}[1]
\Require Data $\by$; maximum latent dimension $\underline{d}$; initial values $\balpha^{(0)}$, $\{\Phi_d^{(0)}, \Upsilon_d^{(0)}\}_{d=1}^{\underline{d}}$, $\btau^{(0)}$, $\br^{(0)}$, $\bz^{(0)}$, $\bbeta^{(0)}$
\Ensure Posterior draws $\{ \balpha^{(h)}, \bx^{(h)}, d^{(h)}, \Phi^{(h)}, \Upsilon^{(h)}, \btau^{(h)}, \br^{(h)}, \bw^{(h)}, \bz^{(h)}, \bbeta^{(h)} \}_{h=1}^H$
\For{$h=1$ \textbf{to} $H$}

    \For{$d=1$ \textbf{to} $\underline{d}$} \Comment{candidate paths at each dimension}
        \State \textsc{Compute candidate $\bx$:} given $d$, 
        \Statex \hspace{32pt} run a Kalman smoother AWOL   \Comment{multivariate Gaussian}
        \begin{equation*}
            \bx_d^{(h)} \sim p\big(\bx \mid \btau, \br, \balpha, d, \Phi_d^{(\text{ref})},\Upsilon_d^{(\text{ref})}\big),
        \end{equation*}
        \Statex \hspace{32pt} where $\Upsilon_d^{(\text{ref})}$ is the most recent value available for dimension $d$ (see line~\ref{step:sigma-bookkeeping}).
        \State \textsc{Compute likelihood:} given $d,\bx$, evaluate the (approximate) 
        \Statex \hspace{32pt} marginal log-likelihood, $\widetilde\ell_d^{(h)}$, in eq.~\eqref{eq:Laplace_approx_d}.    \Comment{Laplace method}
        % \Statex \hspace{32pt} evaluate (approximate) marginal log-likelihood       \Comment{Laplace method}
        % \begin{equation*}
        %     \ell_d^{(h)} \approx \log p\big( \btau, \br \mid d, \balpha,\Phi_d,\Upsilon_d \big)
        % \end{equation*}
    \EndFor

    \State \textsc{Compute probabilities:} evaluate to get the posterior probabilities
    \begin{equation*}
        \kappa_d = \frac{\pi(d) \exp\big(\widetilde\ell_d^{(h)}\big)}{\sum_{k=1}^{\underline{d}} \pi(k) \exp\big(\widetilde\ell_k^{(h)}\big)}, \qquad d=1,\dots,\underline{d}.
    \end{equation*}
    
    % \State \textsc{Compute weights:} set unnormalised weights
    % \begin{equation*}
    %   \widetilde{\varpi}_d \propto \pi(d) \exp\big(\widetilde\ell_d^{(h)}\big), \qquad d=1,\dots,\underline{d},
    % \end{equation*}
    % \Statex \hspace{\algorithmicindent} then normalise $\varpi_d = \widetilde{\varpi}_d / \sum_{k=1}^{\underline{d}} \widetilde{\varpi}_k$.

    \State \textbf{Update $d$ (dimension):} draw    \Comment{categorical}
    \begin{equation*}
        d^{(h)} \mid \btau,\br,\balpha,\Phi,\Upsilon \sim \text{Categorical}( d^{(h)} \mid\varpi_1,\dots,\varpi_{\underline{d}})
    \end{equation*}

    \State \textbf{Update $\bx$ (state):} set $\bx^{(h)} \gets \bx_{d^{(h)}}^{(h)}$

    \algstore{MCMC_dynamic_random_d1}
\end{algorithmic}
\end{algorithm}

\begin{algorithm}[H]
\begin{algorithmic}[1]
    \algrestore{MCMC_dynamic_random_d1}

    \State \textbf{Update $\balpha$ (intercept):} draw $\balpha^{(h)} \sim p(\balpha \mid \btau, \br, \bx, \cdot)$   \Comment{multivariate Gaussian}
    
    \State \textbf{Update $\Phi$ (AR):}
    \Statex \hspace{32pt} (i) [Active] draw $\Phi_{d^{(h)}}^{(h)} \sim p(\Phi \mid \bx, d, \Upsilon)$     \Comment{multivariate Gaussian}
    \Statex \hspace{32pt} (ii) [Inactive] for $\ell \neq d^{(h)}$, set  $\Phi_\ell^{(h)} = \Phi_\ell^{(h-1)}$ or draw from prior
    
    \State \label{step:sigma-bookkeeping}\textbf{Update $\Upsilon$ (covariance):}
    \Statex \hspace{32pt} (i) [Active - conjugate] draw $\Upsilon_{d^{(h)}}^{(h)} \sim p(\Upsilon \mid \bx, d, \Phi)$   \Comment{inverse Wishart}
    \Statex \hspace{32pt} (i) [Active - Graphical HS] draw $\Upsilon^{(h)}$ as in \cite{li2019graphical}, then draw $\bpsi^{(h)} \mid \Upsilon$
    \Statex \hspace{32pt} (ii) [Inactive] for $\ell \neq d^{(h)}$, set  $\Upsilon_\ell^{(h)} = \Upsilon_\ell^{(h-1)}$ or draw from prior

    \State \textbf{Update $\btau,\br$ (DA):} draw  \Comment{beta, exponential, and categorical}
    \begin{align*}
        \btau^{(h)} & \sim p\big(\btau \mid \by, \bx, \balpha \big) \qquad \br^{(h)}  \sim p\big(\br \mid \by, \btau, \bx, \balpha \big)
    \end{align*}

    \bigskip
    
    \State \textit{[IF a ZI specification is used, do also the following]}
    \State \textbf{Update $\bw$ (DA):} draw  \Comment{Bernoulli}
    \begin{align*}
        \bw^{(h)} & \sim p\big(\bw \mid \by, \bx, \balpha, \bz \big)
    \end{align*}
    
    \State \textbf{Update $\bz$ (RU):} draw  \Comment{truncated Gaussian}
    \begin{align*}
        \bz^{(h)} & \sim p\big(\bz \mid \bw, \bbeta \big)
    \end{align*}

    \State \textbf{Update $\bbeta$ (coefficients):} draw  \Comment{multivariate Gaussian}
    \begin{align*}
        \bbeta^{(h)} & \sim p\big(\bbeta \mid \bz \big)
    \end{align*}
\EndFor
\end{algorithmic}
\end{algorithm}

\begin{algorithm}[H]
\caption{PCG sampler for dynamic model with fixed dimension $d$}
\label{alg:MCMC_dynamic_fixed_d}
\begin{algorithmic}[1]
\Require Data $\by$; latent dimension $d$; initial values $\balpha^{(0)}$, $\Phi$, $\Upsilon$, $\btau^{(0)}$, $\br^{(0)}$, $\bz^{(0)}$, $\bbeta^{(0)}$
\Ensure Posterior draws $\{ \balpha^{(h)}, \bx^{(h)}, \Phi^{(h)}, \Upsilon^{(h)}, \btau^{(h)}, \br^{(h)}, \bw^{(h)}, \bz^{(h)}, \bbeta^{(h)} \}_{h=1}^H$
\For{$h=1$ \textbf{to} $H$}    
    \State \textbf{Update $\bx$ (state):} run a Kalman smoother AWOL   \Comment{multivariate Gaussian}
    \begin{equation*}
            \bx^{(h)} \sim p\big(\bx \mid \btau, \br, \balpha, \Phi,\Upsilon\big),
    \end{equation*}

    \State \textbf{Update $\balpha$ (intercept):} draw $\balpha^{(h)} \sim p(\balpha \mid \btau, \br, \bx, \cdot)$  \Comment{multivariate Gaussian}

    \State \textbf{Update $\Phi$ (AR):} draw $\Phi^{(h)} \sim p(\Phi \mid \bx, \Upsilon)$  \Comment{multivariate Gaussian}
    
    \State \textbf{Update $\Upsilon$ (covariance):} draw $\Upsilon^{(h)} \sim p(\Upsilon \mid \bx, \Phi, \bpsi)$
    \Statex \hspace{32pt} (i) [conjugate] $\Upsilon^{(h)} \sim p(\Upsilon \mid \bx, \Phi)$ \Comment{inverse Wishart}
    \Statex \hspace{32pt} (ii) [Graphical HS] draw $\Upsilon^{(h)}$ as in \cite{li2019graphical}, then draw $\bpsi^{(h)} \mid \Upsilon$
    
    \State \textbf{Update $\btau,\br$ (DA):} draw  \Comment{beta, exponential, and categorical}
    \begin{align*}
        \btau^{(h)} & \sim p\big(\btau \mid \by, \bx, \balpha \big) \qquad \br^{(h)}  \sim p\big(\br \mid \by, \btau, \bx, \balpha \big)
    \end{align*}

    \bigskip
    
    \State \textit{[IF a ZI specification is used, do also the following]}
    \State \textbf{Update $\bw$ (DA):} draw  \Comment{Bernoulli}
    \begin{align*}
        \bw^{(h)} & \sim p\big(\bw \mid \by, \bx, \balpha, \bz \big)
    \end{align*}
    
    \State \textbf{Update $\bz$ (RU):} draw  \Comment{truncated Gaussian}
    \begin{align*}
        \bz^{(h)} & \sim p\big(\bz \mid \bw, \bbeta \big)
    \end{align*}

    \State \textbf{Update $\bbeta$ (coefficients):} draw  \Comment{multivariate Gaussian}
    \begin{align*}
        \bbeta^{(h)} & \sim p\big(\bbeta \mid \bz \big)
    \end{align*}
\EndFor
\end{algorithmic}
\end{algorithm}

\begin{algorithm}[H]
\caption{PCG sampler for static model with random dimension $d$}
\label{alg:MCMC_static_random_d}
\begin{algorithmic}[1]
\Require Data $\by$; maximum latent dimension $\underline{d}$; initial values $\balpha^{(0)}$, $\btau^{(0)}$, $\br^{(0)}$, $\bz^{(0)}$, $\bbeta^{(0)}$
\Ensure Posterior draws $\{ \balpha^{(h)}, \bx^{(h)}, d^{(h)}, \btau^{(h)}, \br^{(h)}, \bw^{(h)}, \bz^{(h)}, \bbeta^{(h)} \}_{h=1}^H$
\For{$h=1$ \textbf{to} $H$}

    \For{$d=1$ \textbf{to} $\underline{d}$} \Comment{candidate paths at each dimension}
        \State \textsc{Compute candidate $\bx$:} given $d$, draw   \Comment{multivariate Gaussian}
        \begin{equation*}
            \bx_d^{(h)} \sim p\big(\bx \mid \btau, \br, \balpha, d \big),
        \end{equation*}
        \State \textsc{Compute likelihood:} given $d,\bx$, evaluate the (approximate) 
        \Statex \hspace{32pt} marginal log-likelihood, $\widetilde\ell_d^{(h)}$, in eq.~\eqref{eq:Laplace_approx_d}.   \Comment{Laplace method}
    \EndFor

    \State \textsc{Compute probabilities:} evaluate to get the posterior probabilities
    \begin{equation*}
        \varpi_d = \frac{\pi(d) \exp\big(\widetilde\ell_d^{(h)}\big)}{\sum_{k=1}^{\underline{d}} \pi(k) \exp\big(\widetilde\ell_k^{(h)}\big)}, \qquad d=1,\dots,\underline{d}.
    \end{equation*}

    \State \textbf{Update $d$ (dimension):} draw   \Comment{categorical}
    \begin{equation*}
        d^{(h)} \mid \btau,\br,\balpha \sim \text{Categorical}(\varpi_1,\dots,\varpi_{\underline{d}})
    \end{equation*}
    
    \State \textbf{Update $\bx$ (state):} set $\bx^{(h)} \gets \bx_{d^{(h)}}^{(h)}$

    \State \textbf{Update $\balpha$ (intercept):} draw $\balpha^{(h)} \sim p(\balpha \mid \btau, \br, \bx, \cdot)$ \Comment{multivariate Gaussian}
    
    \State \textbf{Update $\btau,\br$ (DA):} draw  \Comment{beta, exponential, and categorical}
    \begin{align*}
        \btau^{(h)} & \sim p\big(\btau \mid \by, \bx, \balpha \big) \qquad \br^{(h)}  \sim p\big(\br \mid \by, \btau, \bx, \balpha \big)
    \end{align*}

    \bigskip
    
    \State \textit{[IF a ZI specification is used, do also the following]}
    \State \textbf{Update $\bw$ (DA):} draw  \Comment{Bernoulli}
    \begin{align*}
        \bw^{(h)} & \sim p\big(\bw \mid \by, \bx, \balpha, \bz \big)
    \end{align*}
    
    \State \textbf{Update $\bz$ (RU):} draw  \Comment{truncated Gaussian}
    \begin{align*}
        \bz^{(h)} & \sim p\big(\bz \mid \bw, \bbeta \big)
    \end{align*}

    \State \textbf{Update $\bbeta$ (coefficients):} draw  \Comment{multivariate Gaussian}
    \begin{align*}
        \bbeta^{(h)} & \sim p\big(\bbeta \mid \bz \big)
    \end{align*}
\EndFor
\end{algorithmic}
\end{algorithm}

\begin{algorithm}[H]
\caption{PCG sampler for static model with fixed dimension $d$}
\label{alg:MCMC_Static_fixed_d}
\begin{algorithmic}[1]
\Require Data $\by$; latent dimension $d$; initial values $\balpha^{(0)}$, $\btau^{(0)}$, $\br^{(0)}$, $\bz^{(0)}$, $\bbeta^{(0)}$
\Ensure Posterior draws $\{ \balpha^{(h)}, \bx^{(h)}, \btau^{(h)}, \br^{(h)}, \bw^{(h)}, \bz^{(h)}, \bbeta^{(h)} \}_{h=1}^H$
\For{$h=1$ \textbf{to} $H$}    
    \State \textbf{Update $\bx$ (state):} draw $\bx^{(h)} \sim p\big(\bx \mid \btau, \br, \balpha, \big)$  \Comment{multivariate Gaussian}
    
    \State \textbf{Update $\balpha$ (intercept):} draw $\balpha^{(h)} \sim p(\balpha \mid \btau, \br, \bx, \cdot)$ \Comment{multivariate Gaussian}
    
    \State \textbf{Update $\btau,\br$ (DA):} draw  \Comment{beta, exponential, and categorical}
    \begin{align*}
        \btau^{(h)} & \sim p\big(\btau \mid \by, \bx, \balpha \big) \\
        \br^{(h)} & \sim p\big(\br \mid \by, \btau, \bx, \balpha \big)
    \end{align*}

    \bigskip
    
    \State \textit{[IF a ZI specification is used, do also the following]}
    \State \textbf{Update $\bw$ (DA):} draw  \Comment{Bernoulli}
    \begin{align*}
        \bw^{(h)} & \sim p\big(\bw \mid \by, \bx, \balpha, \bz \big)
    \end{align*}
    
    \State \textbf{Update $\bz$ (RU):} draw  \Comment{truncated Gaussian}
    \begin{align*}
        \bz^{(h)} & \sim p\big(\bz \mid \bw, \bbeta \big)
    \end{align*}

    \State \textbf{Update $\bbeta$ (coefficients):} draw  \Comment{multivariate Gaussian}
    \begin{align*}
        \bbeta^{(h)} & \sim p\big(\bbeta \mid \bz \big)
    \end{align*}
\EndFor
\end{algorithmic}
\end{algorithm}

%%%%%%%%%%%%%%%%%%%%%%%%%%%%%%%%%%%%%%%%%%%%%%%%%%%%%%%%%%
\section{Simulation Results}
\label{sec:simulation_further}

\subsection{Remarks on the Laplace approximation}

The role of the prior distribution for $d$ in eq. (15) is typically negligible compared to the magnitude of the (marginal) likelihood, which makes the resulting inference almost insensitive to the prior choice. Conversely, it may be possible to impose strong beliefs by carefully choosing each prior probability $\pi_\ell = p(d = \ell)$, $\ell = 1,\dots,\underline{d}$, such that unlikely or undesired values of $d$ have prior probability close to zero (e.g., $\pi_\ell = 10^{-8}$).

In a static framework, we can obtain a Laplace approximation by replacing $\widehat{\bx}_{i}$ with the MLE obtained from a simplified version of \cite{macdonald2022latent} or \cite{li2023statistical}.
From a computational perspective, we would need to compute $\underline{d}$ MLEs at each iteration of the MCMC.

In simulated experiments, we find evidence of the proposed Laplace approximation's good performance. A possible explanation is that the integral to be approximated is close to a multivariate Gaussian distribution, due to the shapes of the likelihood and prior distributions. In fact, the Gaussian likelihood includes the variable of integration in the mean as the inner product of two vectors, each one having a Gaussian prior density. The nonlinear function of $\bx$ in the likelihood precludes analytical integration, but we argue that it yields a sufficiently accurate Laplace approximation, as empirically shown on synthetic data.

\subsection{Hyperparameter Choice}
\label{app:hyperprior}
We assume a multivariate Gaussian prior distribution for the vector of intercepts $\balpha \sim \mathcal{N}_N(\balpha \mid \mathbf{0}_N, 5 \Id_N)$, and $\bbeta_i \overset{iid}{\sim} \mathcal{N}_N(\bbeta_i \mid \mathbf{0}_N, 5 \Id_N)$, $i=1,\dots,N$, and a multivariate Gaussian prior for the initial latent vector $\bx_{::,0} \sim \mathcal{N}_{dN}(\bx_{::,0} \mid \mathbf{0}_N, \Omega)$. The priors on the auxiliary variables of the Graphical horseshoe are set as:
\begin{align*}
    \varpi_{ij}^2 \mid \eta_{ij}^\varpi & \overset{ind}{\sim} \mathcal{IG}\bigg(\varpi_{ij}^2 \mid \frac{1}{2}, \frac{1}{\eta_{ij}^\varpi} \bigg), &
    \eta_{ij}^\varpi & \overset{iid}{\sim} \mathcal{IG}\bigg( \eta_{ij}^\varpi \mid \frac{1}{2}, 1 \bigg), \\
    \rho^2 \mid \eta^\rho & \sim \mathcal{IG}\bigg(\rho^2 \mid \frac{1}{2}, \frac{1}{\eta^\rho} \bigg), &
    \eta^\rho & \sim \mathcal{IG}\bigg( \eta^\rho \mid \frac{1}{2}, 1 \bigg).
\end{align*}

\subsection{Synthetic Data}
\label{sec:synth_data}

We simulate a sequence of weighted, undirected sparse networks by drawing edges from a zero-inflated Poisson projection model. The probability of not observing a structural zero between nodes $i$ and $j$ at time $t$ is such that the observed edge counts $y_{ijt}$ are generated from a mixture between a Poisson distribution and a Dirac mass at zero. Formally, each edge $y_{ij,t}$ is given by
\begin{equation*}
    y_{ij,t} \mid \lambda_{ij,t}, z_{ij,t} \overset{ind}{\sim} 
\delta_{\{0\}}\I(z_{ij,t} < 0)+\mathcal{P}oi(y_{ij,t} \mid \lambda_{ij,t}) \I(z_{ij,t}\geq 0),
\end{equation*}
where $z_{ij,t} \sim \mathcal{N}(z_{ij,t} \mid \beta_{i}+\beta_{j},1)$ and $\log(\lambda_{ijt}) = \alpha_i + \alpha_j + \bx_{i:,t}'\bx_{j:,t}$ independently $i,j=1,\dots,N$ with $i > j$.
The network's latent coordinates evolve over time according to a dynamic latent-space structure. 
Specifically, we generate synthetic data of dimensions $(N,d,T) = (50,2,20)$. The node-specific parameters $\beta_i \sim \mathcal{N}(\beta_i \mid 0.7,0.5^2)$ control for the probability of observing a draw from the Poisson component. Within this setting, structural zeros account for an average of 13.4\% of entries in each weighted adjacency matrix. Each node $i$ carries also time-invariant node-specific parameters $\alpha_i \sim \mathcal{N}(\alpha_i \mid 2,0.4^2)$ for the Poisson part.
%The latent position of node $i$ at time $t$ is a vector $\bx_{i:,t} \in \R^d$.
%, stacking across nodes gives $\mathbf{X}_t \in \R^{N\times d}$. 
The latent features evolve \textit{feature-wise} as a mean-stationary AR(1) process with a shared cross-sectional covariance:
\begin{equation*}
    \bx_{:\ell,t} = 0.3\bx_{:\ell,t-1} + \bepsilon_{:\ell,t}, \qquad
    \bepsilon_{:\ell,t} \overset{iid}{\sim} \mathcal{N}_n(\bepsilon_{:\ell,t} \mid \mathbf{0},\Sigma), \qquad \ell=1,2, \;\;\; t=1,\dots,15,
\end{equation*}
where $\bx_{:\ell,t}\in\R^N$ collects the $\ell$th latent feature for each node at time $t$.
The innovation covariance $\Sigma$ induces heterogeneous within-group correlation across nodes via a block-diagonal structure. Specifically, $\Sigma = \operatorname{blkdiag}(\mathbf{B}_5,\mathbf{B}_5,\mathbf{B}_5, \mathbf{B}_{10},\mathbf{B}_{10},\mathbf{B}_5, \mathbf{B}_{10})$ so that block sizes sum to $N=40$. Each block $\mathbf{B}_s$ is symmetric with diagonal variance $\sigma^2=0.8$ and common off-diagonal covariance $\rho=0.5$, that is, $\mathbf{B}_s = (\sigma^2-\rho) \Id_s + \rho \biota_s\biota_s'$, $s\in\{5,10\}$.
% We refer to this data-generating process as Scenario $S_1$. 
% Scenario $S_2$ represents a first variation, where $\beta_i \sim \mathcal{N}(0,0.5^2)$, leading to an average of 50\% structural zeros. Scenario $S_3$ corresponds to a second variation in which no structural zeros are present.

\subsection{Simulation results}
\label{sec:sym_results}
We assess the performance of our MCMC algorithm in a series of simulation settings. The first exercise tests the accuracy of our sampler in recovering the ground-truth parameters under the true data-generating process in the presence of model misspecification. In the second exercise, we assess the accuracy of our sampler for different sample sizes and degrees of zero inflation. Finally, we compare the IAMS-based sampler's mixing performance against a well-established alternative in the literature.

Throughout this section, we assume a random-walk dynamics for the latent features. Latent dimensions commonly used in many applications are 2 and, less frequently 3, thus, we assume a uniform prior on $\mathcal{D}=\{1,\dots,\underline{d}\}$, with $\underline{d}$ sufficiently large, i.e., $\underline{d} = 4$. All the parameters are randomly initialized in the MCMC. The set of experiments shows how the procedure can recover the ground-truth parameters and latent variables. The second set of experiments investigates the impact of sparsity level and sample size on inference effectiveness and MCMC efficiency.

\begin{figure}[t!]
\centering
\captionsetup{width=0.95\linewidth}
\begin{tabular}{cc}
\multicolumn{2}{c}{\footnotesize (a) Node effect posterior distribution}\\
  \includegraphics[width=0.48\textwidth]{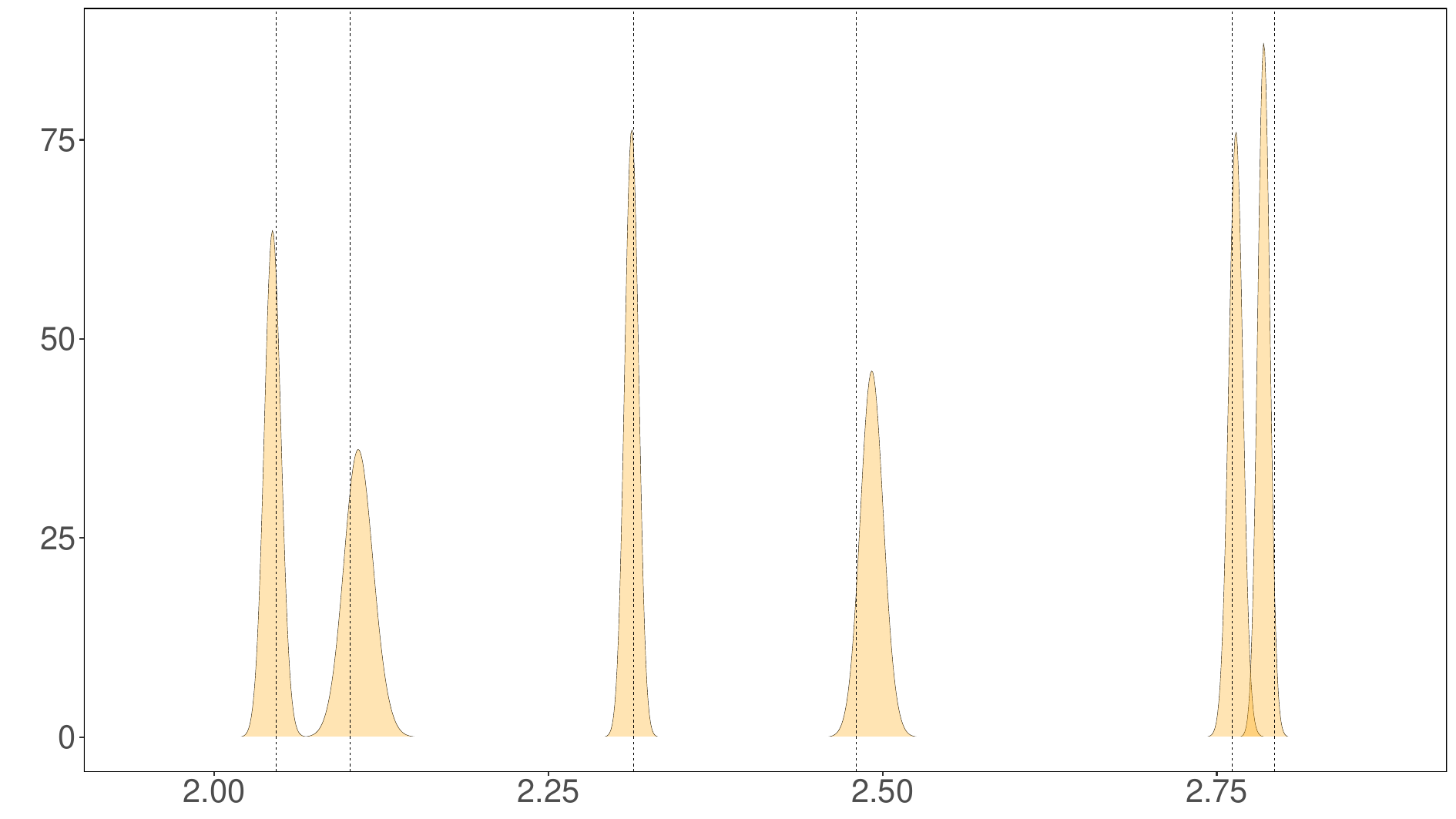} &
  \includegraphics[width=0.48\textwidth]{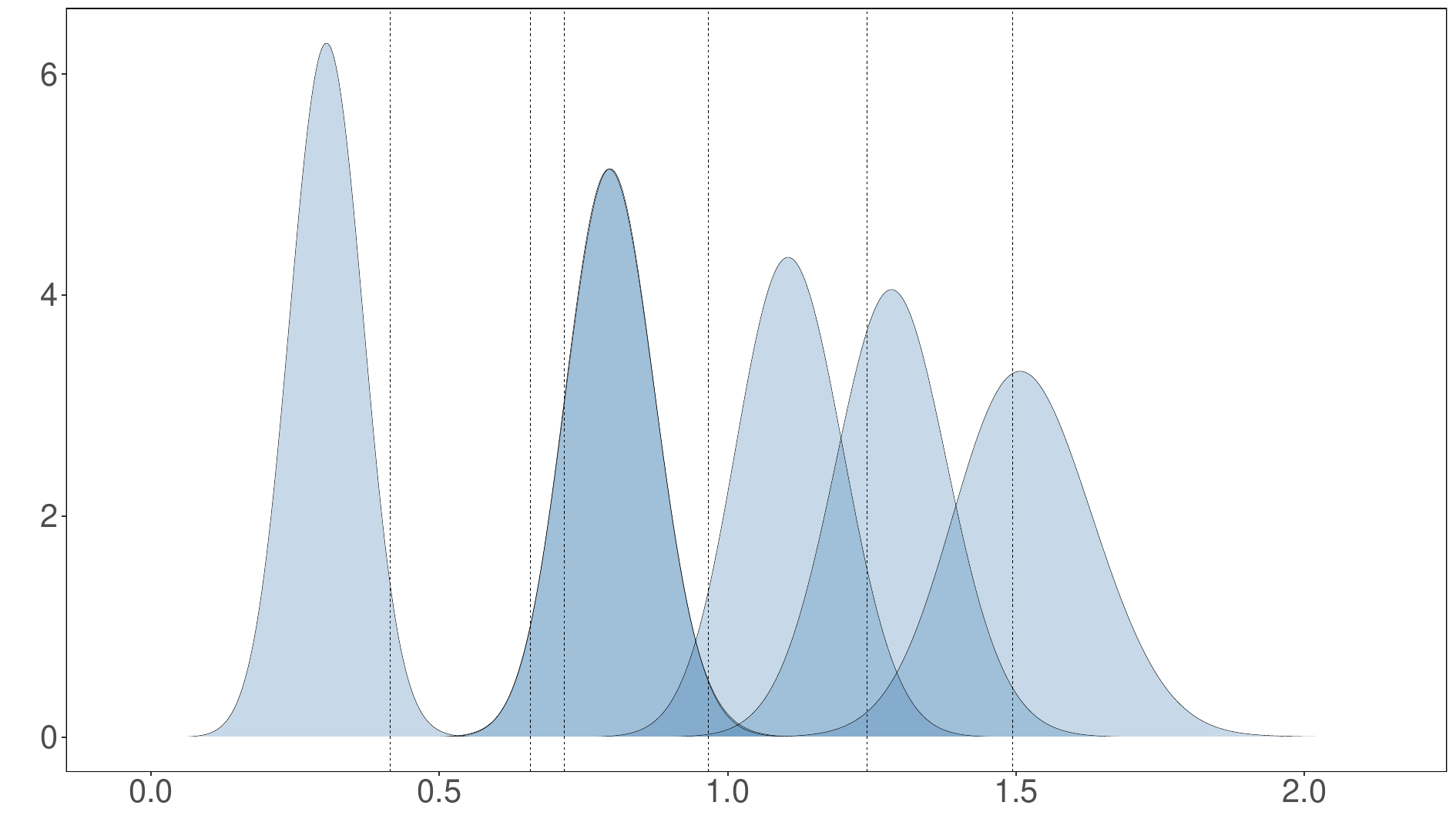}\\
\multicolumn{2}{c}{\footnotesize (b) Latent dimension and coordinate posterior distributions}\\
  \includegraphics[width=0.48\textwidth]{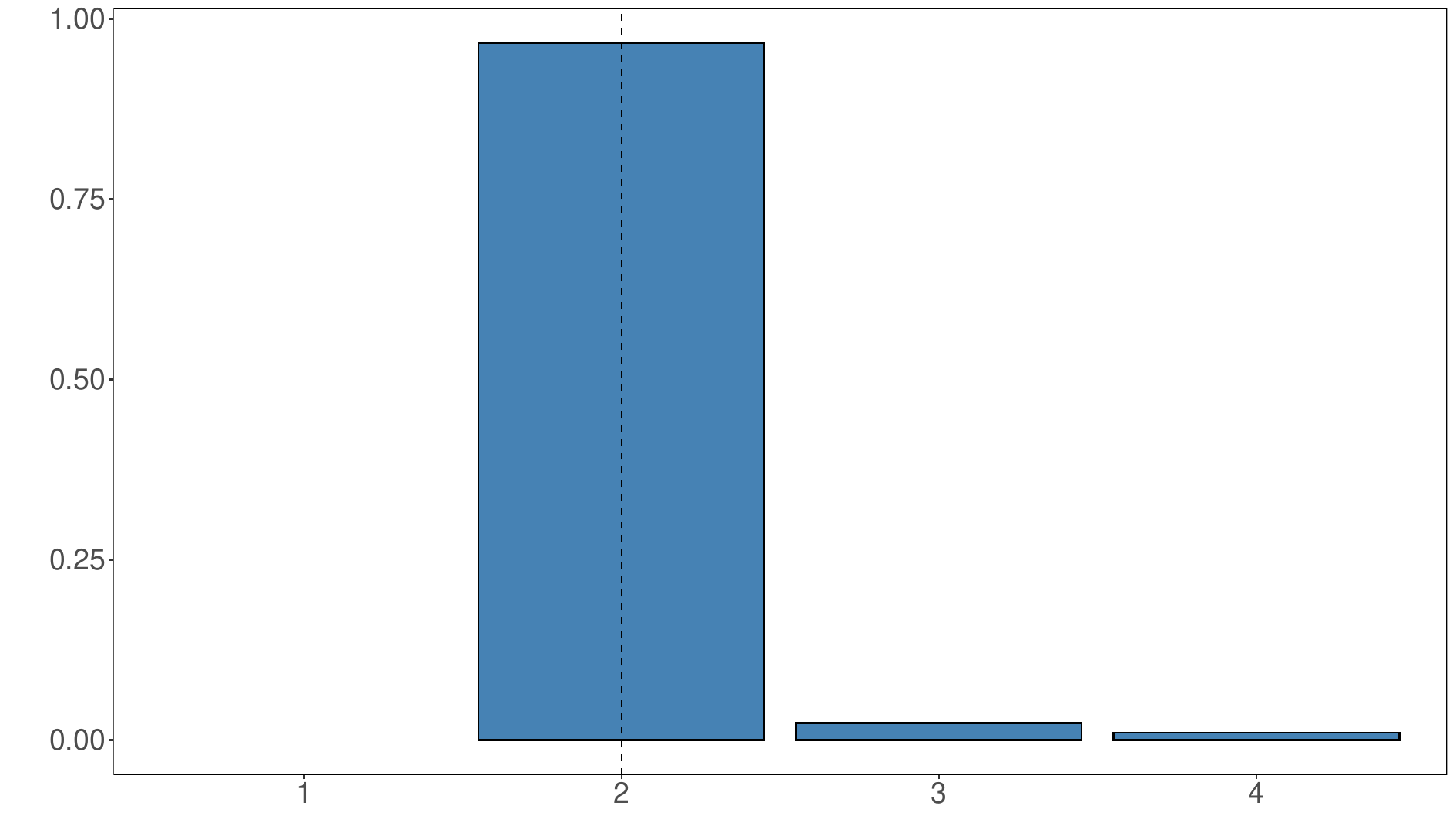} &
  \includegraphics[width=0.48\textwidth]{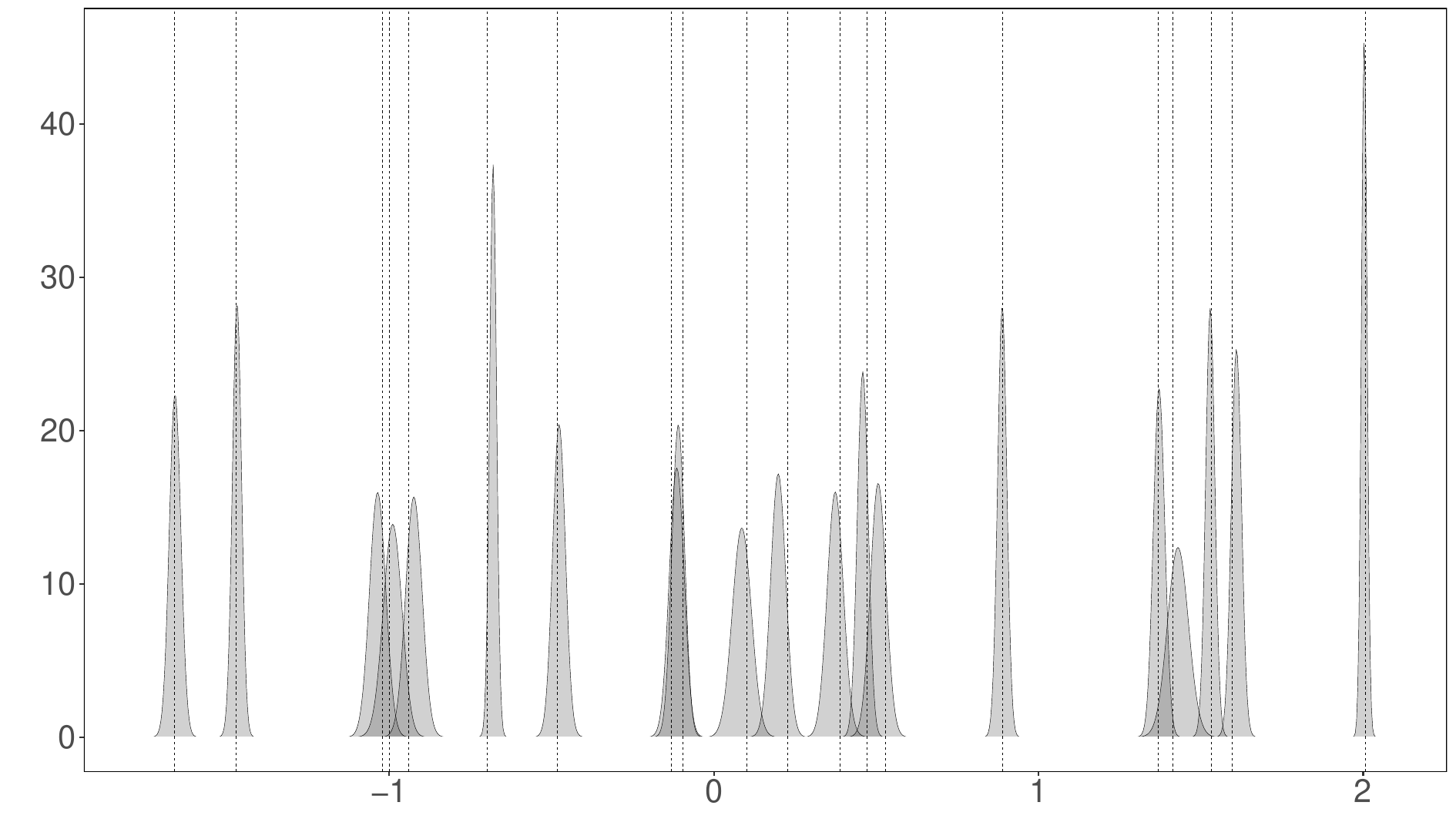} \\
%\end{tabular}
%\setlength{\tabcolsep}{1pt}   % less horizontal space
%\renewcommand{\arraystretch}{0.5} % less vertical space
\multicolumn{2}{c}{\footnotesize (c) Latent position and covariance estimates}\\
\multicolumn{2}{c}{\includegraphics[width = 0.32\textwidth]{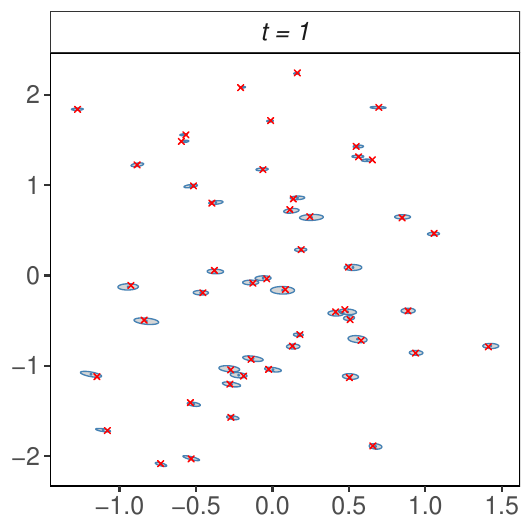} 
\includegraphics[width = 0.32\textwidth]{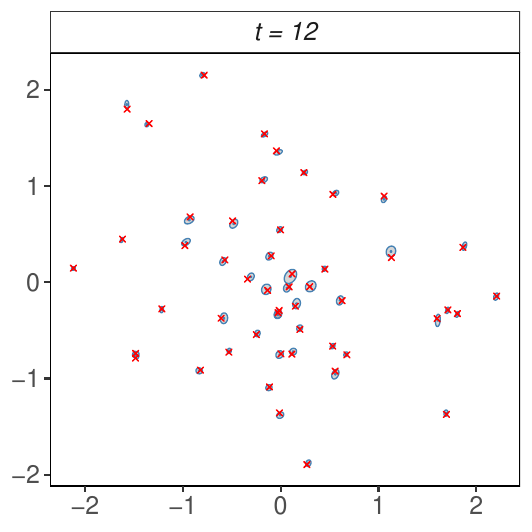}
\includegraphics[width = 0.32\textwidth]{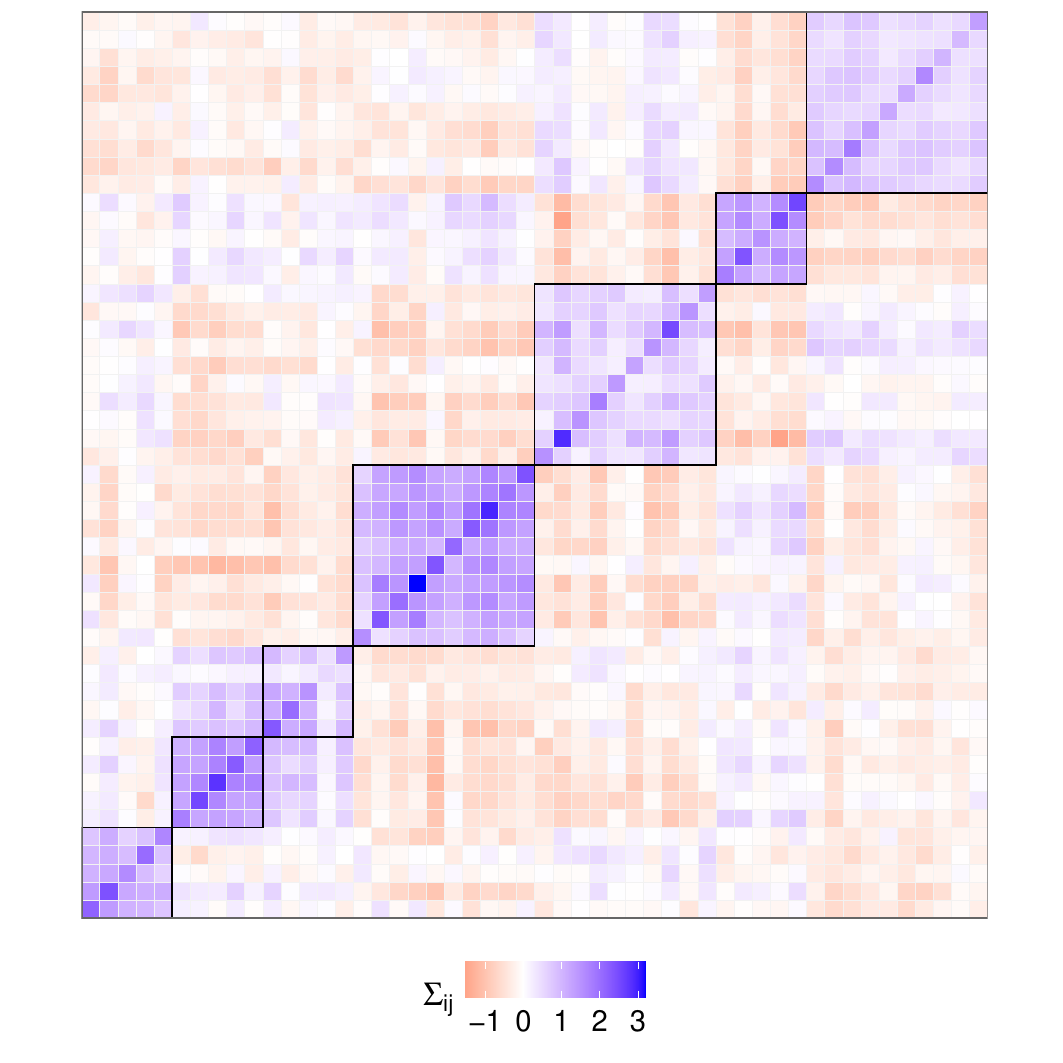}}
\end{tabular}
\caption{(a) Posterior distributions of $\alpha_i$ (left) and $\beta_i$ (right) for node $i \in \{2,4,6,8, 10, 12\}$.
(b) Posterior distribution of the latent space dimension $d$ (left) and posterior distributions of the first latent coordinate of node $i=1$, at selected times (right). Vertical dashed lines indicate the true parameter values. (c) Representation at $t = 1$ and $t = 12$ (left and middle) of the true (red crosses) and estimated (blue dots) latent positions with their posterior credible regions (blue ellipses). Comparison  (right) between the ground truth (black squares) and the estimated variance-covariance matrix (coloured cells).}
% The top panel reports the distributions of $\alpha_i$ (left) and $\beta_i$ (right) for node $i \in \{2,4,6,8, 10, 12\}$.
% The bottom left panel reports the posterior distribution of the latent space dimension $d$. The bottom right panel reports the posterior distributions across time for the first latent coordinate of node $i =1$.
% Vertical dashed lines indicate the true parameter values.}
%, and labels identify the corresponding node indices.}
\label{fig:kde-alpha-beta-zip}
\end{figure}

\paragraph{Setting 1 (baseline)}
In this first exercise, we test the ability of our sampler to retrieve the ground truth parameters as described in Section \ref{sec:synth_data}. In this exercise, we run our sampler for 5000 iterations using the first $500$ iterations as a burn-in. 
The results of this simulation study confirm that the proposed zero-inflated Poisson latent space model provides an accurate estimation of both node-specific parameters and latent coordinates. Panel (a) of Figure~\ref{fig:kde-alpha-beta-zip} shows the posterior kernel density estimates for a subset of $\alpha_i$ and $\beta_i$ parameters. The posterior distributions are centered near the true values (vertical dashed lines) and exhibit relatively narrow dispersion, indicating that the sampler successfully recovers the individual effects governing both the probability of observing a structural zero and the Poisson intensity. Turning to the latent space representation, panel (b) shows that the true latent dimension $d = 2$ (left) and the latent position of node $i = 1$ across time (right) have been fully recovered. The estimated latent positions closely track the ground truth, with credible ellipses generally covering the true positions. See the left and middle plots in Panel (c). Finally, the right plot compares the ground-truth covariance matrix (grid) with the estimated variance-covariance matrix. The estimated matrix correctly recovers the block-diagonal structure, despite some inaccuracies in the remaining covariance elements. The Supplement reports the estimated positions for all 12 time instances and convergence diagnostics for the parameters of interest.

%\begin{figure}[t]
%\centering
%\everyrow{\tabucline[0.4pt]-}
%\setlength{\tabcolsep}{1pt}   % less horizontal space
%\renewcommand{\arraystretch}{0.5} % less vertical space
%\resizebox{0.9\textwidth}{!}{ 
%\begin{tabular}{ccc}
%\includegraphics[width = 0.20\textwidth]{Figures/Simulation/ZIP/PlanePlot1_ZIP.pdf} &
%\includegraphics[width = 0.20\textwidth]{Figures/Simulation/ZIP/PlanePlot12_ZIP.pdf}
%\includegraphics[width = 0.20\textwidth]{Figures/Simulation/ZIP/VarCovar_ZIP.pdf}
%\end{tabular}}
%\caption{Left and middle panels: Latent space representation at $t = 1$ and $t = 12$. Red crosses indicate the true latent positions, blue dots represent the posterior means, and blue ellipses denote the posterior credible regions. Right panel: Comparison between the ground truth block variance covariance matrix (black squares) and the estimated variance covariance matrix (coloured cells).}
%\label{fig:planeplot}
%\end{figure}

% Figure~\ref{fig:latent-dynamics2} in the Supplementary Materials examines the temporal dynamics of the latent coordinates for two representative nodes. The posterior mean trajectories follow the true latent path very closely, while the shaded credible intervals remain narrow across time. 

% Note that the wide scale of the y-axis prevents a clear assessment of the dispersion of the posterior distribution around the true value. The Supplementary Material provides the posterior distributions for the selected nodes $i \in {1,20}$ at different time points $t$. 

\paragraph{Setting 2  (varying sparsity and sample size)}
In this second exercise, we first assess the ability to retrieve the structural zeros over 8 different scenarios along three main dimensions: proportion of zeros (10\%, 30\%), network size ($N = 25$, $N = 50$), and time span ($T = 15$, $T = 30$). The data-generating process is the one described in Section \ref{sec:synth_data}.
We run our algorithms for 10000 iterations over 10 random initialisation settings. 

% , defined as $F_1 = 2(precision \times recall)/(precision + recall)$

% defined as $MCC=\frac{T P \times T N-F P \times F N}{\sqrt{(T P+F P)(T P+F N)(T N+F P)(T N+F N)}}$

\begin{figure}[t]
 \centering
  \begin{tabular}{cc}
    \includegraphics[width=0.48\textwidth]{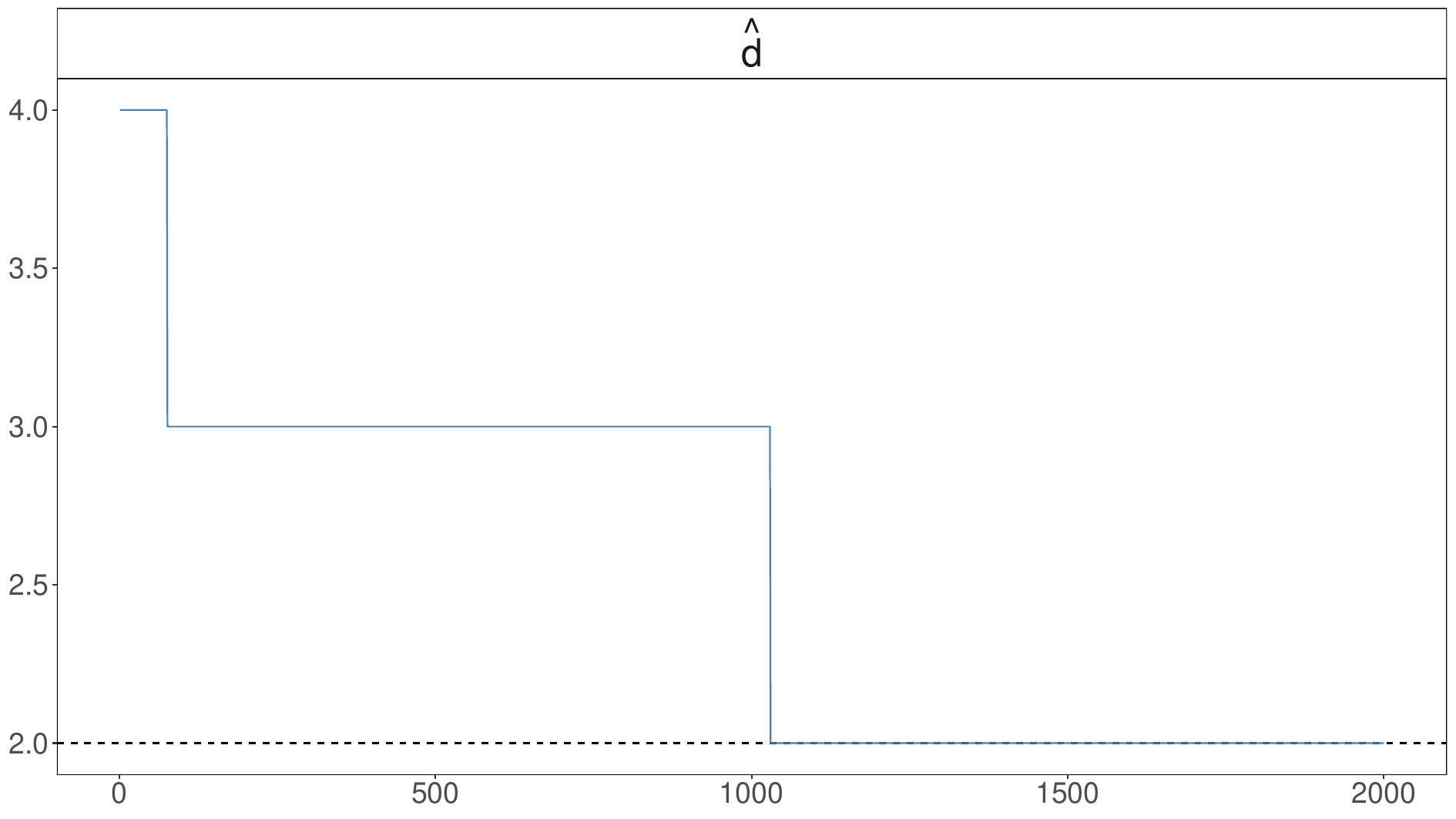} &
    \includegraphics[width=0.48\textwidth]{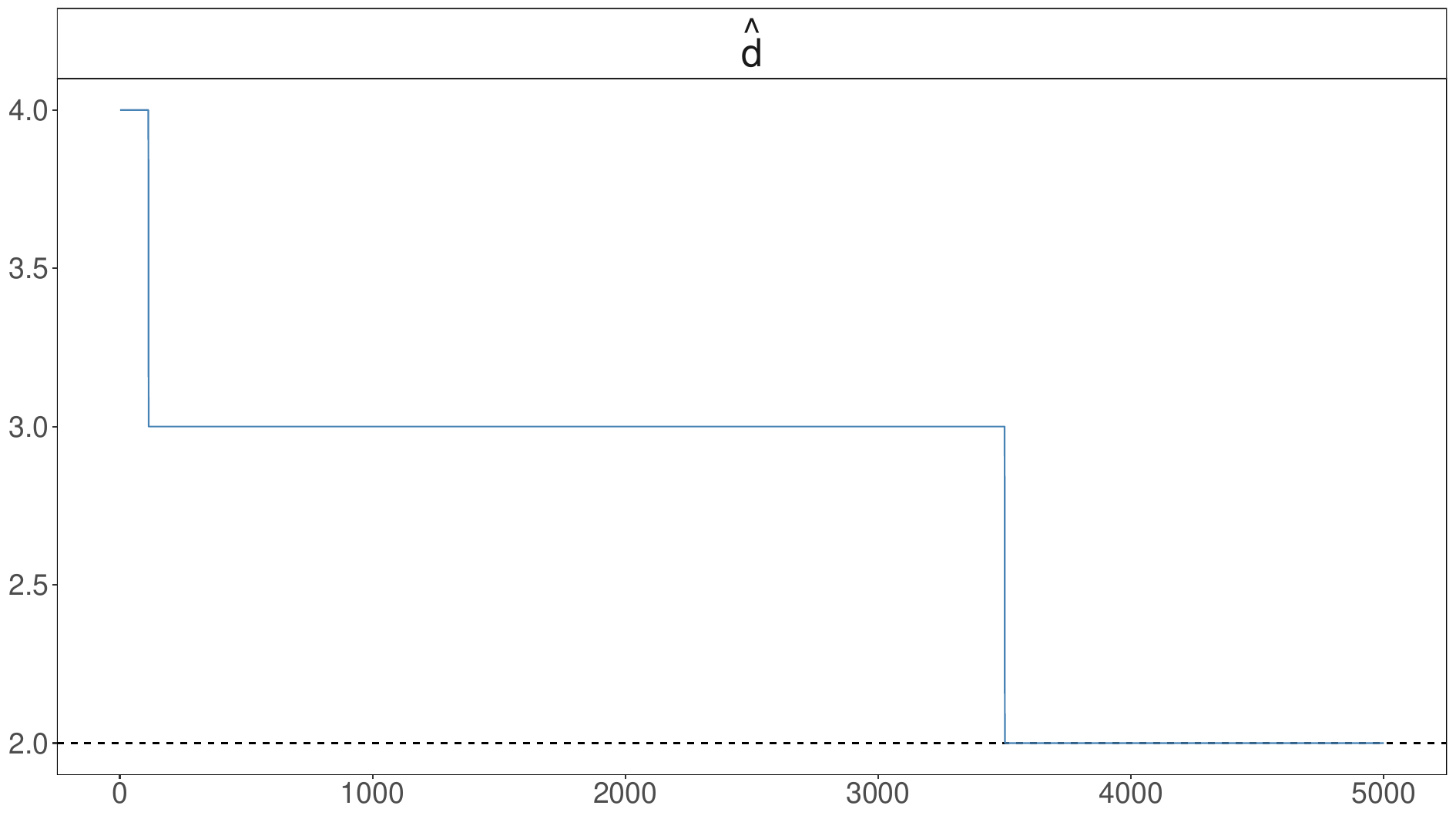}\\
  \end{tabular}
  \caption{
  The left panel reports the draws of the parameter $d$ in the case $N = 50$, $T = 30$ and $\operatorname{Pr}(w_{ijt} = 0)=$ 10\%. The right panel reports the draws of the parameter $d$ in the case $N = 50$, $T = 30$ and $\operatorname{Pr}(w_{ijt} = 0)=$ 30\%.}
  \label{fig:d_estim}
\end{figure}

Panel (a) of Table~\ref{tab:zeros_stats} reports the classification performance, where we test the ability of the sampler to identify the structural zeros. We focus on two standard classification metrics: the $F_1$ Score and the Matthews Correlation Coefficient (MCC). Both metrics are close to 1 across all settings, indicating good performance of our model in retrieving structural zeros. Table~\ref{tab:zeros_stats} also reports the estimated dimension $d$. We observe that we retrieve the correct latent dimension, $ d=2$, in approximately 90\% of cases. Figure \ref{fig:d_estim} shows how convergence is achieved after $\approx$ 1100 iterations for the case $N = 50$, $T = 30$, $\operatorname{Pr}(w_{ij,t} = 0)=$10\% and after $\approx$ 3500 iterations for the case $N = 50$, $T = 30$, $\operatorname{Pr}(w_{ij,t} = 0)=30$\%. In conclusion, different setups may require different numbers of iterations to achieve convergence. We notice some stickiness in the trace plot of the latent dimension parameter $d$. The MCMC behaviour is due to a highly concentrated posterior distribution, which makes it very unlikely for a $\pm 1$ jump to occur once the ground truth value is visited by the MCMC chain. 

\begin{table}[t!h]
\centering
\resizebox{0.95\textwidth}{!}{
\begin{tabular}{cccccc}
\multicolumn{6}{c}{(a) Classification performance}\\
\toprule
 &  & $(N,T)=(25,15)$ & $(N,T)=(50,15)$ & $(N,T)=(25,30)$ & $(N,T)=(50,30)$ \\
\midrule
\multirow{2}{*}{$F_1$ Score}
  & $\operatorname{Pr}(w_{ijt}=0)=10\%$ & 0.987 & 0.992 & 0.992 & 0.991 \\
  & $\operatorname{Pr}(w_{ijt}=0)=30\%$ & 0.994 & 0.995 & 0.994 & 0.994 \\
\midrule
\multirow{2}{*}{$MCC$} 
  & $\operatorname{Pr}(w_{ijt}=0)=10\%$ & 0.986 & 0.991 & 0.991 & 0.990 \\
  & $\operatorname{Pr}(w_{ijt}=0)=30\%$ & 0.991 & 0.992 & 0.991 & 0.992 \\
\midrule
\multirow{2}{*}{\% iters with $d=2$} 
  & $\operatorname{Pr}(w_{ijt}=0)=10\%$ & 90.0\% & 90.0\% & 90.0\% & 90.0\% \\
  & $\operatorname{Pr}(w_{ijt}=0)=30\%$ & 80.0\% & 100.0\% & 100.0\% & 80.0\% \\
\bottomrule
\end{tabular}}
\resizebox{0.95\textwidth}{!}{
\begin{tabular}{cccccccc}
\multicolumn{8}{c}{(b) Estimated $d$ and estimation errors}\\
\toprule
DGP zero-infl.                  & Model zero-infl.                 & Estimated $d$ & Avg. MSE $\boldsymbol{\alpha}$ & Avg. SD $\boldsymbol{\alpha}$ & Avg. MSE $\lambda_{ijt}$ & Avg. SD $\lambda_{ijt}$ & $F_1$   \\
\midrule
YES   &YES     &    2             &    0.00008          &    0.00608               &    0.00202                       &  0.02979 & 0.99359   \\
YES    & NO &   4            &  0.04877                               &  0.01781                             & 1.8535                         &      0.06267  & 0.95529   \\
NO & YES    &  2  &      0.00007           &        0.00574                                                    &     0.00164                      &       0.02684   & 0.98990         \\
NO & NO &     2          &     0.00007                               &        0.00577                       &     0.00163                     & 0.02681  & $-$   \\
\bottomrule
\end{tabular}}
\caption{(a) Average $F_1$, $MCC$, and percentage of simulations for which $d=2$ for different $(N,T)$ and share of structural zeros over 10 simulations. 10000 iterations. (b) Estimated $d$, average mean squared error (MSE) and standard deviation (SD) of $\alpha_i$ across $i = 1, \dots, N$ and of $\log \lambda_{ijt}$ across $i,j = 1,\dots N$, $i >j$ and $t = 1,\dots,T$ for different DGPs and model specifications.}
\label{tab:zeros_stats}
\end{table}

%\FloatBarrier

We then test the accuracy of the estimates of $\lambda_{ijt}$ and $\alpha_i$ when the data-generating process exhibits zero inflation. The data-generating process is the one described in Section \ref{sec:synth_data}. Panel (b) in Table \ref{tab:zeros_stats} reports simulation results across four scenarios. We observe that failing to account for zero inflation leads to a higher MSE and an upward-biased estimate of the latent dimension $d$, whereas accounting for it unnecessarily yields satisfactory results and an accurate estimate of the network features.
 
%\begin{table}[t!h]
%\centering
%\resizebox{0.95\textwidth}{!}{
%\begin{tabular}{cccccccc}
%\toprule
%DGP zero-infl.                  & Model zero-infl.                 & Estimated $d$ & Avg. MSE $\boldsymbol{\alpha}$ & Avg. SD $\boldsymbol{\alpha}$ & Avg. MSE $\lambda_{ijt}$ & Avg. SD $\lambda_{ijt}$ & $F_1$   \\
%\midrule
%YES   &YES     &    2             &    0.00008          &    0.00608               &    0.00202                       &  0.02979 & 0.99359   \\
%YES    & NO &   4            &  0.04877                               &  0.01781                             & 1.8535                         &      0.06267  & 0.95529   \\
%NO & YES    &  2  &      0.00007           &        0.00574                                                    &     0.00164                      &       0.02684   & 0.98990         \\
%NO & NO &     2          &     0.00007                               &        0.00577                       &     0.00163                     & 0.02681  & $-$   \\
%\bottomrule
%\end{tabular}}
%\caption{Estimated $d$, average mean squared error (MSE) and standard deviation (SD) of $\alpha_i$ across $i = 1, \dots, N$ and of $\log \lambda_{ijt}$ across $i,j = 1,\dots N$, $i >j$ and $t = 1,\dots,T$ for different DGPs and model specifications.}
%\label{tab:accuracy}
%\end{table}

\clearpage
\paragraph{Efficiency of IAMS}
Finally, the proposed model and sampling algorithm have also been tested against the benchmark method for static networks proposed by \cite{handcock2008fitting} and implemented in the \texttt{R} package \texttt{latentnet}. The latter designs a static version of a Poisson LSM and implements an MH algorithm to sample from the (exact) posterior distribution. The simulation study (described in the Supplement) shows that our IAMS-based strategy yields results indistinguishable from the exact posterior, with a substantial improvement in efficiency (ESS increased by a factor of 10).

% \begin{table}[htb!]
% \centering
% \begin{tabular}{ccccc}
% \hline\hline
% $d$       & Avg. MSE $\alpha_i$ & Avg. SD $\alpha_i$ & Avg. MSE $\log \lambda_{ijt}$ & Avg. SD $\lambda_{ijt}$ \\\hline
% $d = 1$   & 0.09563 &  0.06768 & 0.03207 & 0.11\\
% $d = 2$   & 0.00819 &  0.00709 & 0.01624  & 0.03 \\
% $d = 3$   &  0.00011 & 0.00634 & 0.00016 & 0.03\\
% $\hat{d} = 3$ &0.00028 & 0.00816  & 0.00277 & 0.04\\        
% \hline\hline
% \end{tabular}

% \caption{The table reports the average mean squared error ($MSE$) and standard deviation ($SD$) of $\alpha_i$ across $i = 1, \dots, N$ and of $\log \lambda_{ijt}$ across $i,j = 1, \dots, N, \quad i >j$  and $t = 1,\dots,T$ under different specifications of $d$. The 
%   }
%   \label{fig:ex3}
% \end{table}

\subsection{Convergence Diagnositcs Checks}

We present further diagnostic results for the zero-inflated Poisson model under the scenario described in the main text. Table \ref{tab:mcmc-diag-summary} reports the averaged diagnostics results for the individual effects $\alpha_i$ and $\beta_i$ and for the latent coordinates $x_{ijt}$. Overall, our diagnostics support convergence and satisfactory mixing for the ZIP model in $S_1$. The convergence diagnostics (CD) of \cite{geweke1991evaluating} provide evidence of overall convergence as shown by the average $p$-value across coefficients (last column in Tab. \ref{tab:mcmc-diag-summary}).
We notice that jointly sampling the entire vector $\bx_{i:,:}$ from its smoothed posterior distribution has beneficial effects on the mixing of the corresponding chains, which appears particularly striking when compared to the ESS of $\alpha_i$ and $\bbeta_i$. Once more, this finding highlights one key advantage of the proposed approach as opposed to the existing dynamic latent space models, where the dynamic latent features are drawn using a single-move sampler conditioning on the entire cross-section and time series \citep[e.g.,][]{sewell2015analysis}.

\begin{table}[htbp]
\centering
\begin{tabular}{lrrrrr}
\toprule
Type & Mean ESS & ACF(1) & ACF(10) & Mean CD $z$ & Mean CD $p$ \\
\midrule
$\alpha_i$  & 1019.873 & 0.599 & 0.221 &  -0.405 & 0.377 \\

$\beta_i$  & 1020.187 & 0.671 & 0.065  &  0.195 & 0.458 \\

$x_{1it}$ & 2283.621 & 0.321 & 0.019  &  -0.006 & 0.463     \\
$x_{2it}$ & 2314.334 & 0.319 & 0.013 & -0.014  & 0.466  \\
\bottomrule
\end{tabular}
\caption{MCMC diagnostics summary (means across series). Effective Sample Size over the
number of draws (ESS), Mean Auto-correlation (ACF) at lag 1 and 10, and mean Convergence Diagnostic and its $p$-value (CD) as defined in \citet{geweke1991evaluating}. 5000 iterations, no burn-in and no thinning.}
\label{tab:mcmc-diag-summary}
\end{table}

\begin{figure}[h!t]
\centering
\hspace*{-10pt}
\begin{tabular}{cc}
{\footnotesize feature $1$} & {\footnotesize feature $2$} \\
\includegraphics[width=0.48\textwidth]{Figures/Simulation/ZIP/KDE_x11_ZIP.pdf} &
\includegraphics[width=0.48\textwidth]{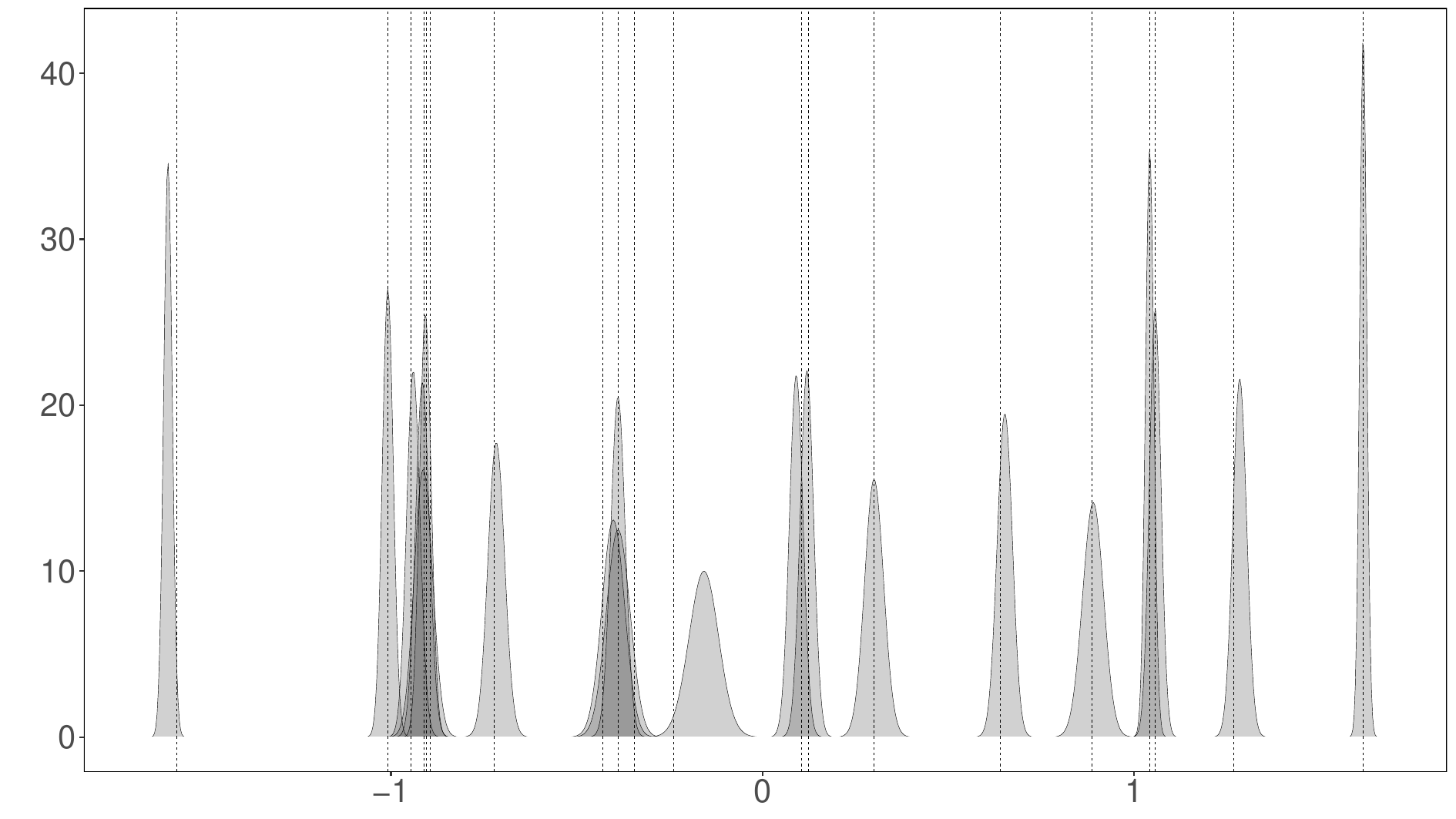} \\
\includegraphics[width=0.48\textwidth]{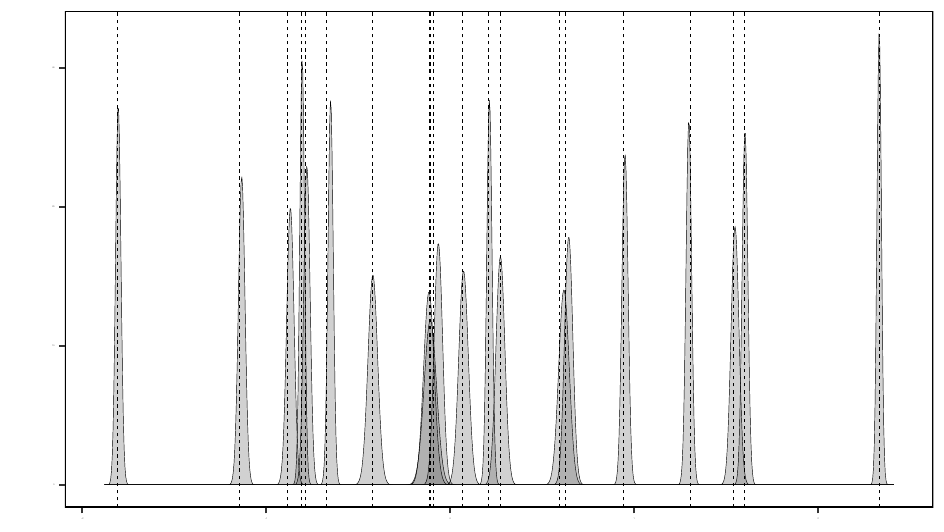} &
\includegraphics[width=0.48\textwidth]{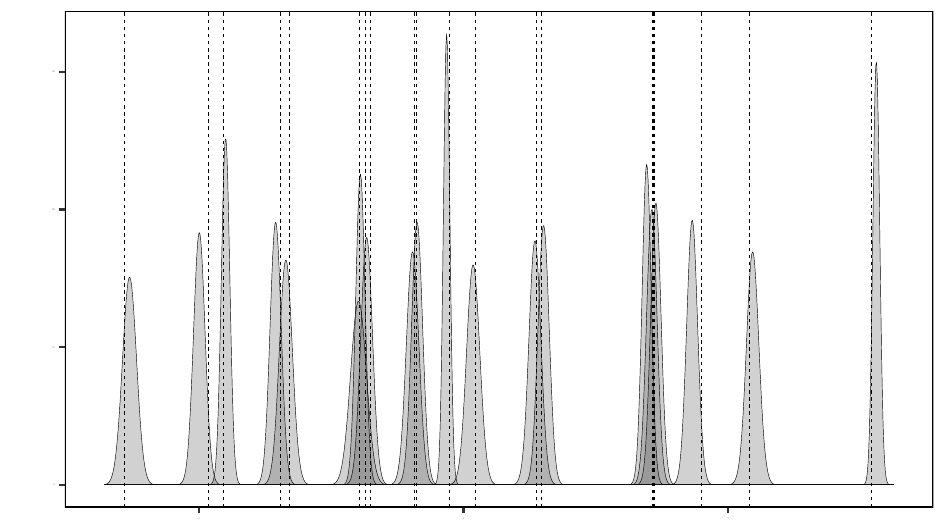} 
\end{tabular}
\caption{Posterior distributions per time for selected latent coordinates under the ZIP model for node 1 (top row) and node 20 (bottom row).}
\label{fig:kde-x-nodes-1-20}
\end{figure}

\begin{figure}[h!t]
\centering
\setlength{\tabcolsep}{4pt}
\renewcommand{\arraystretch}{1}
\begin{tabular}{ccc}
\includegraphics[width=0.31\textwidth]{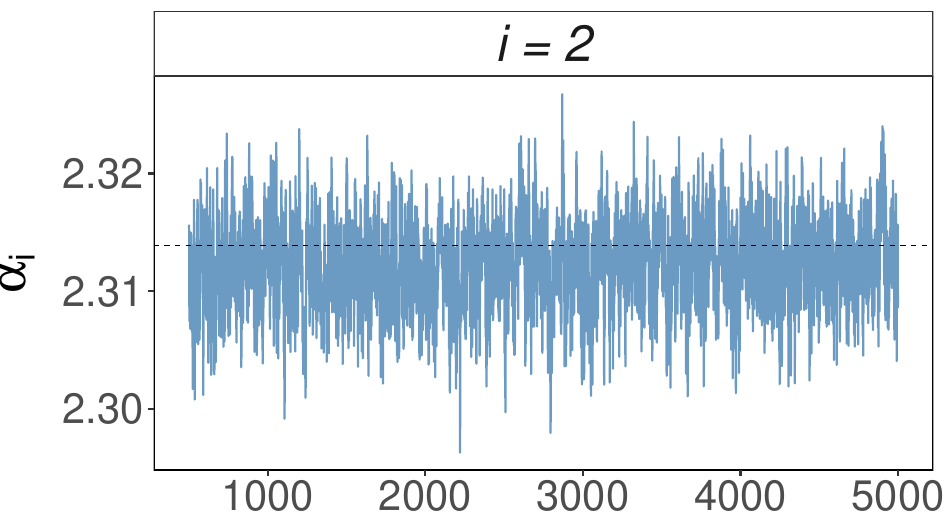} &
\includegraphics[width=0.31\textwidth]{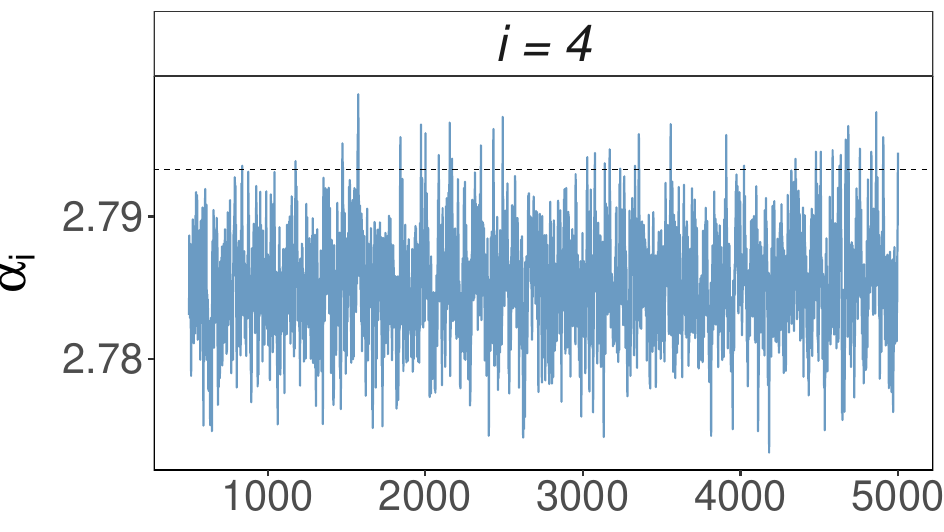} &
\includegraphics[width=0.31\textwidth]{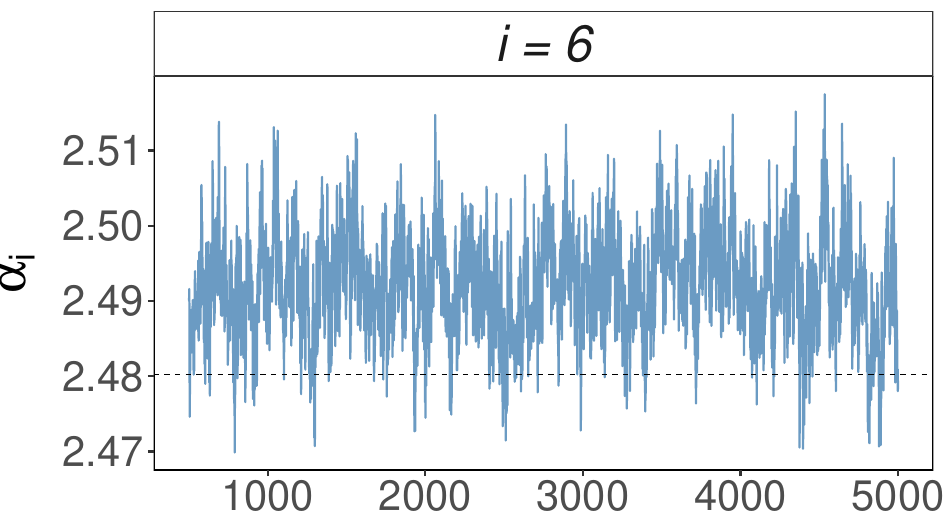} \\
\includegraphics[width=0.31\textwidth]{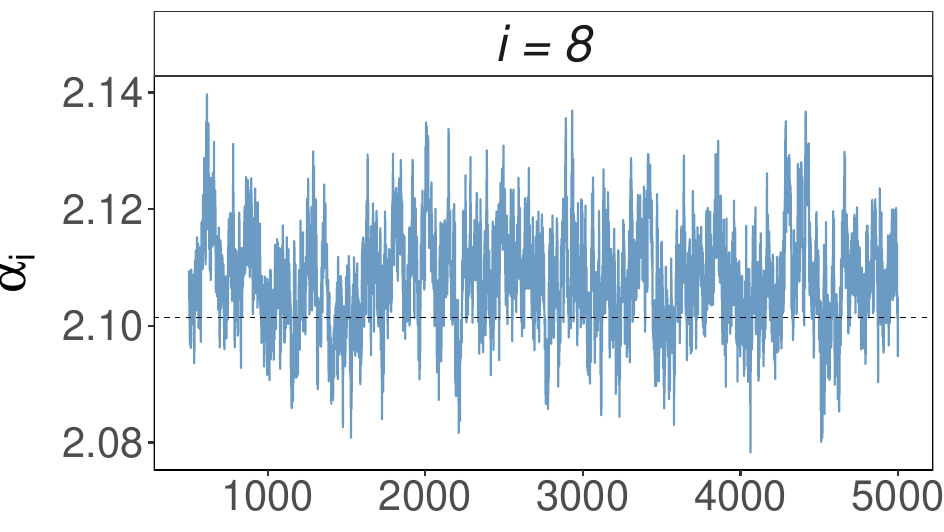} &
\includegraphics[width=0.31\textwidth]{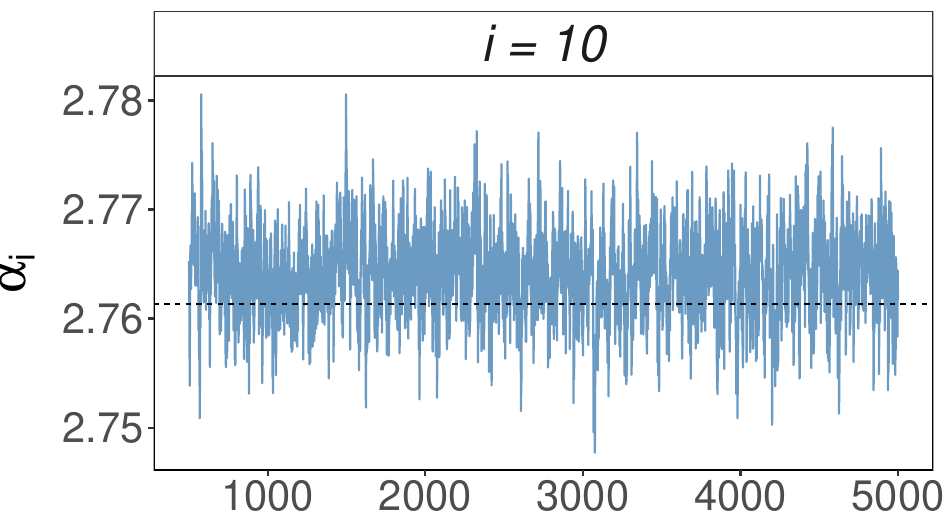} &
\includegraphics[width=0.31\textwidth]{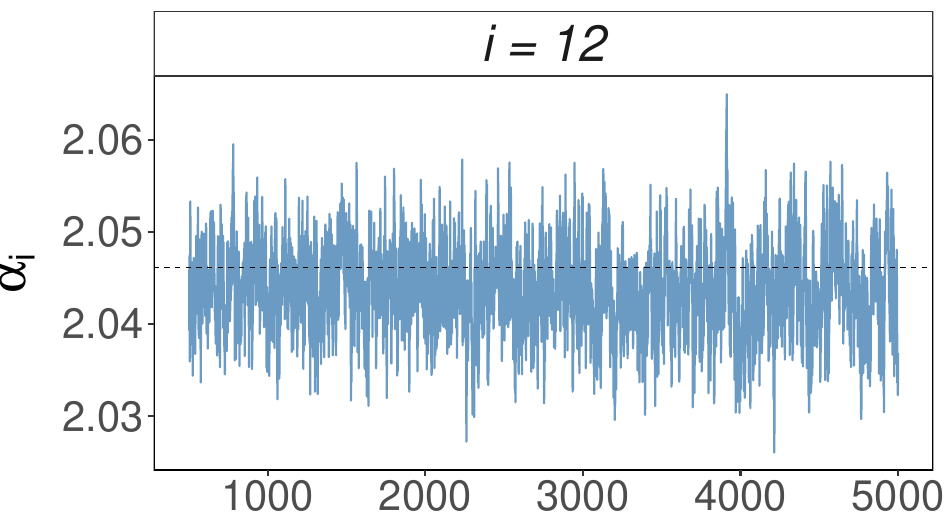} \\
\end{tabular}
\caption{Trace plots of $\alpha_i$ under the ZIP model for nodes $i\in\{2,4,6,8,10,12\}$. The dashed line represents the true value under $S_1$.}
\label{fig:trace-alpha-zip}
\end{figure}

\begin{figure}[h!t]
\centering
\setlength{\tabcolsep}{4pt}
\renewcommand{\arraystretch}{1}
\begin{tabular}{ccc}
\includegraphics[width=0.31\textwidth]{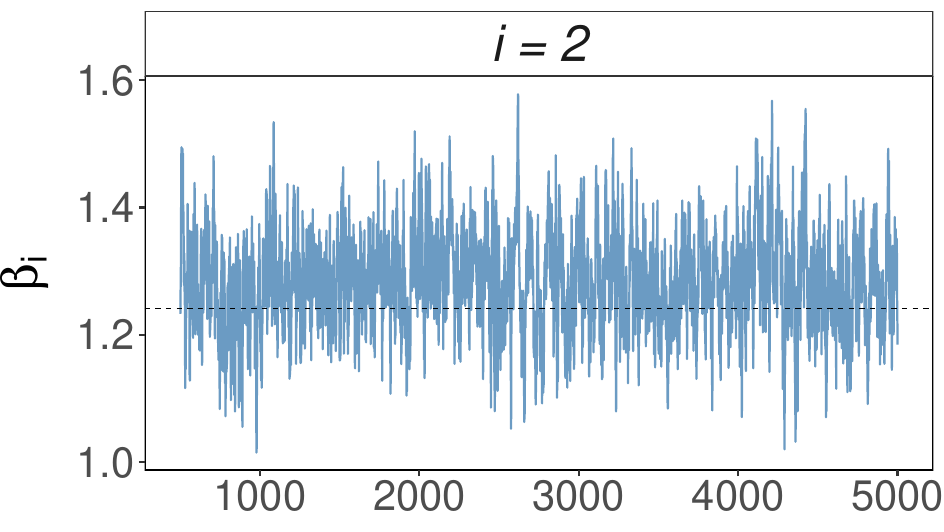} &
\includegraphics[width=0.31\textwidth]{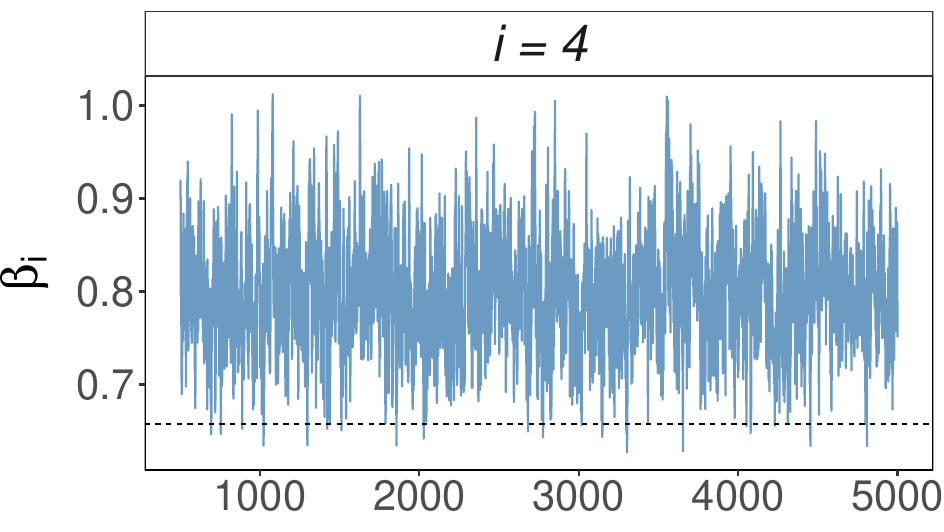} &
\includegraphics[width=0.31\textwidth]{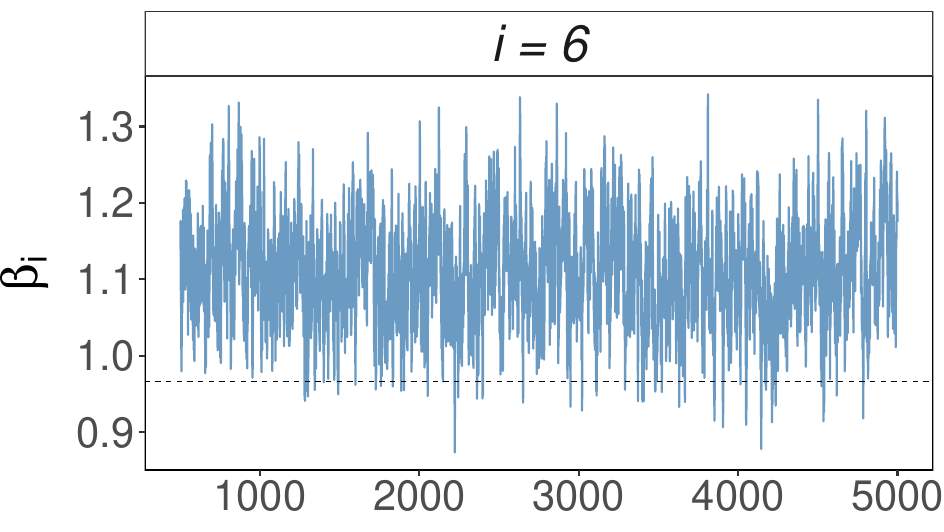} \\
\includegraphics[width=0.31\textwidth]{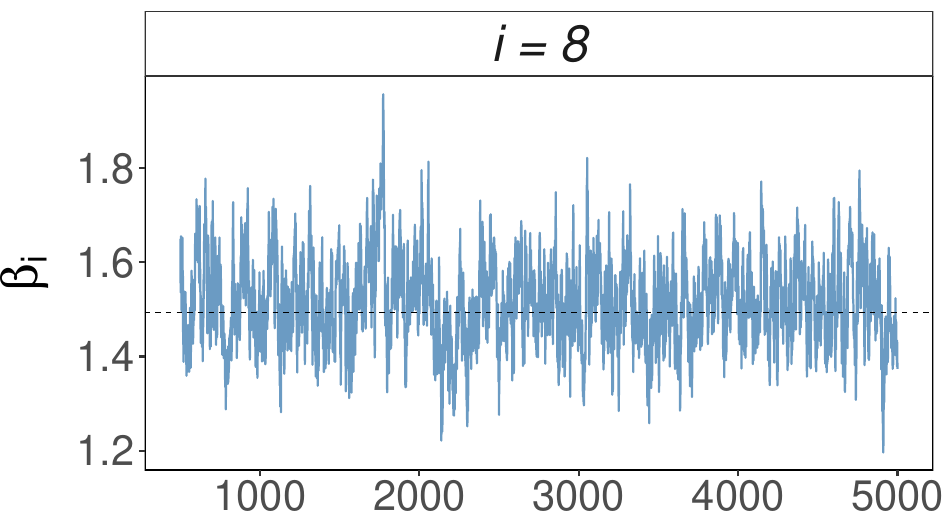} &
\includegraphics[width=0.31\textwidth]{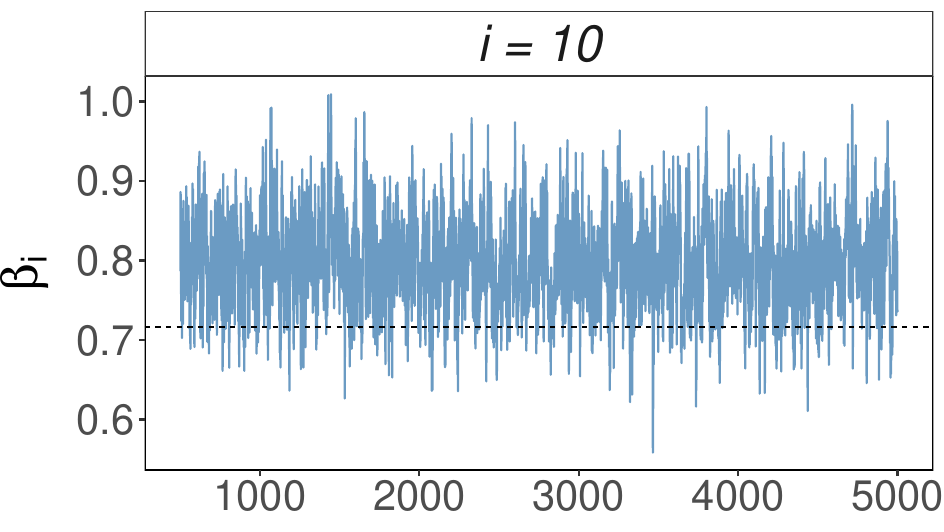} &
\includegraphics[width=0.31\textwidth]{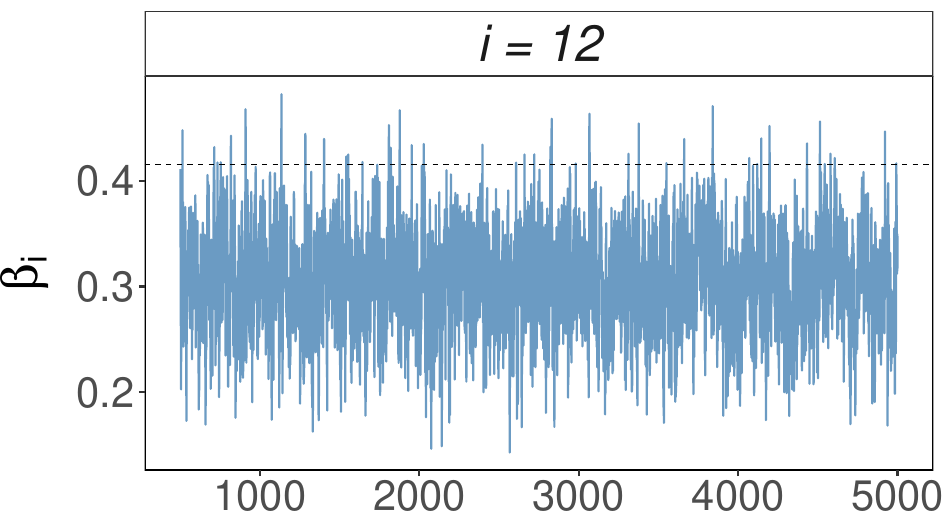} \\
\end{tabular}
\caption{Trace plots of $\beta_i$ under the ZIP model for nodes $i\in\{2,4,6,8,10,12\}$. The dashed line represents the true value under $S_1$.}
\label{fig:trace-beta-zip}
\end{figure}

\begin{figure}[h!t]
\centering
\setlength{\tabcolsep}{4pt}
\renewcommand{\arraystretch}{1}
\begin{tabular}{ccc}
\includegraphics[width=0.31\textwidth]{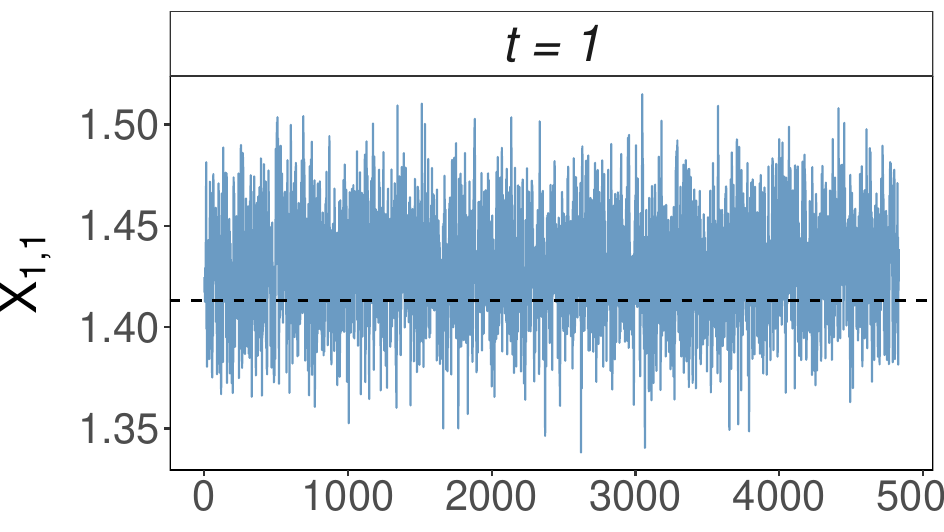} &
\includegraphics[width=0.31\textwidth]{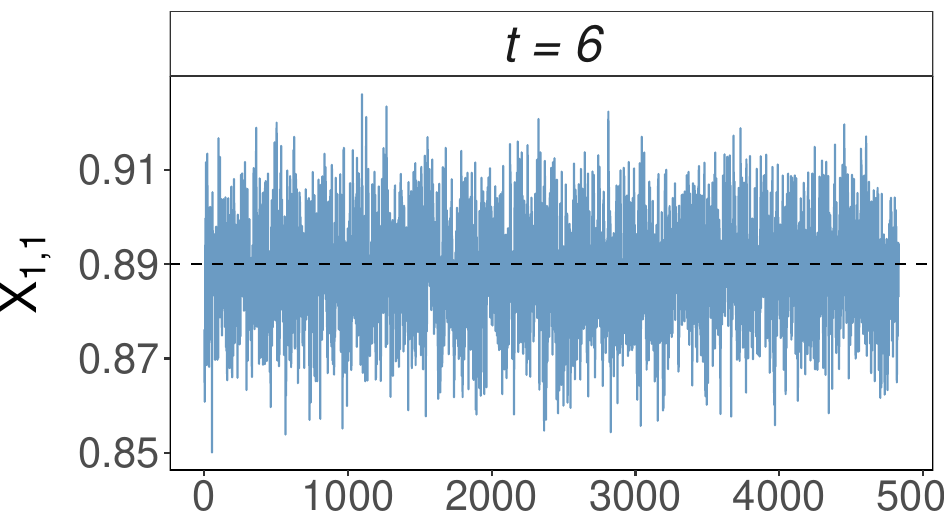} &
\includegraphics[width=0.31\textwidth]{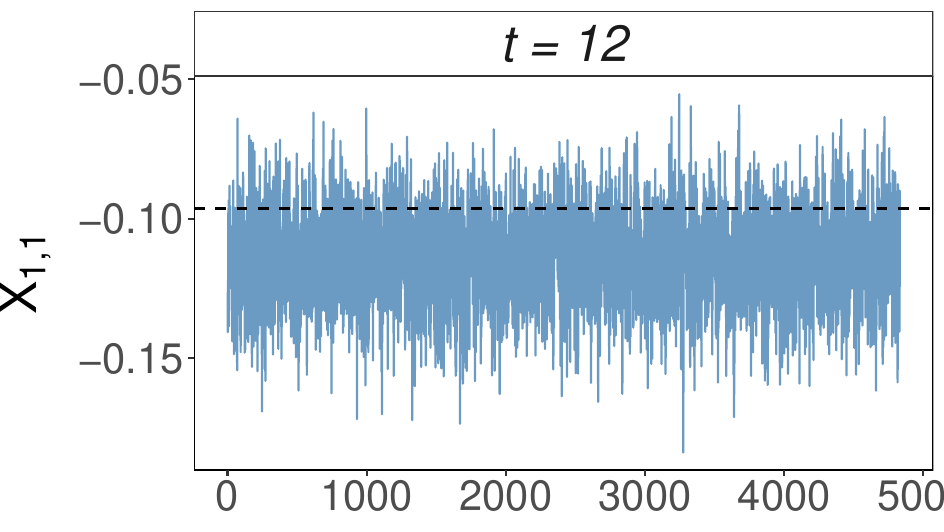} \\
\includegraphics[width=0.31\textwidth]{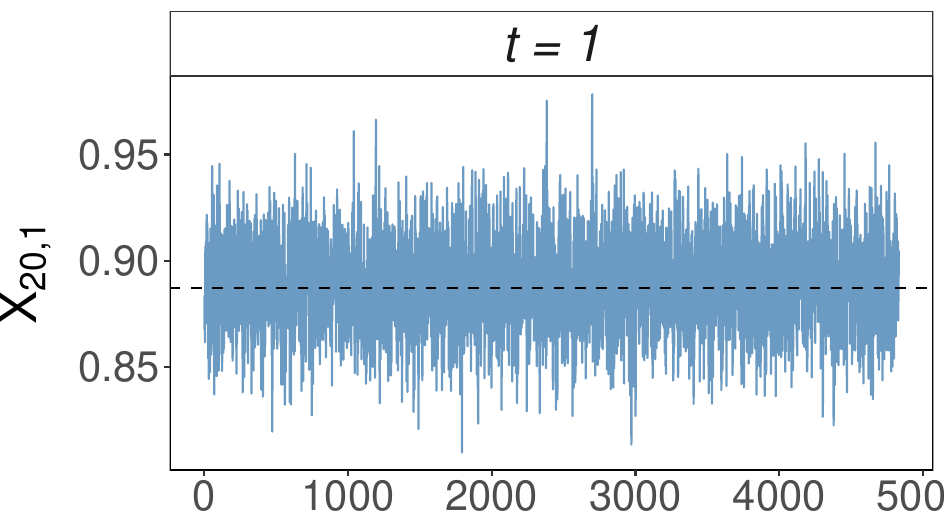} &
\includegraphics[width=0.31\textwidth]{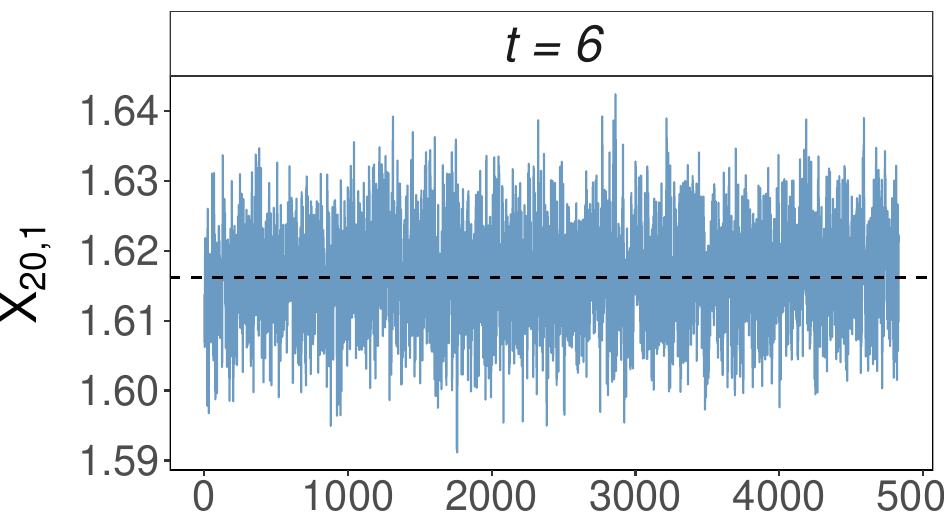} &
\includegraphics[width=0.31\textwidth]{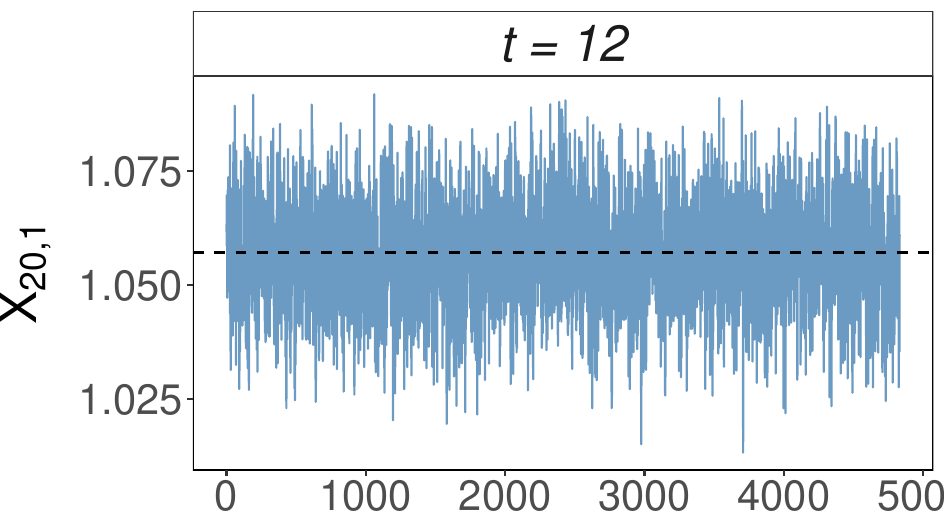} \\
\end{tabular}
\caption{Trace plots of latent coordinates $x_{1,i,t}$ (feature 1) under the ZIP model for nodes $i = 1$ (top panel) and $i = 20$ (bottom panel) at time $t\in\{1,6,12\}$. The dashed line represents the true value under $S_1$.}
\label{fig:trace-x-dim1-zip}
\end{figure}

\begin{figure}[h!t]
\centering
\setlength{\tabcolsep}{4pt}
\renewcommand{\arraystretch}{1}
\begin{tabular}{ccc}
\includegraphics[width=0.31\textwidth]{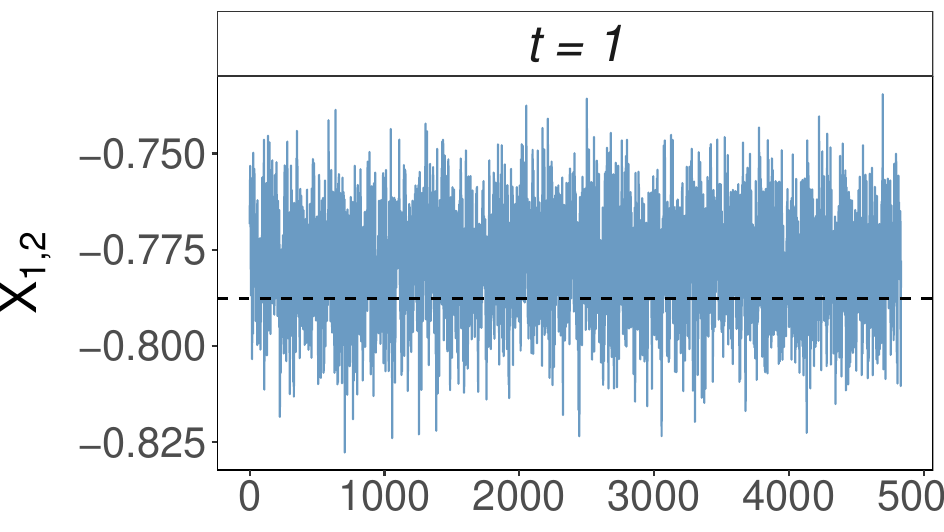} &
\includegraphics[width=0.31\textwidth]{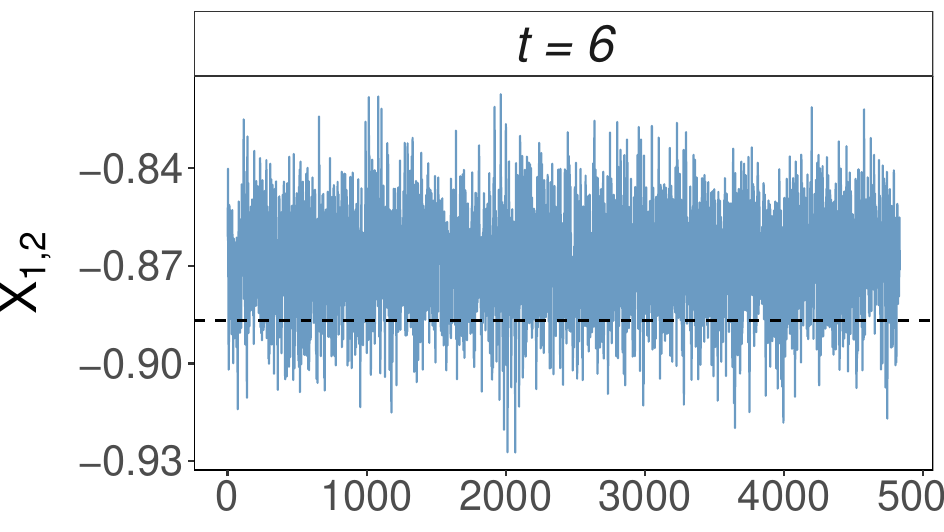} &
\includegraphics[width=0.31\textwidth]{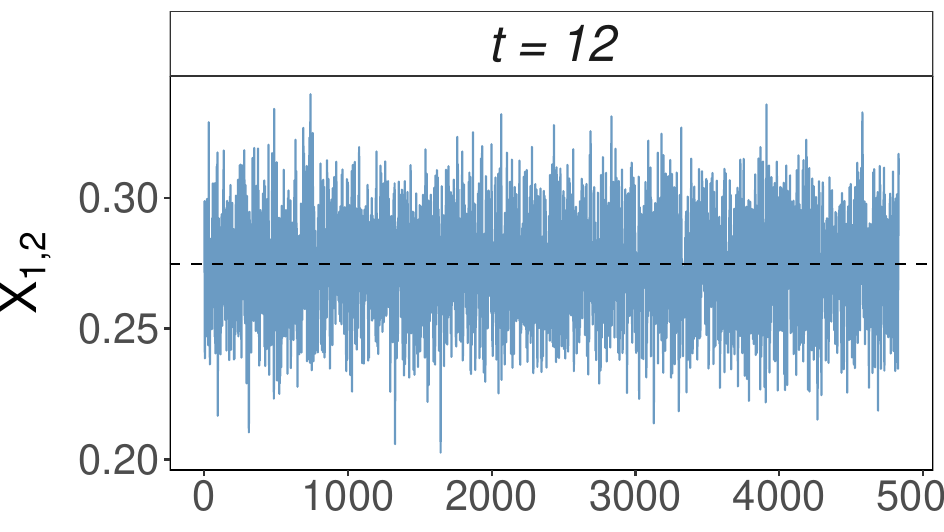} \\
\includegraphics[width=0.31\textwidth]{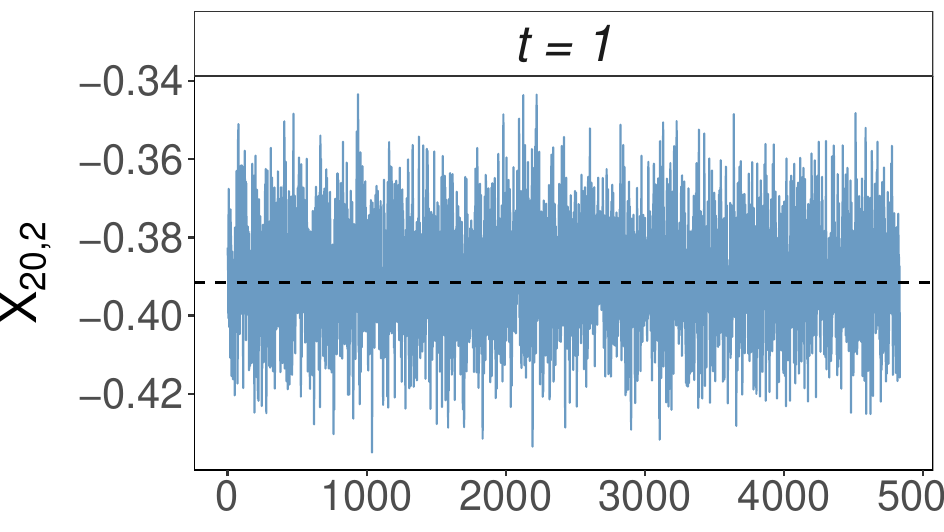} &
\includegraphics[width=0.31\textwidth]{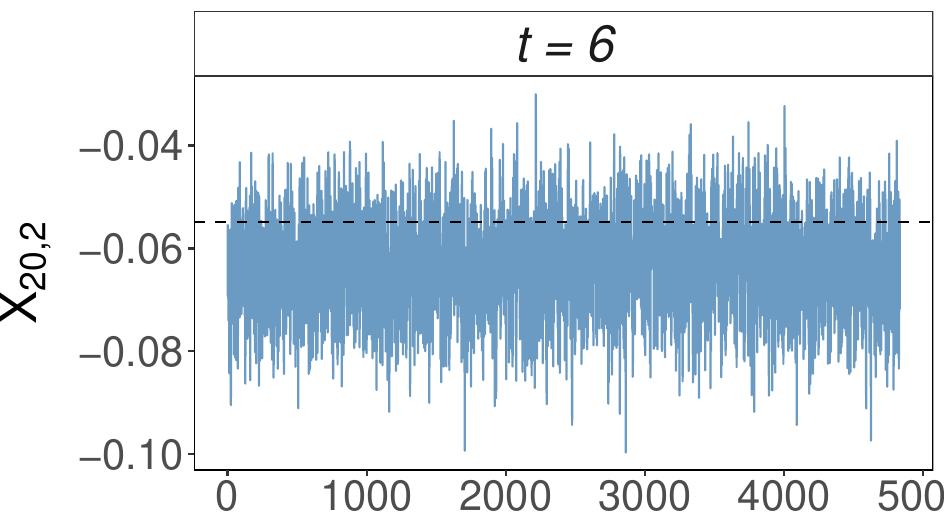} &
\includegraphics[width=0.31\textwidth]{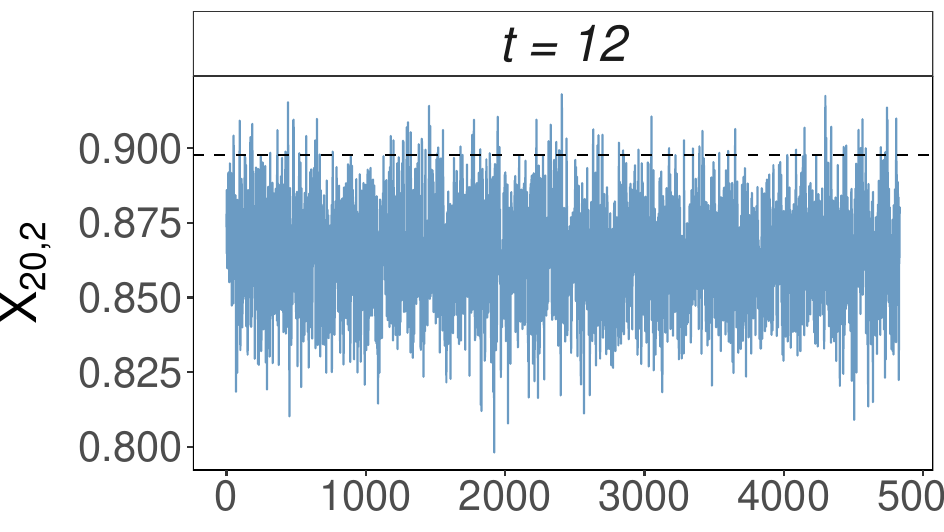} \\
\end{tabular}
\caption{Trace plots of latent coordinates $x_{2,i,t}$ (feature 2) under the ZIP model for nodes $i = 1$ (top panel) and $i = 20$ (bottom panel) at time $t\in\{1,6,12\}$. The dashed line represents the true value under $S_1$.}
\label{fig:trace-x-dim2-zip}
\end{figure}

For illustration purposes, we report in Figure \ref{fig:trace-alpha-zip}, \ref{fig:trace-beta-zip} ,  \ref{fig:trace-beta-zip}, \ref{fig:trace-x-dim1-zip} and \ref{fig:trace-x-dim2-zip} the trace plots (without any thinning) of the aforementioned parameters for some of the nodes.

\begin{figure}[t]
  \centering
  %\everyrow{\tabucline[0.4pt]-}
\setlength{\tabcolsep}{1pt}   % less horizontal space
\renewcommand{\arraystretch}{0.5} % less vertical space
\resizebox{0.9\textwidth}{!}{ 
\begin{tabular}{cccc}
  \includegraphics[width = 0.20\textwidth]{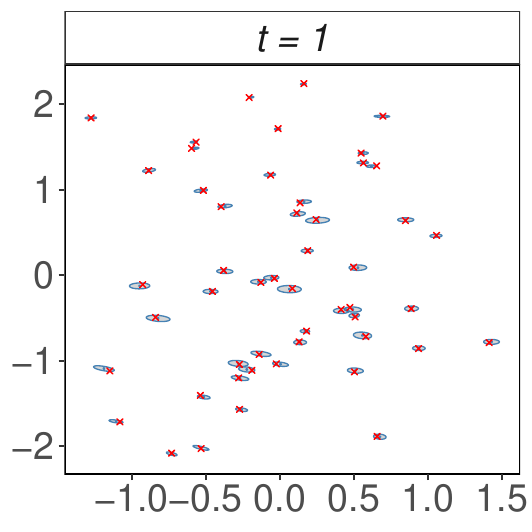} &
 \includegraphics[width = 0.20\textwidth]{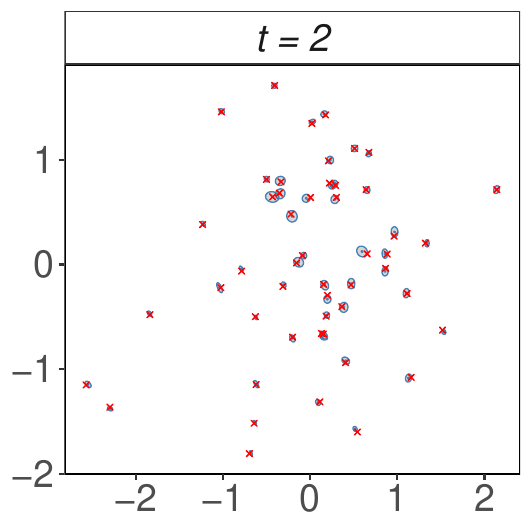} &
\includegraphics[width = 0.20\textwidth]{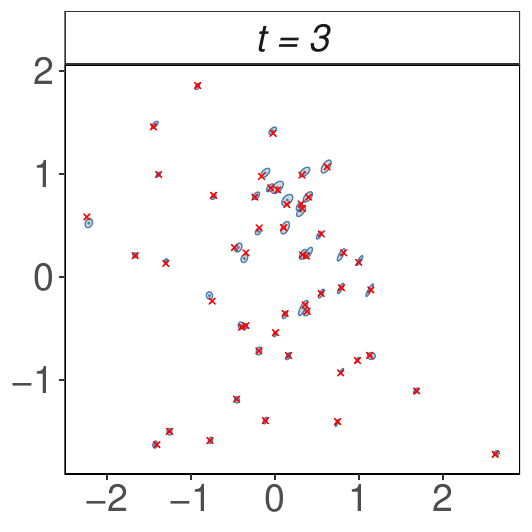} &
\includegraphics[width = 0.20\textwidth]{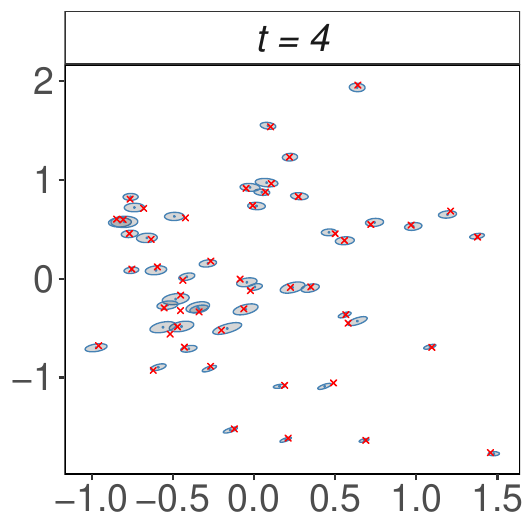} \\
\includegraphics[width = 0.20\textwidth]
{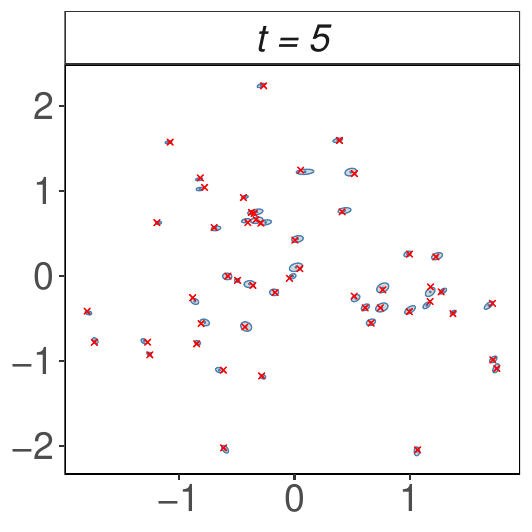} &
 \includegraphics[width = 0.20\textwidth]{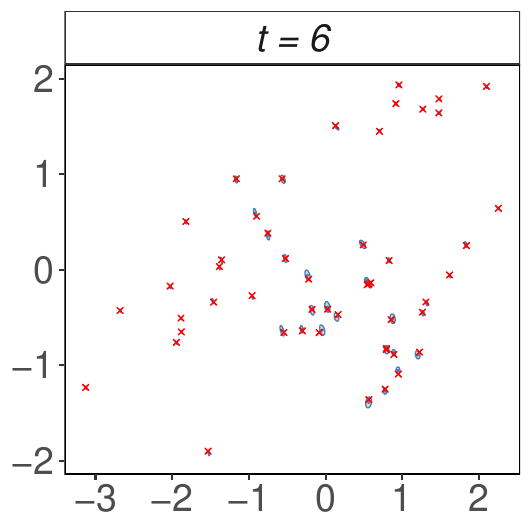} &
\includegraphics[width = 0.20\textwidth]{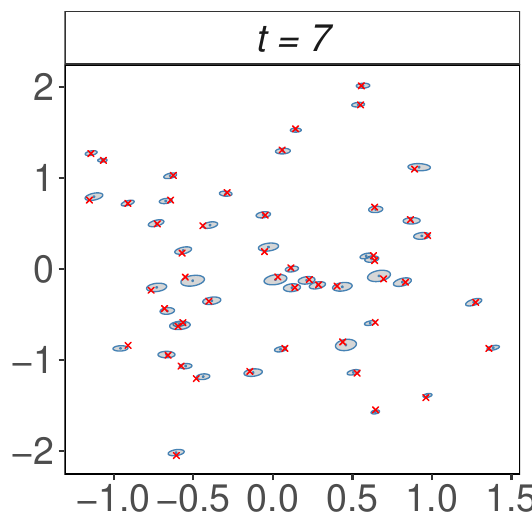} &
\includegraphics[width = 0.20\textwidth]{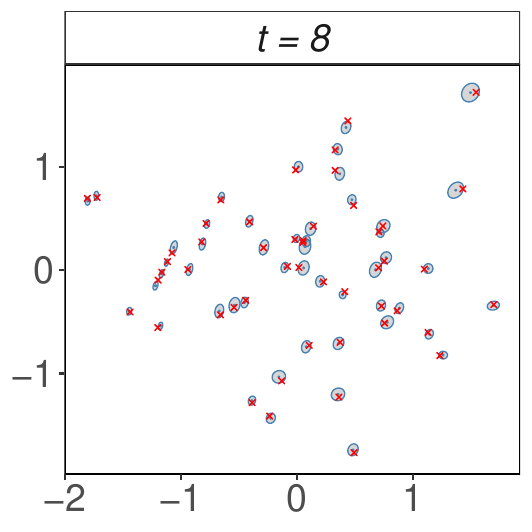} \\
\includegraphics[width = 0.20\textwidth]
{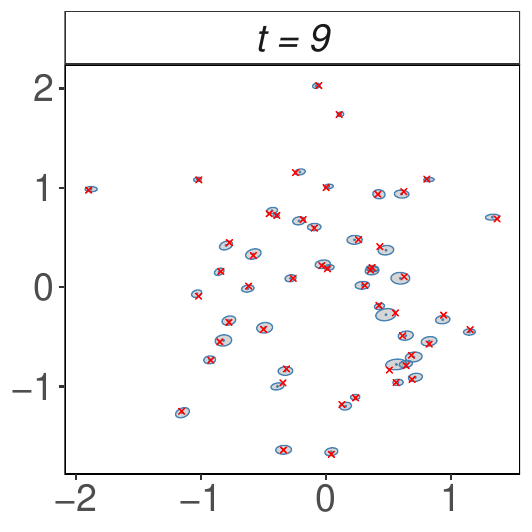} &
 \includegraphics[width = 0.20\textwidth]{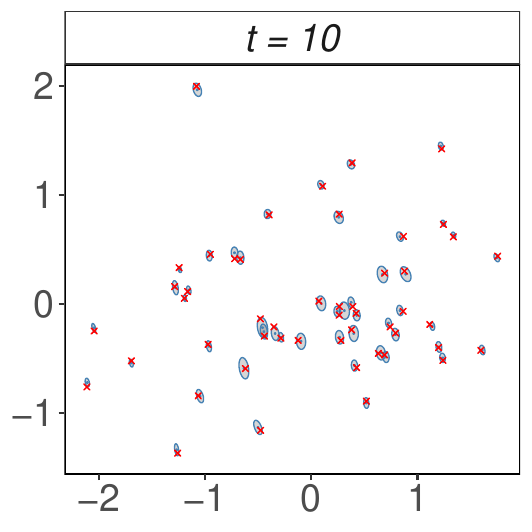} &
\includegraphics[width = 0.20\textwidth]{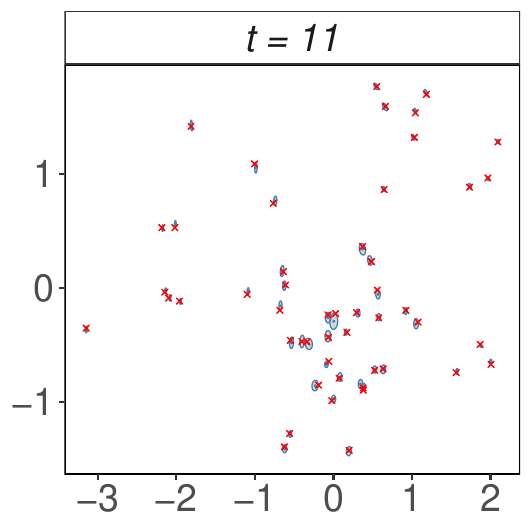} &
\includegraphics[width = 0.20\textwidth]{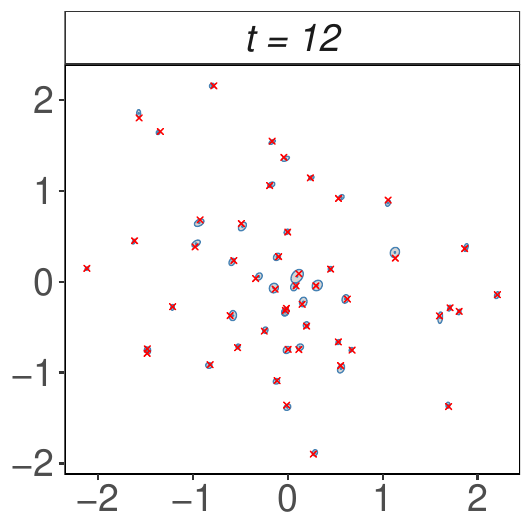}\end{tabular}}
  \caption{Latent space representation across time. Each panel corresponds to a time point  $t=1,\dots ,12$. Red crosses indicate the true latent positions, blue dots represent the posterior means, and blue ellipses denote the posterior credible regions. %The plots illustrate the ability of the model to recover the underlying latent positions over time.
  }
  \label{fig:planeplot2}
\end{figure}

\begin{figure}[t!h]
  \centering
  %\everyrow{\tabucline[0.4pt]-}
\resizebox{0.9\textwidth}{!}{ 
\begin{tabular}{cc}
  \includegraphics[width = 0.45\textwidth]{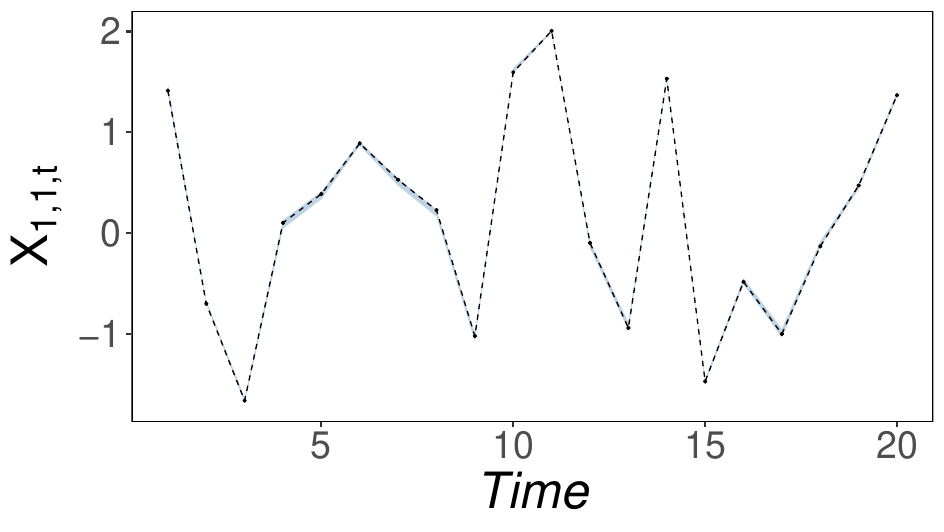} &
 \includegraphics[width = 0.45\textwidth]{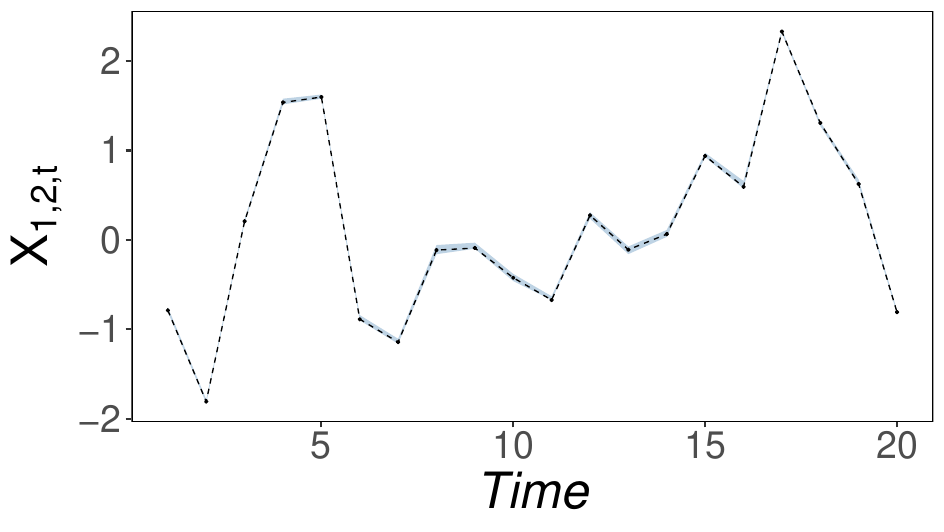} \\
\includegraphics[width = 0.45\textwidth]{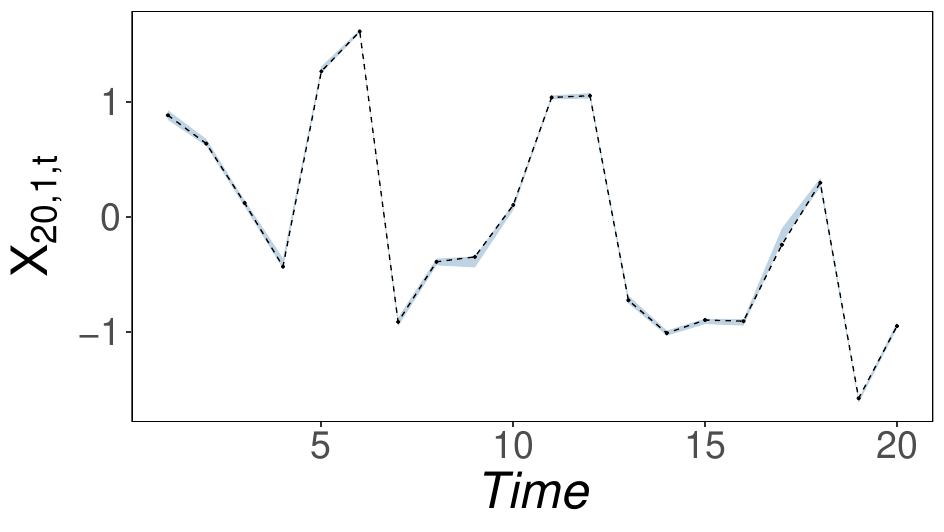} &
\includegraphics[width = 0.45\textwidth]{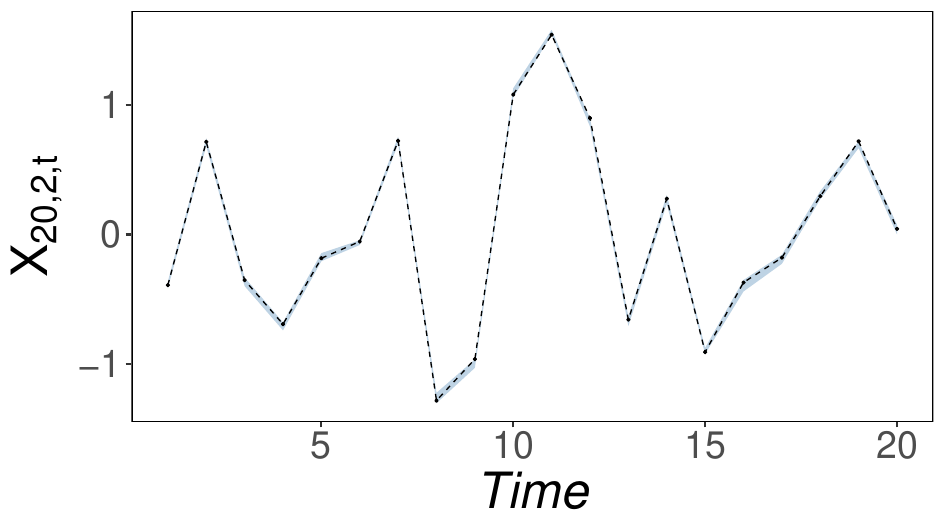} 
\end{tabular}}
\caption{Dynamics of the latent coordinates for nodes 1 and 20. Each panel reports the temporal evolution of one latent coordinate: the top row corresponds to node 1 ($x_{1,1,t}$ and $x_{1,2,t}$), and the bottom row to node 20 ($x_{20,1,t}$ and $x_{20,2,t}$). The black dashed line denotes the true trajectory, while the shaded red area indicates the associated 95\% posterior credible intervals.}
\label{fig:latent-dynamics2}
\end{figure}

\clearpage

\subsection{Comparison with Latentnet}

We compare a static implementation of our MCMC algorithm for Poisson latent space (LS) models,
\begin{equation*}
    \log \lambda_{ij} = \alpha_i + \alpha_j + \bx_{i:}^\prime \bx_{j:},
\end{equation*}
which exploits the IAMS data-augmentation scheme, with the MCMC implementation adopted in the \texttt{latentnet} R package by \citet{handcock2008fitting}. This comparison assesses the accuracy of the IAMS data-augmentation scheme and the mixing of the chains.

The simulated dataset is generated from a data-generating process in which $\alpha_i \overset{iid}{\sim} \mathcal{N}(\alpha_i \mid 2,\,0.2^2)$ for $i = 1,\dots,N$, and $\bx_{:\ell} \overset{iid}{\sim} \mathcal{N}(\bx_{:\ell} \mid \boldsymbol{0},\,\Sigma)$ for $\ell = 1,\dots,d$, with $N = 40$, $d = 3$, and $\Sigma$ is the block-diagonal variance-covariance matrix defined as in Section~\ref{sec:simulation_further}. We run both algorithms (\texttt{latentnet} and our IAMS-based algorithm) for 5000 iterations, using the first 1000 as a burn-in.

\begin{figure}[t!h]
\centering
\begin{tabular}{cc}
\includegraphics[width=0.48\textwidth]{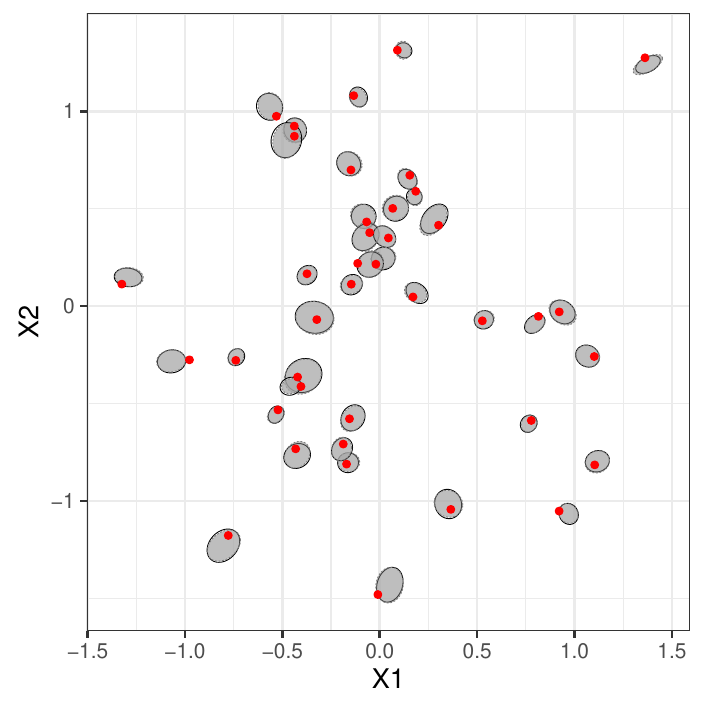} &
\includegraphics[width=0.48\textwidth]{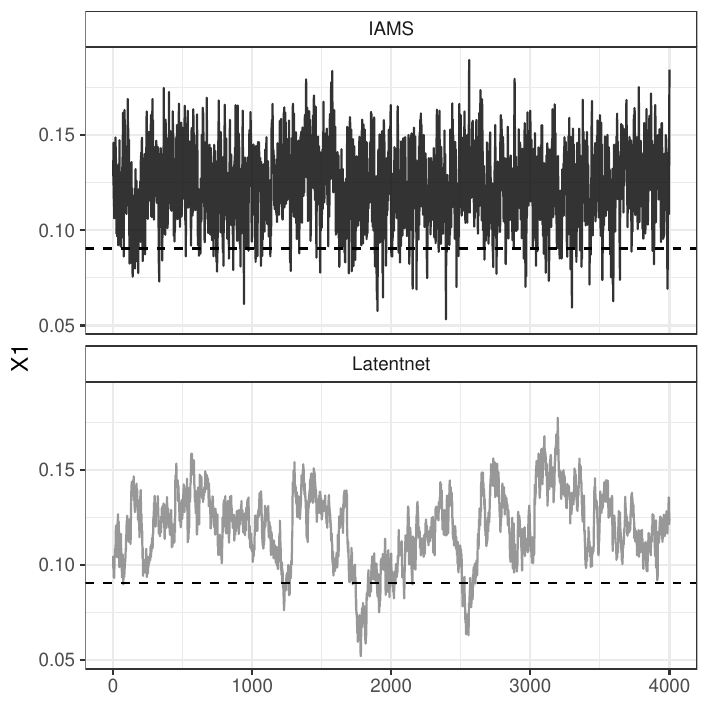}
\end{tabular}
\caption{Comparison between \texttt{latentnet} and IAMS. The left panel displays the 95\% confidence ellipses of the estimated first and second latent coordinates under IAMS (in black) and \texttt{latentnet} (in grey) with red full dots representing the true values. The right panel shows the MCMC traces for the first latent coordinate of node 40 from the two chains.}
\label{fig:latentnet_iams_comparison}
\end{figure}

Figure~\ref{fig:latentnet_iams_comparison} presents a comparison of the estimation results obtained using \texttt{latentnet} and our IAMS-based sampler. The left panel reports confidence ellipses for the estimated latent positions under IAMS (black) and \texttt{latentnet} (grey). In both cases, the true latent positions are accurately recovered, and the area of the posterior draws overlaps. The right panel displays the MCMC trace plots for the first latent coordinate of node~40 from the two samplers. Consistent with the theoretical motivations of IAMS, the proposed algorithm exhibits substantially improved mixing relative to \texttt{latentnet}, which exploits Adaptive Metropolis steps.

Table \ref{tab:ess_latent_coordinates} reports the effective sample size of each latent coordinate averaged across nodes. On average, our algorithm provides ESS that are one order of magnitude higher in terms of mixing.

\begin{table}[t!h]
\centering
\renewcommand{\arraystretch}{1.2}
\begin{tabular}{lccc}
\hline
Sampler & \%ESS($\bx_1$) & \%ESS($\bx_2$) &\%ESS($\bx_3$) \\
\hline
IAMS       & 45.2 & 40.1 & 41.8 \\
latentnet & 3.6 & 3.7 & 4.4 \\
\hline
\end{tabular}
\caption{Average \% effective sample size (ESS) with total number of MCMC iterations ($4000$) for each latent coordinate across nodes. Results are reported for the proposed IAMS sampler and for \texttt{latentnet}.}
\label{tab:ess_latent_coordinates}
\end{table}

\clearpage
%%%%%%%%%%%%%%%%%%%%%%%%%%%%%%%%%%%%%%%%%%%%%%%%%
\section{Real-Data Applications: Further Results}
\label{sec:applicaiton_further}

\subsection{Inner-product LSM interpretation and circular projection}\label{sec:circular_projection}

As noticed also in \citet{ma2020universal}, the latent space inner-product term depends on squared Euclidean distances and on vector magnitudes. Using the identity
\begin{equation}
    \bx_{i:,t}' \bx_{j:,t}
= \frac{1}{2} \big(\|\bx_{i:,t}\|^2 + \|\bx_{j:,t}\|^2 - \|\bx_{i:,t}- \bx_{j:,t}\|^2\big),
\end{equation}
the linear predictor in an inner-product latent space model may be rewritten as a sum of node-specific magnitude effects, $\tfrac12\|\bx_{i:,t}\|^2 + \tfrac12\|\bx_{j:,t}\|^2$, and a proximity term, $-\tfrac12\|\bx_{i:,t} - \bx_{j:,t}\|^2$, which depends solely on the squared Euclidean distance between latent positions. We can thus rewrite the intensity parameter in the Poisson component as:
\begin{equation}
\lambda_{ijt} = \exp\Big( \alpha_i +\alpha_j + \frac{1}{2}\big(\|\bx_{i:,t}\|^2 + \|\bx_{j:,t}\|^2 - \|\bx_{i:,t} - \bx_{j:,t}\|^2\big) \Big),
\end{equation}
This suggests that the specification accounts not only for homophily but also for additional node-level heterogeneity.

Moreover, recall that, for $d$-dimensional vectors $\bx_{i:,t}$ and $\bx_{j:,t}$, the cosine similarity between them is defined as
\begin{equation}
    \cos(\theta_{ij,t}) = \frac{\bx_{i:,t}' \bx_{j:,t}}{\|\bx_{i:,t}\|\, \|\bx_{j:,t}\|},
\end{equation}
which implies $\bx_{i:,t}' \bx_{j:,t} = \|\bx_{i:,t}\|\,\|\bx_{j:,t}\| \cos(\theta_{ij,t})$, showing that the inner-product model jointly captures angular similarity and vector magnitudes. In the special case where latent vectors are constrained to have unit norm, $\|\bx_{i:,t}\| = \|\bx_{j:,t}\| = 1$, the squared Euclidean distance and cosine similarity are directly related by
\begin{align*}
    \|\bx_{i:,t}-\bx_{j:,t}\|^2 & = 2\big(1-\cos(\theta_{ij,t})\big),
    & \bx_{i:,t}'\bx_{j:,t} & = \cos(\theta_{ij,t})
    % = 2\big( 1 - \bx_i' \bx_j \big),
\end{align*}
so that ordering dyads by inner product, cosine similarity, or squared Euclidean distance is equivalent up to a monotone transformation.

\subsection{Application to UN Co-voting Networks}
\label{sec:applicaiton_un_sup}

%\begin{figure}[t!h]
%\centering
%\resizebox{0.85\textwidth}{!}{
%\begin{tabular}{cc}
%\includegraphics[width=0.36\textwidth]{Figures/Application3/Barplot_d.pdf} & 
%\includegraphics[width=0.64\textwidth]{Figures/Application3/logPost_all.pdf}
%\end{tabular}}
%\caption{
%Latent dimensions selection. Left: Posterior distribution of $d$. Right: Log-posterior MCMC trace plot for each $d\in\mathcal{D}$. The higher $d$ values, the darker the colour.}
%\label{fig:d_estim_app_un}
%\end{figure}

\begin{figure}[p]
\centering
\begin{tabular}{cc}
\multicolumn{2}{c}{\footnotesize (a) Posterior distribution of $d$} \\
\includegraphics[width=0.4\textwidth]{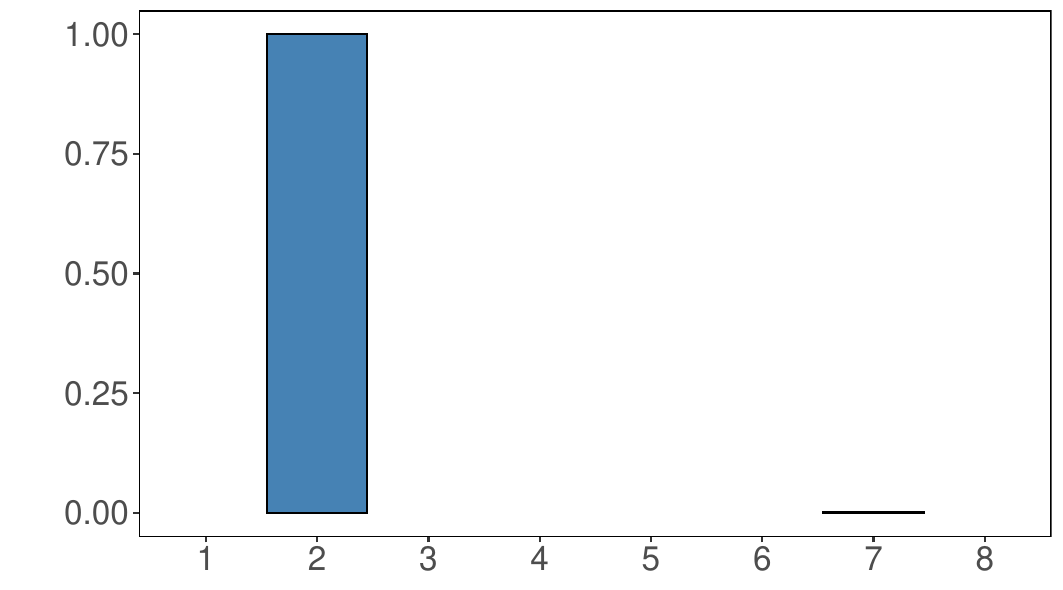} & 
\includegraphics[width=0.4\textwidth]{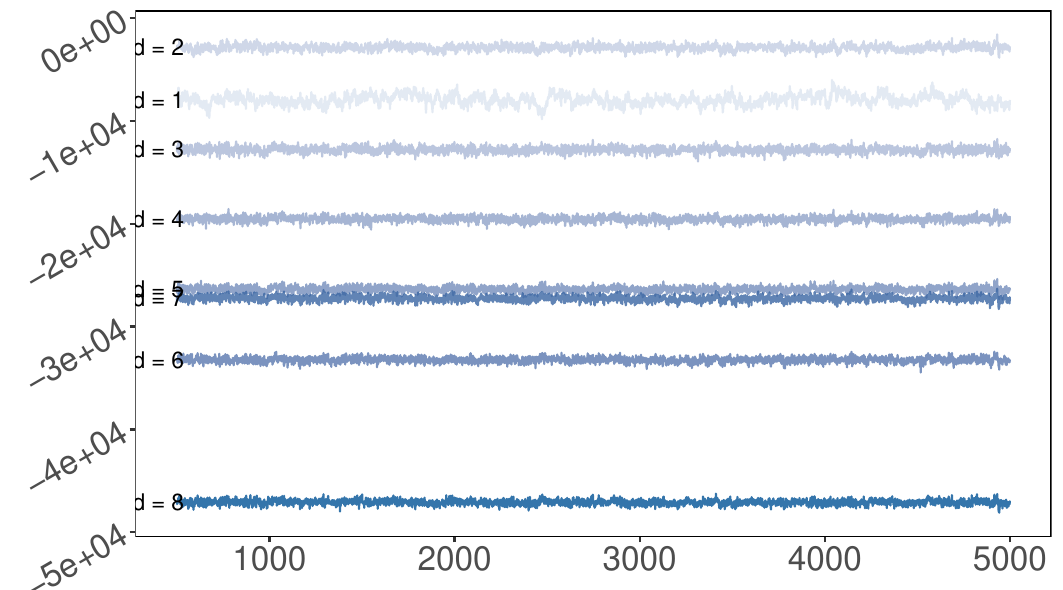}\\
\multicolumn{2}{c}{\footnotesize (b) Latent space} \\
\footnotesize 2014 & \footnotesize 2024  \\
\includegraphics[width=0.35\textwidth]{Figures/Application3/Circular_2014_only.pdf} &
\includegraphics[width=0.35\textwidth]{Figures/Application3/Circular_2024_only.pdf}\\
\multicolumn{2}{c}{\footnotesize (c) Temporal and contemporaneous dependence} \\
\raisebox{0.5\height}{
  \includegraphics[width=0.35\textwidth]{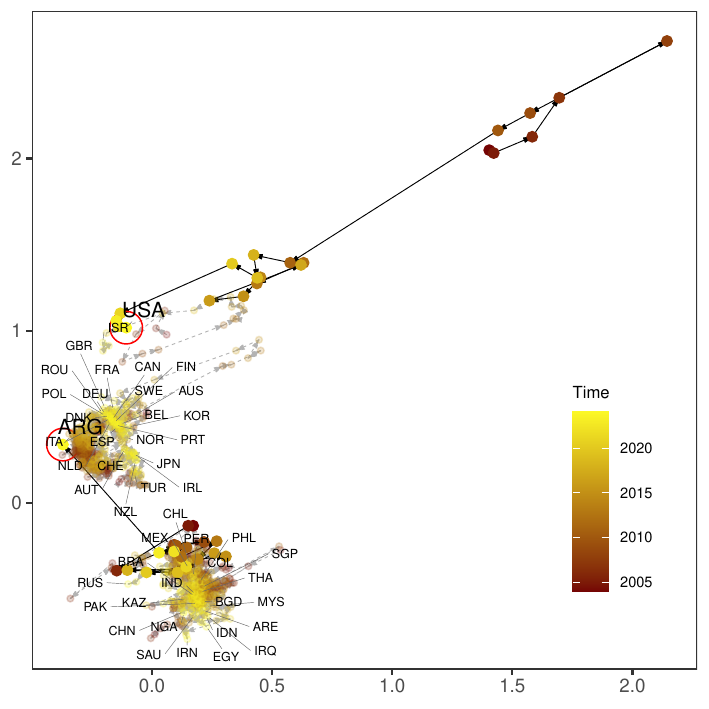}
} &\raisebox{0.5\height}{
  \includegraphics[width=0.35\textwidth]{Figures/Application3/Cor_ZIP.pdf}}
\end{tabular}
\vspace{-2.5cm}
\caption{UN General Assembly co-voting network. (a) $d\in\mathcal{D}$ posterior distribution (left) and its log-posterior MCMC trace plot (right). (b) Circular projection of the latent space representation for years 2014 (left) and 2024 (right) with the country size proportional to the posterior mean of $\alpha_i$. (c) Country trajectories (left) and node correlation posterior mean (right). Trajectories indicate the temporal evolution of the posterior mean, with lighter node colours denoting more recent years.}
\label{fig:plane_time_pol}
\end{figure}

Panel (a) of Figure~\ref{fig:plane_time_pol} reports the estimated posterior probability of the number of latent dimensions, which concentrates on $\widehat{d} = 2$. The strong peak is due to large differences in the log-posterior for different values of $d$ (right panel).

\begin{figure}[t]
\centering
\begin{tabular}{c}
   \footnotesize World Focus \\
    \includegraphics[width= 0.6\textwidth]{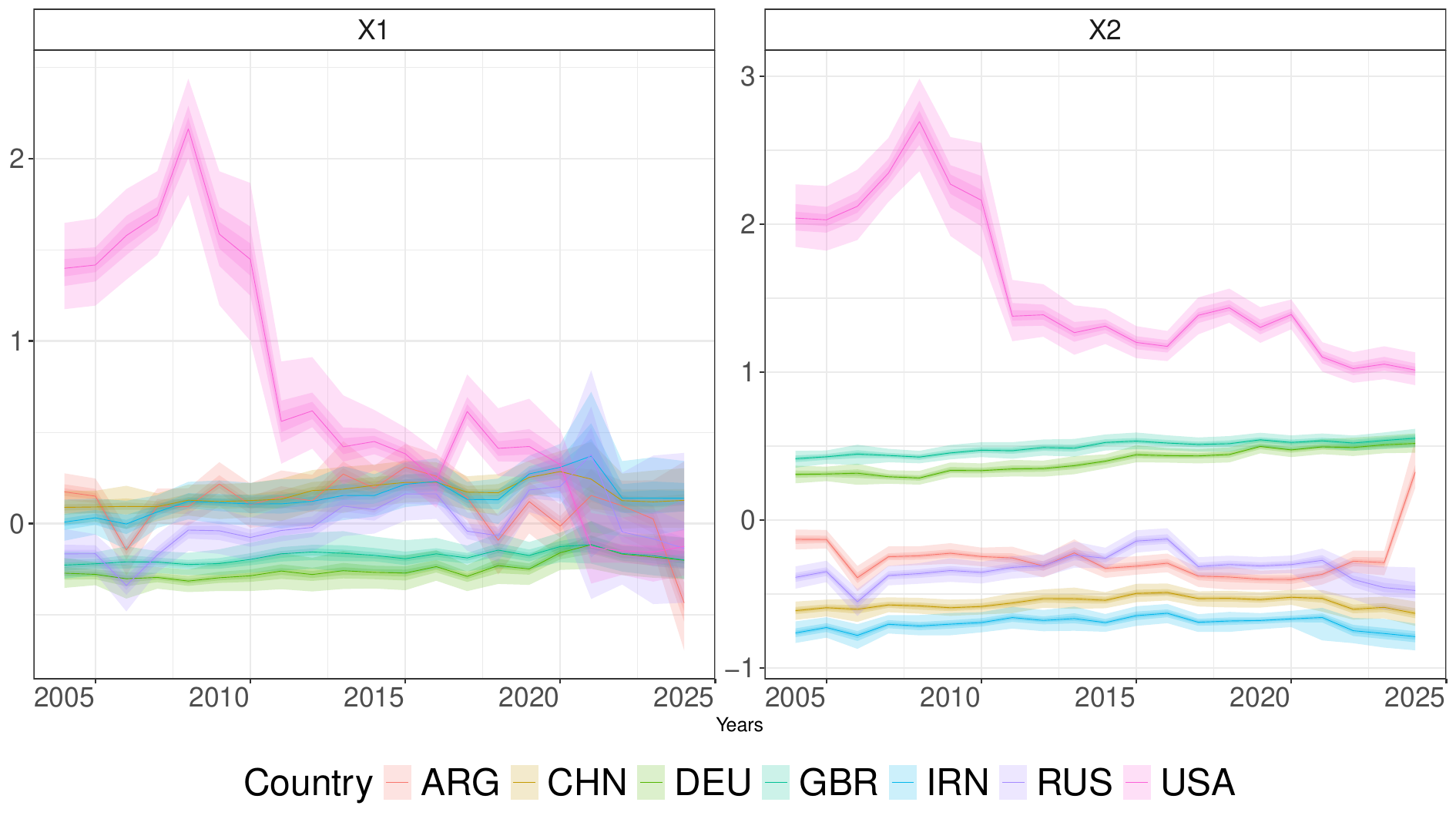} \\
     \footnotesize Europe Focus\\
     \includegraphics[width= 0.6\textwidth]{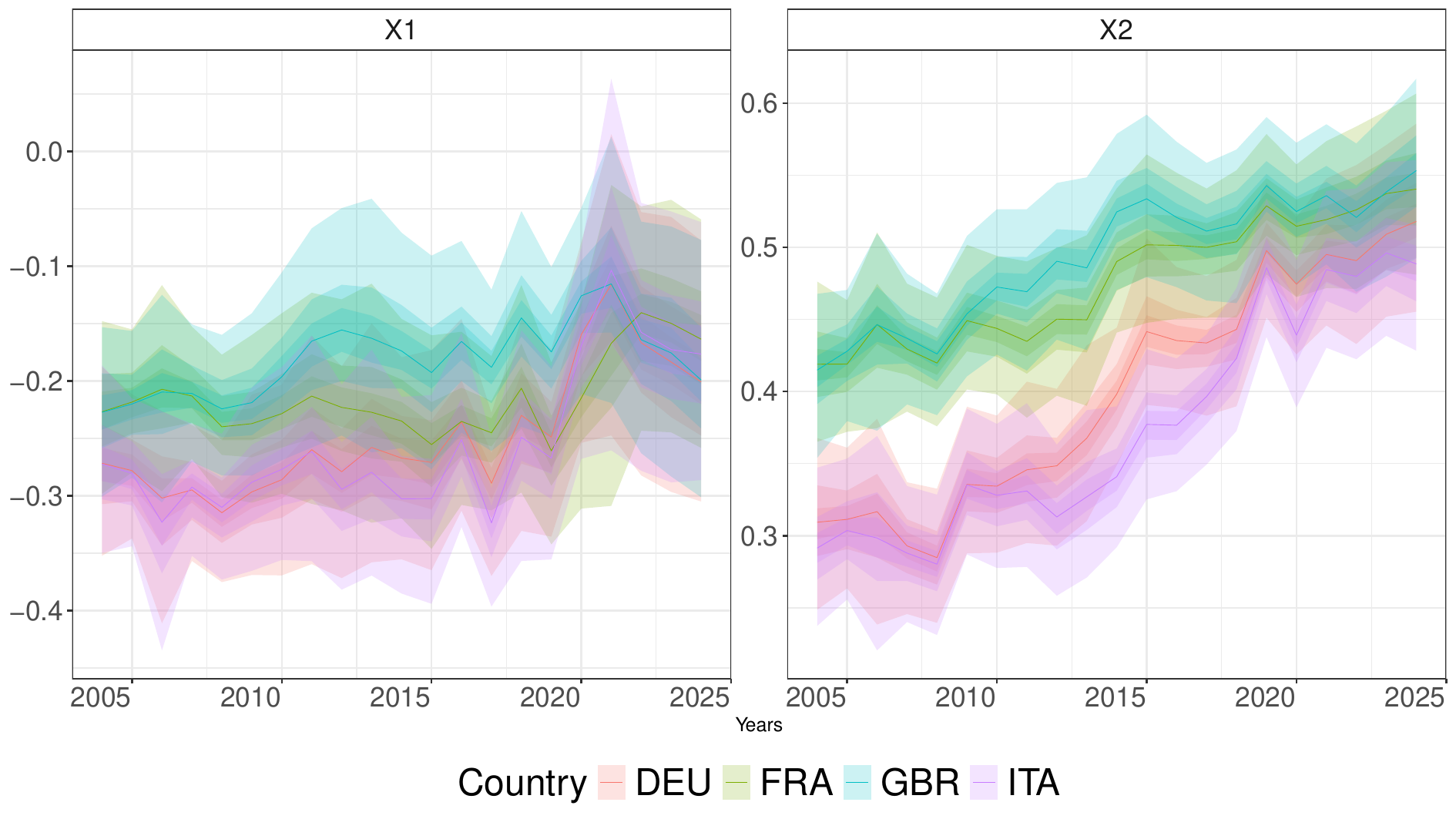}
\end{tabular}
\caption{Evolution of countries’ latent coordinates over time for Argentina, China, Germany, Great Britain, Iran, Russia, and the U.S. (left panel) and for France, Germany, Great Britain, and Italy (right panel). Solid lines denote posterior means, and shaded fan charts represent 90\% credible intervals. 
The left panel focuses on major world actors, while the right zooms in on European countries.}
\label{fig:latent-ts_sub_pol}
\end{figure}
Our model can capture non-trivial network topologies and highlights the presence of time-varying communities and islands. The circular projections in Panel (b) for the nodes' latent features in 2014 (left) and 2024 (right), show that two countries have a stronger co-voting relationship with latent features pointing in the same direction. We notice the emergence of two distinct clusters. The first is mostly related to \emph{NATO} and \emph{EU}, while the second one is mostly related to \emph{BRICS}.
The United States, Israel, and, to a lesser extent, Canada were detached from the rest of the \emph{Nato} group in 2014, but this distance has decreased over time. This is consistent with the breakthrough of major events, such as the war in Ukraine or the instability in the Middle East. We also observe that Argentina underwent a radical change in orientation in 2023-2024. This comes as no surprise, as President of Argentina Javier Milei explicitly re-aligned the Argentine foreign policy towards the US and \emph{NATO}.

\begin{figure}[ht!]
\centering
\setlength{\tabcolsep}{3pt} % riduce lo spazio tra le colonne
\renewcommand{\arraystretch}{1.0}
\begin{tabular}{ccc}
\includegraphics[width=0.32\textwidth]{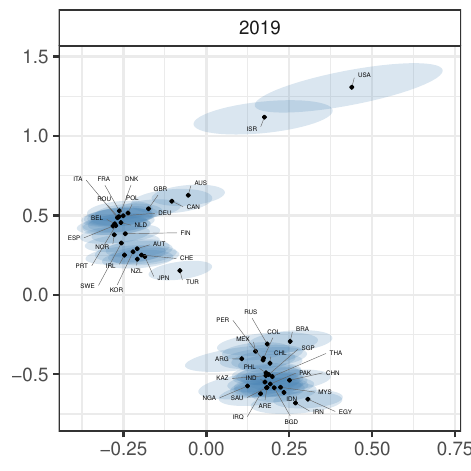} &
\includegraphics[width=0.32\textwidth]{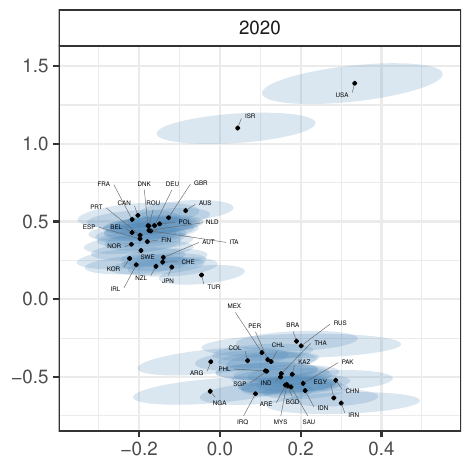} &
\includegraphics[width=0.32\textwidth]{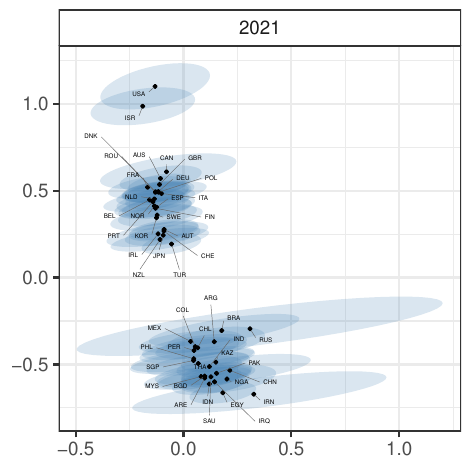} \\[-0.5em]
\includegraphics[width=0.32\textwidth]{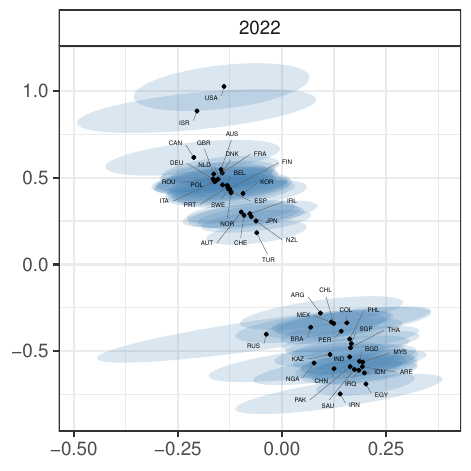} &
\includegraphics[width=0.32\textwidth]{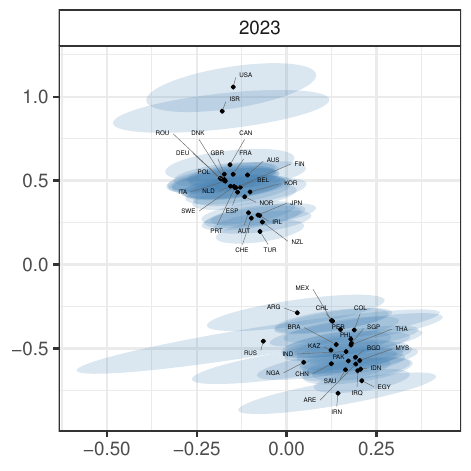} &
\includegraphics[width=0.32\textwidth]{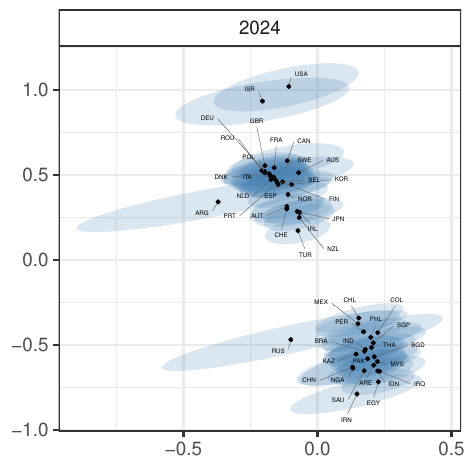} \\[-0.5em]
\end{tabular}
\caption{Latent positions $\bx_{:1,:}$ and $\bx_{:2,:}$ of the countries over time. Black dots denote posterior means, and blue ellipses denote 95\% credible regions.}
\label{fig:latent-positions1_pol}
\end{figure}

The bottom-left panel in Figure \ref{fig:plane_time_pol} reports the trajectories over time of the countries in the latent space. One can clearly see that the US is progressively realigning with the rest of the \emph{NATO} group, while Argentina moves from the \emph{BRICS} group to the \emph{NATO} group. The bottom-right panel in Figure \ref{fig:plane_time_pol} reports the posterior estimate of the cross-country correlation.
The matrix suggests a block structure which is finer than the \emph{NATO}-\emph{BRICS} bi-polarity. For instance, we identify a tightly connected China–Iran–Russia block, more loosely linked to other \emph{BRICS} members, two distinct EU blocks, and a cohesive Latin American–Asian group showing partial alignment with the United States. See Figure \ref{fig:latent-positions1_pol} for a more detailed representation.

Figure \ref{fig:latent-ts_sub_pol} reports the time evolution of the latent coordinates for two subsets of countries. The left panels illustrate the trajectories of Argentina, China, Germany, Great Britain, Iran, Russia, and the U.S. We notice that the position of Germany and Great Britain, and of China, Iran, and Russia, remain relatively stable and respectively coherent with the \emph{NATO} and \emph{BRICS} alignments. We further notice the gradual movement of the US toward the \emph{NATO} group and the sudden shift of Argentina in 2024. The right panel illustrates the trajectories of France, Germany, Great Britain, and Italy. The latent positions appeared to be consistently aligned overall. We observe a progressive closer alignment between Germany and Italy, on the one hand, and France and Great Britain, on the other, after 2010.

\FloatBarrier
\clearpage
%%%%%%%%%%%%%%%%%%%%%%%%%%%%%%%%%%%%

\subsection{Application to Trade Networks}
\label{sec:applicaiton_trade_sup}
% In this section, we apply our zero-inflated latent space model to a dynamic trade network constructed from the \textit{CEPII--BACI} dataset of international trade flows \citep{de2014network}. We consider a time series of $T = 21$ trade networks from 2004 to 2024. In these networks, each node represents one of the top-50 countries by GDP, and each edge weight represents the number of distinct HS2 product categories exchanged between countries $i$ and $j$ in year $t$. Considering trade variety rather than its economic value is of interest because it offers a different perspective on bilateral relations between countries. This measure can be considered a proxy for the extensive margin of trade \citep{hummels2005variety}, reflecting the diversity and strength of bilateral exchange relationships in the global trade system \citep{gemmetto2016multiplexity}.

\begin{figure}[p]
\centering
\setlength{\tabcolsep}{1pt}   % less horizontal 
\begin{tabular}{cc}
\multicolumn{2}{c}{\footnotesize (a) Posterior distribution of $d$}\\
\includegraphics[width=0.42\textwidth]{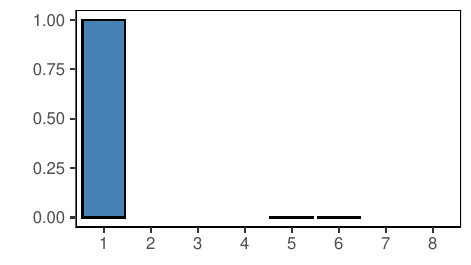} & 
\includegraphics[width=0.42\textwidth]{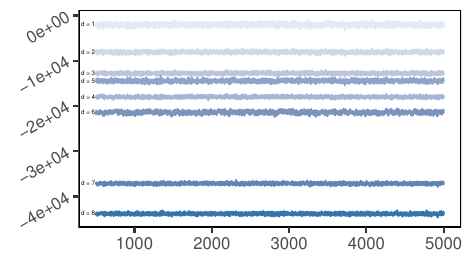}
\\
\multicolumn{2}{c}{\footnotesize (b) Latent space over time}\\
\includegraphics[width=0.40\textwidth]{Figures/Application2/planeplot1_2024pl.pdf} &\includegraphics[width=0.40\textwidth]{Figures/Application2/Plane_Plot_th_12.pdf}\\
\end{tabular}
\caption{(a) $d\in\mathcal{D}$ posterior distribution and its MCMC trace plot. (b) Latent space representations of trade networks. The left panel displays the latent space representation on the $(\bx_{:1,t}, \balpha)$ plane in 2024 with 95\% credible ellipses reported in blue. The right panel shows the latent space trajectory over time. Trajectories indicate the temporal evolution of the posterior mean of $\bx_{:1,t}$, with lighter node colours denoting more recent years.}
\label{fig:plane_time}
\end{figure}

The top-left panel of Figure~\ref{fig:plane_time} reports the estimated posterior probability of the number of latent dimensions, which concentrates on $\widehat{d} = 1$. The strong peak is due to large differences in the log-posterior for different values of $d$ (top-right panel).

Our model can capture non-trivial network topologies, such as those in the commercial trade, which exhibit a core-periphery structure, communities, and islands. The bottom-left panel of Figure ~\ref{fig:plane_time} reports the $(\bx_{:1,t}, \balpha)$-plane for the year 2024. Two countries have latent features pointing in the same horizontal direction if they trade many goods with one another relative to their total number of goods exchanged, and/or if their trade with other countries is similar. Countries with higher $\alpha_i$ are globally more connected. We find evidence of a clear, stable core comprising European countries, the United States, and a few other advanced economies that remain closely clustered throughout the sample period. Some clear clusters emerge as those related to Latin American countries (Argentina, Chile, Colombia, Mexico, and Peru).  

The model allows the study of the dynamics of relative positions in the latent space and the detection of differences in node trajectories. For example, from 2022 onward, Russia shows a displacement away from the core group, possibly reflecting trade disruptions and geopolitical fragmentation following the outbreak of the war in Ukraine.  Another notable example is Iran's progressive drift away from the core after some alternating periods. See the bottom-right panel of Figure \ref{fig:plane_time} for the trajectory of all the countries and the red circles for the more recent position of Russia and Iran.

\begin{figure}[t]
\centering
\color{red}
\begin{tabular}{cc}
    \includegraphics[width=0.5\textwidth]{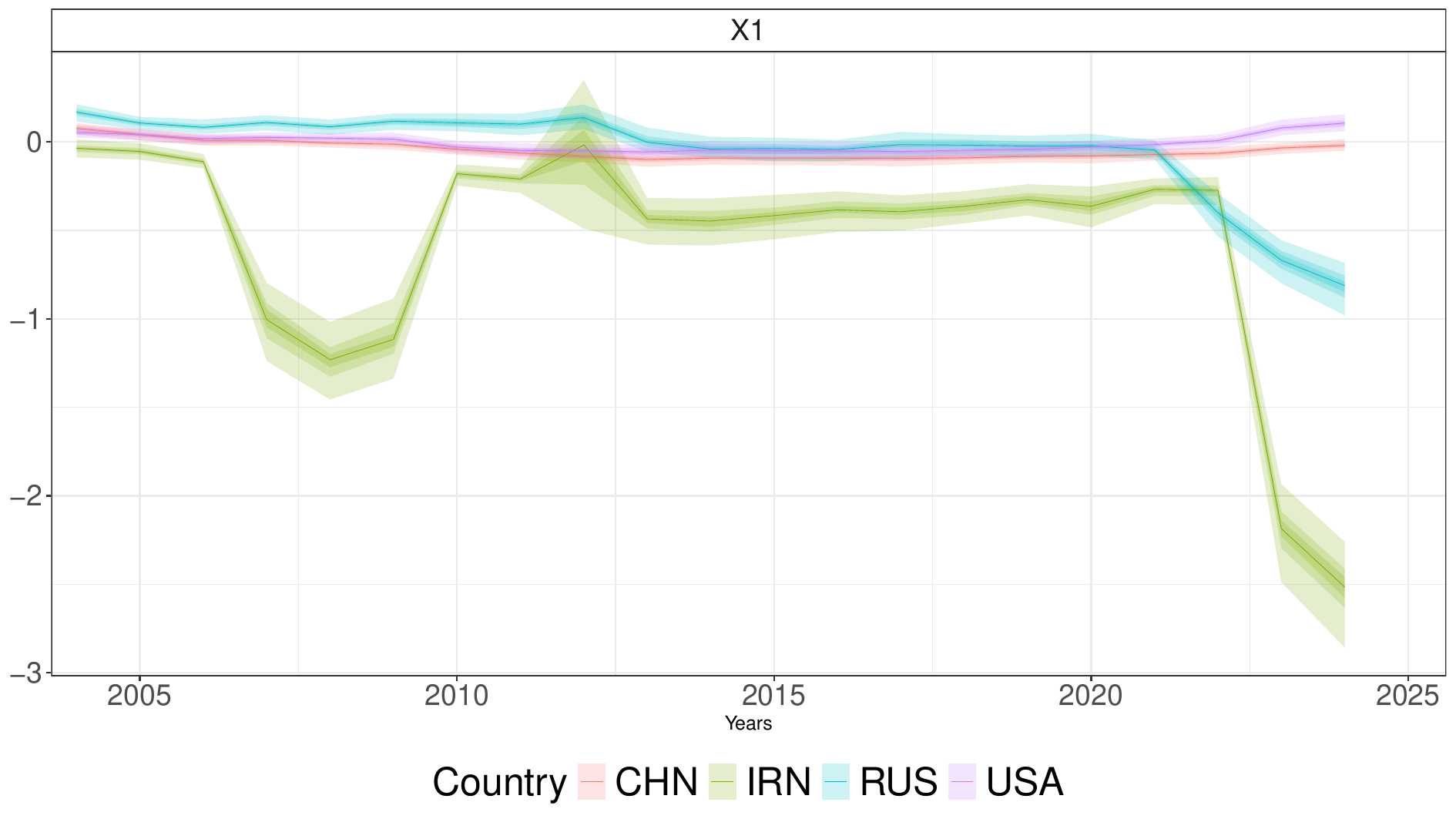} &
     \includegraphics[width= 0.5\textwidth]{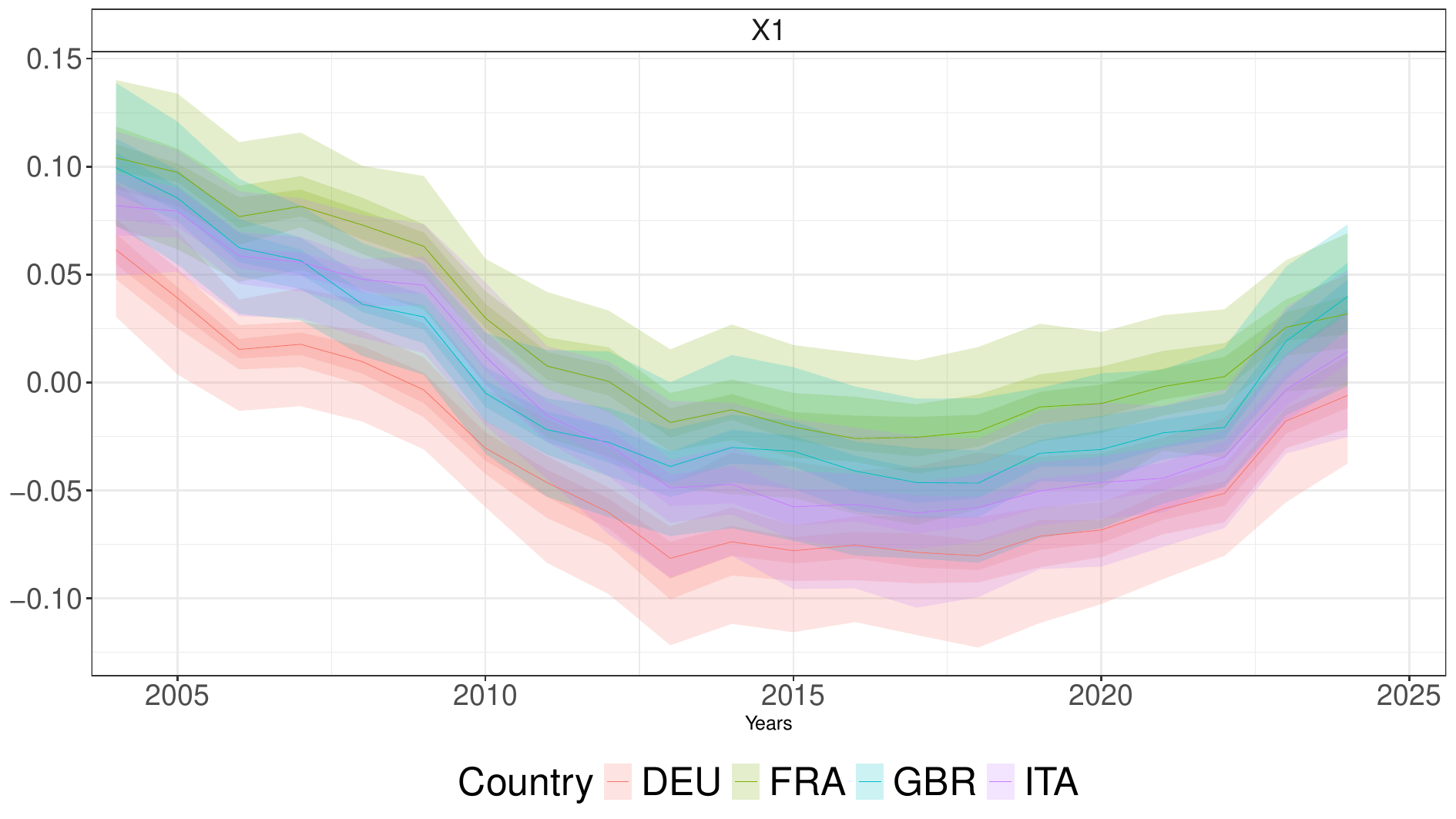}
\end{tabular}
\caption{Evolution of countries’ latent coordinate $\bx_{1:,t}$ over time for China, Iran, Russia, and the U.S. (left panel) and for France, Germany, Great Britain, and Italy (right panel) 
Solid lines denote posterior means, and shaded fan charts represent 90\% credible intervals. 
The left panel focuses on major world actors, while the right on European countries.}
\label{fig:latent-ts_sub}
\end{figure}

Figure \ref{fig:latent-ts_sub} reports the time evolution of the latent coordinate $\bx_{1:,t}$ for two subsets of countries. The top panel illustrates the trajectories of China, Iran, Russia, and the United States. The positions of the United States and China remain relatively stable over time, whereas Iran and Russia exhibit progressive detachment from the core. Iran experiences a pronounced period of separation between 2007 and 2009, followed by a phase of increasing connectivity through 2023, at which point its detachment from the core intensifies sharply. In contrast, Russia remains well integrated within the global trade network until the onset of the Ukraine conflict in 2022, after which its level of integration declines markedly. The bottom panel illustrates the trajectories of France, Germany, Italy, and the United Kingdom. Overall, the positions of these countries remain remarkably stable over time.

% \begin{figure}[t!h]
% \centering
% \setlength{\tabcolsep}{3pt}
% \begin{tabular}{c}
% World Focus \\[-0.2em]
% \includegraphics[width=0.95\linewidth]{Figures/Application2/TimePlots/timeplot_world.pdf}\\
% Europe Focus \\[-0.2em]
% \includegraphics[width=0.95\linewidth]{Figures/Application2/TimePlots/timeplot_europe.pdf} \\
% \end{tabular}
% \caption{Evolution of countries’ latent coordinates over time. 
% Each panel shows the temporal dynamics of the estimated latent dimensions. 
% Solid lines denote posterior means, and shaded fan charts represent 90\% credible intervals. 
% The left panel focuses on major world actors, while the right zooms in on European countries.}
% \label{fig:latent-ts}
% \end{figure}

Table \ref{tab:ranking} reports a comparison of the top-ten partners of China, the United States, and the United Kingdom in 2023 in terms of extensive margin (top panel) and intensive margin (bottom panel). One can observe that the ranking varies depending on whether the extensive rather than the intensive margin is considered.

\begin{table}[t!h]
\resizebox{\textwidth}{!}{
\begin{tabular}{ccccccc}
\toprule
\multicolumn{1}{l}{} & \multicolumn{2}{c}{China} & \multicolumn{2}{c}{USA}      & \multicolumn{2}{c}{United Kingdom} \\
\midrule
Rank                 & Partner    & \# Products  & Partner        & \# Products & Partner          & \# Products     \\ \hline
1                    & USA        & 4322     & Canada         & 4490    & Netherlands      & 4451.00         \\
2                    & Canada     & 4256      & Mexico         & 4359     & Ireland          & 4409         \\
3                    & Malaysia   & 4192.00      & China          & 4322    & France           & 4388.00         \\
4                    & Germany    & 4184.00      & United Kingdom & 4267    & Germany          & 4326.00         \\
5                    & Indonesia  & 4164      & France         & 4253   & USA              & 4267         \\ \hline
Rank                 & Partner    & Value        & Partner        & Value       & Partner          & Value           \\ \hline
1                    & USA        & 589834445    & Mexico         & 699866810   & USA              & 138019776       \\
2                    & Korea      & 308714106    & Canada         & 679017818   & China            & 133642120       \\
3                    & Japan      & 306763835    & China          & 589834445   & Germany          & 110953712       \\
4                    & Hong Kong  & 284785511    & Germany        & 252128763   & France           & 65863347        \\
5                    & Germany    & 268066332    & Japan          & 223269922   & Netherlands      & 57760762        \\
\bottomrule
\end{tabular}}
\caption{Top-ten trade partners in terms of variety of exchanged products in terms of HS6 (extensive margin, top panel) and in terms of traded value (intensive margin, bottom panel) for China, USA, and United Kingdom.}
\label{tab:ranking}
\end{table}

\begin{figure}[t!h]
\centering
\setlength{\tabcolsep}{3pt} % riduce lo spazio tra le colonne
\renewcommand{\arraystretch}{1.0}
\begin{tabular}{ccc}
\includegraphics[width=0.32\textwidth]{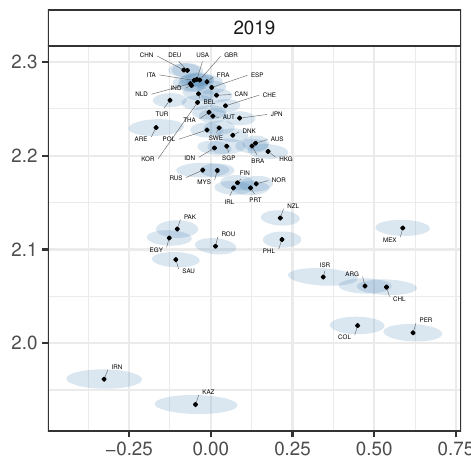} &
\includegraphics[width=0.32\textwidth]{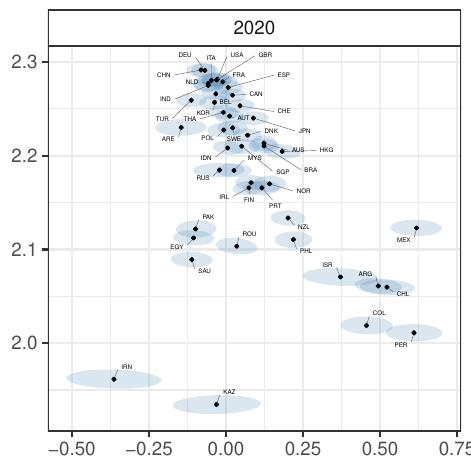} &
\includegraphics[width=0.32\textwidth]{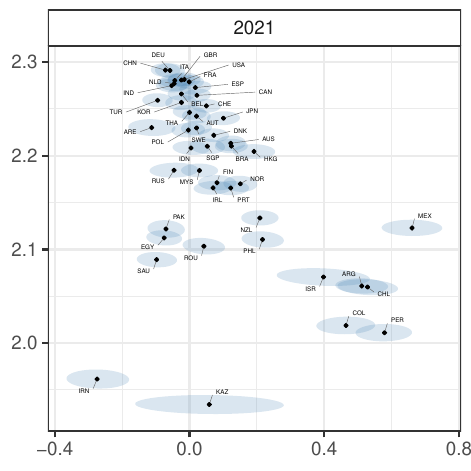} \\[-0.5em]
\includegraphics[width=0.32\textwidth]{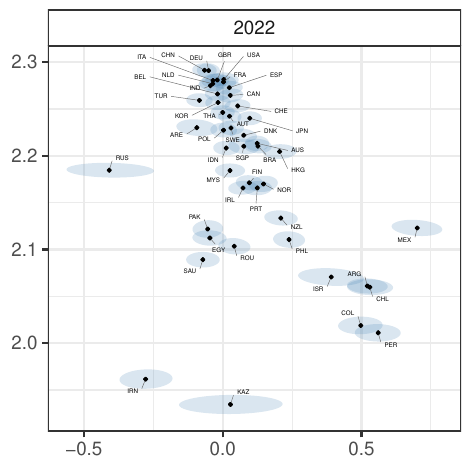} &
\includegraphics[width=0.32\textwidth]{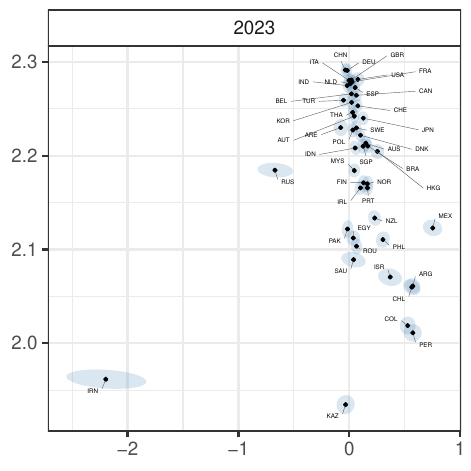} &
\includegraphics[width=0.32\textwidth]{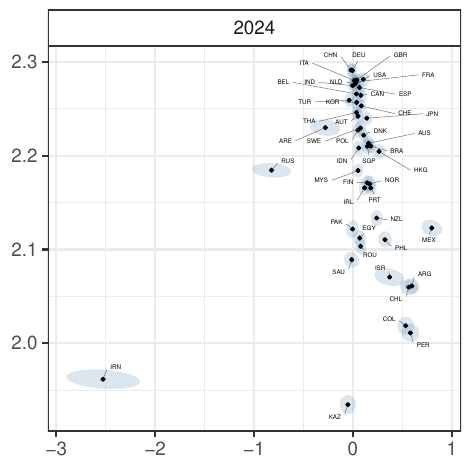} \\[-0.5em]
\end{tabular}
\caption{Latent representation $\bx_{:1,:}$ (x-axis) and $\balpha$ (y-axis) of the countries over time. Black dots denote posterior means, and blue ellipses denote 95\% credible regions.}
\label{fig:latent-positions1}
\end{figure}

Figures \ref{fig:latent-positions1} reports the $(\bx_{:1,t}, \balpha)$-plane from 2018 to 2023.  Countries that are closer in the horizontal direction have stronger relationships in terms of the number of exchanged products. 

\begin{figure}[t!h]
\centering
\setlength{\tabcolsep}{3pt} % riduce lo spazio tra le colonne
\renewcommand{\arraystretch}{1.0}
\begin{tabular}{ccc}
\includegraphics[width=0.28\textwidth]{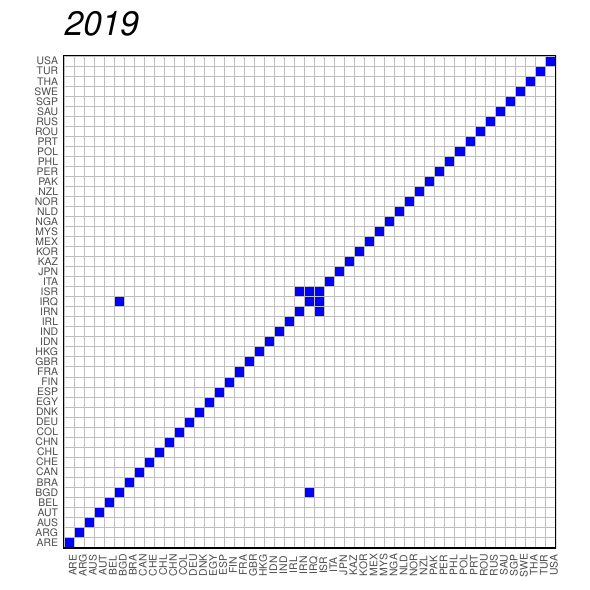} &
\includegraphics[width=0.28\textwidth]{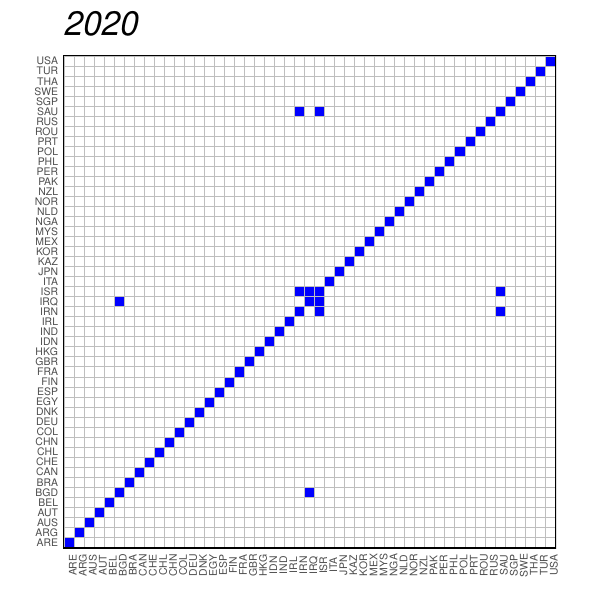} &
\includegraphics[width=0.28\textwidth]{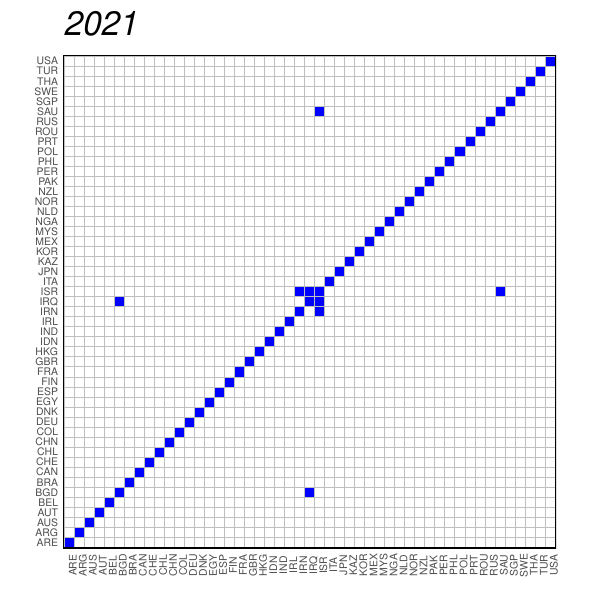} \\[-0.5em]
\includegraphics[width=0.28\textwidth]{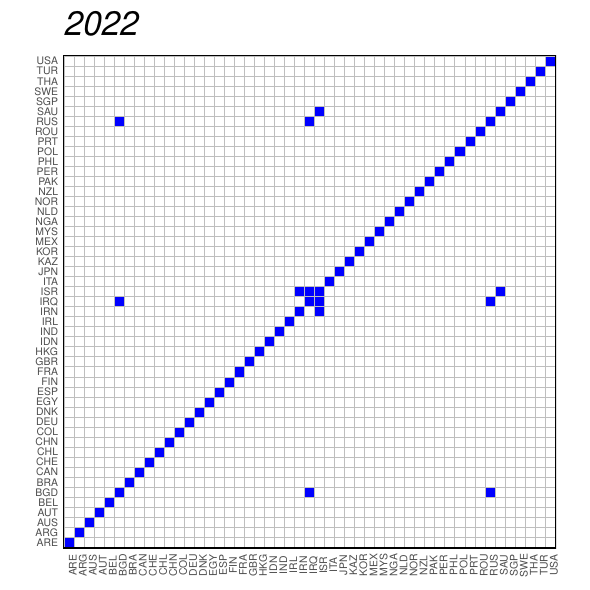} &
\includegraphics[width=0.28\textwidth]{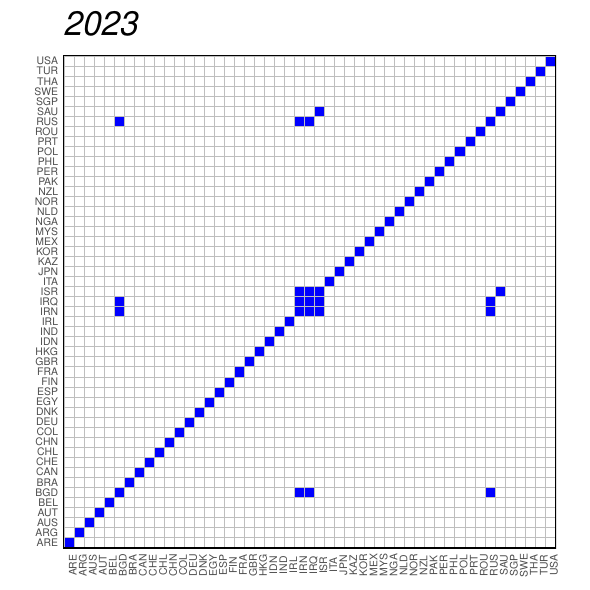} &
\includegraphics[width=0.28\textwidth]{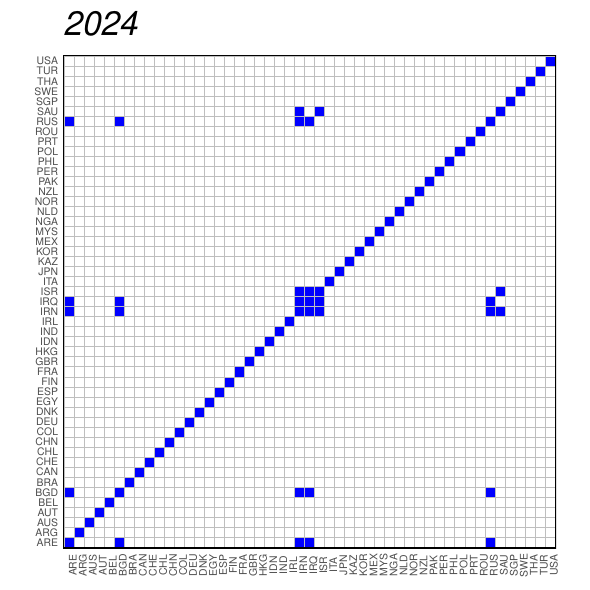} \\[-0.5em]
\end{tabular}
\caption{Structural zeros across countries over time. Lighter values denote higher posterior probability of structural zeros.}
\label{fig:zeros}
\end{figure}

Figure \ref{fig:zeros} reports the estimated $\hat{W}$ matrix of structural zeros. Each panel highlights pairs of countries whose trade relationships are estimated to be structurally absent, i.e., dyads for which the probability of observing any exchange is close to zero even after accounting for latent similarity and random fluctuations.
The pattern of zeros remains remarkably stable over time, suggesting a persistent segmentation of the global trade network.
Most of the structural absences concern peripheral or politically isolated countries. The stability of these zeros indicates that such missing links are not random absences but reflect enduring geopolitical, institutional, or economic barriers. A mild increase in structural zeros in 2022–2023 can be observed, possibly linked to the reconfiguration of trade relations following recent geopolitical tensions and sanctions.

\FloatBarrier
\clearpage

\subsection{Application to Brain Networks}

\begin{figure}[ht!]
\centering
\includegraphics[width=\textwidth]{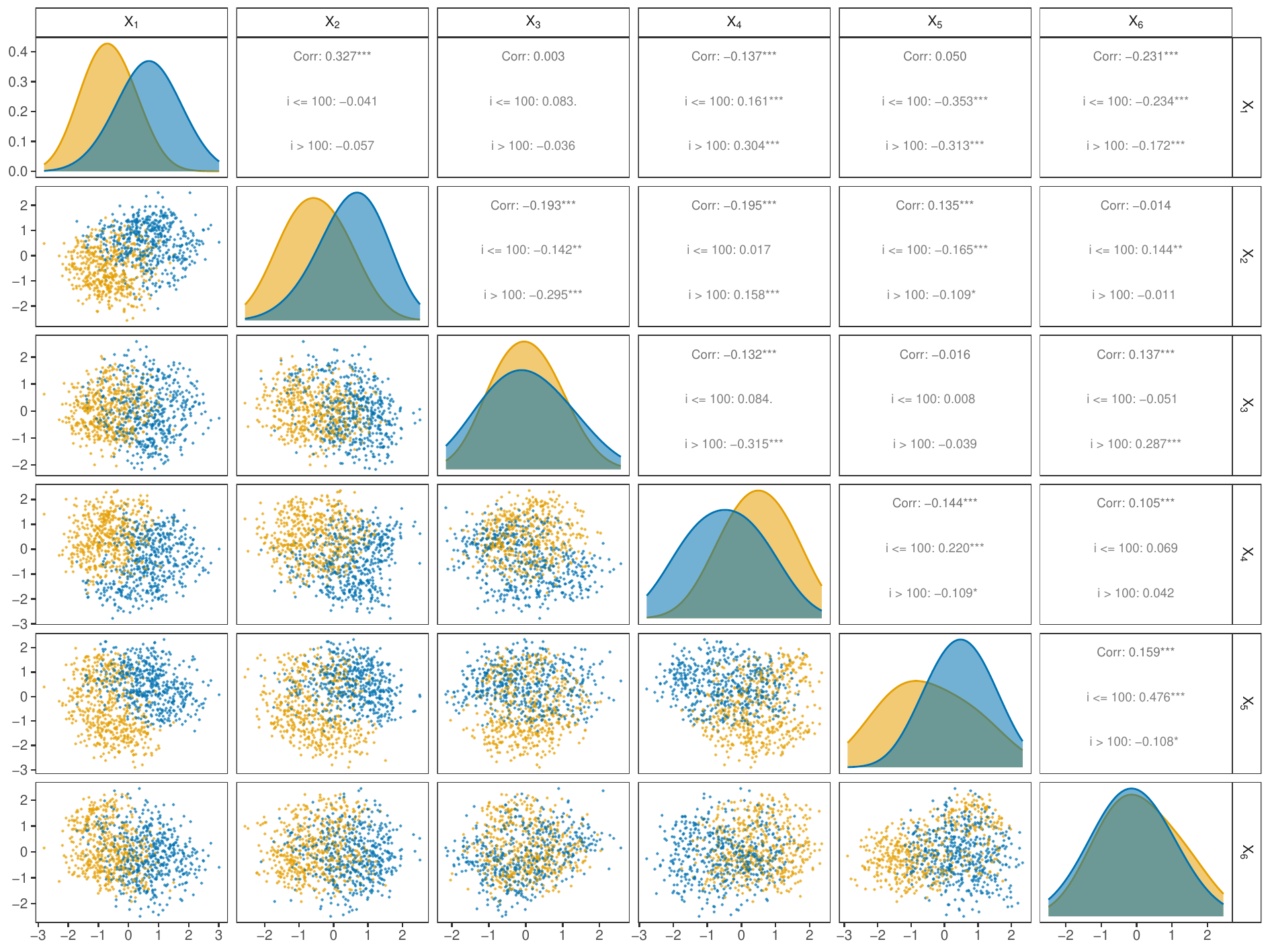}
\caption{Pairwise scatterplots and marginal posterior densities for the six latent coordinates $(\bx_{:1,:},\dots,\bx_{:6,:})$ of the latent space model. Each point corresponds to the posterior mean of a node-specific latent coordinate, with regions in the left hemisphere ($i \le 100$) shown in orange and regions in the right hemisphere ($i > 100$) shown in blue. The lower triangle reports bivariate scatterplots, the diagonal panels show kernel density estimates, and the upper triangle reports overall Pearson correlations and group-specific correlations, with significance levels indicated by stars.}
\label{fig:ls_pairs}
\end{figure}

Figure \ref{fig:ls_pairs} reports the pairwise scatterplots and marginal posterior densities for the six latent coordinates $(\bx_{:1,:},\dots,\bx_{:6,:})$. The figures highlight that while some latent coordinates capture the distinction between the two hemispheres, others, such as $\bx_{:3,:}$ and $\bx_{:6,:}$, capture more nuanced relationships across brain regions.

\newpage

\end{document}